\documentclass[useAMS,usenatbib,letterpaper]{mn2e}
\usepackage[pass]{geometry}
\usepackage{natbib}
\usepackage{amssymb,amsmath}
\usepackage{graphicx}
\usepackage{verbatim}
\usepackage{color,hyperref}
\usepackage{lastpage}
\definecolor{linkcolor}{rgb}{0,0,0.25}
\hypersetup{
  colorlinks=true,        
  linkcolor=linkcolor,    
  citecolor=linkcolor,    
  filecolor=linkcolor,    
  urlcolor=linkcolor      
}
\newcommand{\etal}{et al.}
\newcommand{\dd}{\mathrm{d}}
\newcommand{\eg}{e.g.}

\newcommand{\Eqnname}{Equation}
\newcommand{\equationname}{\Eqnname}
\newcommand{\tablename}{Table}
\newcommand{\figurename}{Figure}

\newcommand{\sectionname}{$\mathsection$}
\newcommand{\appendixname}{Appendix}

\newcommand{\Gyr}{\ensuremath{\,\mathrm{Gyr}}}
\newcommand{\kpc}{\ensuremath{\,\mathrm{kpc}}}
\newcommand{\pc}{\ensuremath{\,\mathrm{pc}}}

\newcommand{\msun}{\ensuremath{\,\mathrm{M}_{\odot}}}
\newcommand{\mas}{\ensuremath{\,\mathrm{mas}}}

\newcommand{\magunit}{\,\mbox{mag}}

\newcommand{\dens}{\ensuremath{\nu_*}}
\newcommand{\denscmd}{\ensuremath{\rho_{\mbox{\tiny CMD}}}}
\newcommand{\zsun}{\ensuremath{Z_\odot}}
\newcommand{\ra}{\ensuremath{\alpha}}
\newcommand{\dec}{\ensuremath{\delta}}
\newcommand{\plx}{\ensuremath{\varpi}}
\usepackage{yfonts}
\newcommand{\essf}{\ensuremath{\textswab{S}}}
\newcommand{\sech}{\ensuremath{\mathrm{sech}}}
\newcommand{\Gaia}{\emph{Gaia}}

\newcommand{\tgas}{\emph{TGAS}}
\newcommand{\tyctwo}{\emph{Tycho-2}}
\newcommand{\nside}{\ensuremath{N_{\mathrm{side}}}}

\def\aj{AJ}
\def\apj{ApJ}
\def\apjl{ApJL}
\def\apjs{ApJS}
\def\mnras{MNRAS}

\def\aap{A \& A}

\def\pasp{PASP}
\def\pasj{PASJ}
\def\araa{ARA\&A}

\title{Stellar Inventory of the Solar Neighborhood using \protect\emph{Gaia} DR1}
\author[Jo Bovy]{Jo Bovy\thanks{E-mail: bovy@astro.utoronto.ca}\thanks{Alfred~P.~Sloan~Fellow}\\
Department of Astronomy and Astrophysics, University of Toronto, 50 St. George Street, Toronto, ON M5S 3H4, Canada\\
and\\
Center for Computational Astrophysics, Flatiron Institute, 162 5th Ave, New York, NY 10010, USA}
\pagerange{\pageref{firstpage}--\pageref{LastPage}} \pubyear{2017}
\date{May 18, 2017}

\begin{document}
\maketitle
\label{firstpage}
\begin{abstract}
  The absolute number and the density profiles of different types of
  stars in the solar neighborhood are a fundamental anchor for studies
  of the initial mass function, stellar evolution, and galactic
  structure. Using data from \Gaia\ DR1, we reconstruct \Gaia's
  selection function and determine \Gaia's volume completeness, the
  local number density, and the vertical profiles of different
  spectral types along the main sequence from early A stars to late K
  stars as well as along the giant branch. We clearly detect the
  expected flattening of the stellar density profile near the
  mid-plane: All vertical profiles are well represented by sech$^2$
  profiles, with scale heights ranging from $\approx50\,\pc$ for A
  stars to $\approx150\pc$ for G and K dwarfs and giants. We determine
  the luminosity function along the main sequence for $M_V < 7$ ($M
  \gtrsim 0.72\msun$) and along the giant branch for $M_J \gtrsim-2.5$
  in detail. Converting this to a mass function, we find that the
  high-mass ($M > 1\,M_\odot$) present-day mass function along the
  main sequence is $\dd n / \dd M =
  0.016\,(M/M_\odot)^{-4.7}\,\mathrm{stars\,\pc}^{-3}\,M_\odot^{-1}$.
  Extrapolating below $M = 0.72\msun$, we find a total mid-plane
  stellar density of $0.040\pm0.002\msun\pc^{-3}$. Giants contribute
  $0.00039\pm0.00001$ stars\,$\pc^{-3}$ or about
  $0.00046\pm0.00005\msun\pc^{-3}$. The star-formation rate surface
  density is $\Sigma(t) =
  7\pm1\,\exp(-t/7\pm1\,\mathrm{Gyr})\,M_\odot\,\mathrm{pc}^{-2}\,\mathrm{Gyr}^{-1}$. Overall,
  we find that \Gaia\ DR1's selection biases are manageable and allow
  a detailed new inventory of the solar neighborhood to be made that
  agrees with and extends previous studies. This bodes well for
  mapping the Milky Way with the full \Gaia\ data set.
\end{abstract}

\section{Introduction}

The Milky Way is a cornerstone in our understanding of the structure
and evolution of galaxies. And within the Milky Way, the solar
neighborhood provides a fundamental basis for studies of Galactic
structure, star formation and stellar evolution, and Galactic
dynamics. Within a few hundred parsec from the Sun, we can see stars
that span the extremes from the most luminous main-sequence and giant
stars to the faintest M dwarfs. This provides an essential basis for
understanding the baryonic content of the Milky Way and external
galaxies as it allows the initial-mass function (IMF; \eg,
\citealt{Gould96a,Kroupa01a,Chabrier01a}), the mass-to-light ratio
\citep[\eg,][]{Flynn06a}, and the local star-formation history
\citep[\eg,][]{Binney00a} to be determined directly. A complete
baryonic census of the solar neighborhood is also important for
comparing with dynamical determinations of the local mass
distribution, which are only sensitive to the combined mass in baryons
and dark matter \citep[\eg,][]{Holmberg00a,Bovy13a}.

The most precise censuses of stars in the solar neighborhood are
largely based on small, volume-complete surveys of local stars (for
example, within 25\pc, \citealt{Reid02a}; within 50\pc,
\citealt{Jahreiss98a}), which contain only a handful of the most
luminous stars. While this does not matter for determining the IMF
from the number density of long-lived stars, for the purpose of, \eg,
measuring the local stellar density distribution and comparing it to
dynamical estimates \citep{McKee15a}, small volumes centered on the
Sun are potentially biased, because the Sun is likely offset from the
mid-plane of the Galaxy by 15 to 25\pc\ (\eg,
\citealt{Binney97a,Chen01a,Juric08a}). Extending the local stellar
census to a few hundred parsecs would allow a much better
determination of the bright end of the stellar luminosity function and
of the present-day mass function.

The ESA \Gaia\ mission has been designed to investigate the luminous
and dark-matter distribution within the Milky Way to constrain its
formation and evolution \citep{GaiaMission}. While \Gaia\ will observe
more than 1 billion stars, these constitute only about 1\,\% of all of
the stars in the Milky Way and the volume completeness for different
types of stars will be complex even in the final data release due to
variations in stellar colors, interstellar extinction, stellar
density, and the observation pattern in different parts of the
sky. \Gaia\ released its first data in the Fall of 2016
\citep{GaiaDR1}, which consists of the primary \emph{Tycho-Gaia
  Astrometric Solution} (\tgas) containing positions, parallaxes, and
proper motions for a subset of the \tyctwo\ catalog \citep{Hog00a} and
the secondary data set with approximate positions for stars brighter
than $G \approx 20.7$ in the wide \Gaia\ photometric bandpass $G$
\citep{Lindegren16a}. Based on 14 months of data and a preliminary
astrometric processing, the \Gaia\ DR1 \tgas\ catalog contains
2,057,050 stars with parallaxes with typical uncertainties of
$\approx0.3\mas$. As such, \tgas\ is by far the largest catalog of
trigonometric, high-precision parallaxes for a uniformly selected
sample of stars. However, no estimate of the completeness of \tgas\ is
provided in the data products. This makes it difficult to use the
\tgas\ data to investigate the spatial and dynamical structure of the
solar neighborhood and many of the first uses of the \tgas\ data have
therefore focused on purely kinematic studies that do not require
knowledge of the selection function
\citep[\eg,][]{AllendePrieto16a,Hunt16a,Bovy17a,Helmi17a}.

This paper has two main parts. The first part consists of an extensive
discussion of the volume completeness of the \tgas\ catalog for
different stellar types. This may be useful to other studies employing
\tgas\ data to study various aspects of the Milky Way's structure. It
can also be straightforwardly extended to the \Gaia\ DR2 data when
they appear. The second part uses the volume completeness combined
with star counts for different classes of stars to perform a new
stellar census of the solar neighborhood, covering main-sequence stars
from early A-type to late K-type dwarfs and giant stars from the
subgiant branch to the upper red-giant branch. We also determine the
vertical density profiles for these stellar types up to a maximum of
400\pc\ above the Galactic mid-plane. The resulting census is
consistent with previous work and constitutes the most precise
determination of the mass distribution in the solar neighborhood
within the mass range observed in \tgas.

The structure of this paper is as follows. We introduce our approach
to determining stellar densities and their spatial dependence from
incomplete surveys in \sectionname~\ref{sec:method}. In particular, we
introduce the effective volume completeness, which for a given small
volume in the Galaxy and a given stellar type represents the fraction
of stars observed by the survey. In \sectionname~\ref{sec:complete} we
apply this formalism to the \Gaia\ DR1 catalog and determine the
completeness of the \tgas\ catalog for different stellar types:
spectral types along the main sequence and giants in various
luminosity ranges. The results from \sectionname~\ref{sec:complete}
depend on a detailed understanding of the ``raw'' \tgas\ selection
function: the fraction of true stars contained in the catalog as a
function of sky position, color, apparent magnitude, etc. We determine
this raw selection function directly from the \tgas\ catalog by
comparing it to the 2MASS catalog \citep{Skrutskie06a}; this is
discussed in detail in Appendix~\ref{sec:tgascomplete}.

In \sectionname~\ref{sec:dens_ms}, we apply the formalism from
\sectionname~\ref{sec:method} to measure the mid-plane densities and
the vertical density profile of different stellar types along the main
sequence and fit parametric density laws. From this we determine the
luminosity function, present-day mass function, total mid-plane number
and mass densities, and the star-formation history of the solar
neighborhood. In \sectionname~\ref{sec:dens_giants}, we similarly
measure the mid-plane densities and the vertical density profile for
different types of giant stars and determine the luminosity function
along the giant branch and the total mid-plane number and mass density
of giants. One of the parameters that we fit for each stellar type is
the Sun's offset from the mid-plane as defined by that stellar type
and we discuss the resulting offsets in detail in
\sectionname~\ref{sec:zsun}. We discuss some additional aspects of our
results in \sectionname~\ref{sec:discussion} and conclude in
\sectionname~\ref{sec:conclusion}. The basic data that we use
throughout this paper comes from the \tgas\ catalog \citep{GaiaDR1}
matched to photometry from the 2MASS catalog using a $4''$
nearest-neighbor search. Various cuts on this basic data set for
different applications are described in the text below. We express the
volume completeness in rectangular Galactic coordinates $(X,Y,Z)$,
centered on the Sun with $X$ toward the Galactic center, $Y$ in the
direction of Galactic rotation, and $Z$ directed toward the North
Galactic Pole.  We do not make any use of Galactocentric coordinates
or any sort of kinematics and our results therefore do not depend on
any assumed value of the standard Galactic constants (\eg, the
distance to the Galactic center or the Sun's velocity).

\section{Stellar Number Density Laws from Incomplete Surveys}\label{sec:method}

Our goal is to determine the intrinsic stellar number density
distribution $\dens(X,Y,Z)$, where $(X,Y,Z)$ is a set of heliocentric
cartesian coordinates, of different stellar populations from
\tgas\ observations of their three-dimensional position ($\ra,\dec,D$)
[right ascension, declination, and distance]. If \tgas\ observed all
stars in the Galaxy, we would simply count the stars in small volumes
and divide this number by the volume to get the density in terms of
stars$\,\pc^{-3}$. However, because \tgas\ only observes stars in a
finite magnitude range and because even at peak completeness it does
not contain all stars, the situation is not quite this simple.

The problem of determining stellar number density laws $\dens(X,Y,Z)$
of stellar populations in the Milky Way has a long history
\citep[\eg,][]{Bok37a,Bahcall86a,Juric08a} and has been discussed
recently by \citet{Bovy12b} and \citet{Bovy16a}. The discussion there
focused primarily on spectroscopic surveys and on the problem of
fitting a parametric density law to star count data. The focus here is
different, in that we aim to determine the density $\dens(X,Y,Z)$
non-parametrically in bins in $(X,Y,Z)$ from data that cover a
substantial fraction of the sky. Much of the discussion, especially
about the effect of dust extinction, in this section follows that in
\citet{Bovy16a}.

We discuss how we determine the \tgas\ completeness as a function of
$(J,J-K_s,\ra,\dec)$ by comparing the number counts in \tgas\ to those
in 2MASS in \appendixname~\ref{sec:tgascomplete}. Thus, we have a
function $S(J,J-K_s,\ra,\dec)$ that gives the fraction of stars at a
given $(J,J-K_s,\ra,\dec)$ contained in the \tgas\ catalog. As
discussed in \appendixname~\ref{sec:tgascomplete}, we only determine
the completeness in the 48\,\% of the sky with `good'
\tgas\ observations and within this region, the completeness is
independent of $(\ra,\dec)$; therefore essentially
$S(J,J-K_s,\ra,\dec) =
S(J,J-K_s)\,\Theta([\ra,\dec]\ \mbox{in\ `good'\ part\ of\ the\ sky})$,
where $\Theta(a)$ is the function that is one when $a$ holds true and
zero otherwise. However, for generality, we will write all expressions
in terms of the full $S(J,J-K_s,\ra,\dec)$.

\subsection{Generalities}

We determine $\dens(X,Y,Z)$ from a data set of stars that is not
complete in any (simple) geometric sense and thus we need to take the
selection function $S(J,J-K_s,\ra,\dec)$ into account. The connection
between the selection function $S(J,J-K_s,\ra,\dec)$ and the
three-dimensional position $(X,Y,Z)$ is made through a color-magnitude
density (CMD) $\denscmd(M_J,[J-K_s]_0|X,Y,Z)$ that gives the
distribution in (absolute magnitude, unreddened color), potentially a
function of position, that allows us to relate the observed
$(J,J-K_s,\ra,\dec)$ to $(X,Y,Z)$ through the distance and
three-dimensional extinction map. While these could (and should) in
principle be inferred simultaneously with the stellar density, we
assume in what follows that the CMD $\denscmd(M_J,[J-K_s]_0|X,Y,Z)$
and the three-dimensional extinction map $(A_J,E(J-K_s))[X,Y,Z]$ are
known a priori.

To determine $\dens(X,Y,Z)$ we use a likelihood approach that models
the full rate function $\lambda(O|\theta)$ that gives the number of
stars as a function of all observables $O$ of interest for a set of
model parameters $\theta$. These observables $O$ are in this case
$(\ra,\dec,D,J,J-K_s)$---we will use $(X,Y,Z)$ and $(\ra,\dec,D)$
interchangeably because they are related by coordinate transformation,
but will keep track of the Jacobian---and we can write
\begin{align}
  \lambda(O|\theta) & = \lambda(\ra,\dec,D,J,J-K_s)\,,\\
  &  = \dens(X,Y,Z|\theta)\,D^2\,\cos \dec\nonumber\\ & \qquad \denscmd(M_J,[J-K_s]_0|X,Y,Z)\,S(J,J-K_s,\ra,\dec)\,,\nonumber
\end{align}
where we have assumed that the model parameters only affect
$\dens$. In this decomposition, the factor $D^2\,\cos \dec$ comes from
the Jacobian of the transformation between $(\ra,\dec,D)$ and
$(X,Y,Z)$.

An observed set of stars indexed by $i$ is a draw from a Poisson
process with rate function $\lambda(O|\theta)$ with the likelihood
$\mathcal{L}(\theta)$ of the parameters $\theta$ describing the
density law given by {\footnotesize
\begin{align}\label{eq:lnl}
  & \ln \mathcal{L}(\theta)\nonumber\\ &\ = \sum_i \ln \lambda(O_i|\theta) -\int \dd O
  \lambda(O|\theta)\,,\\ &\ = \sum_i \ln \dens(X_i,Y_i,Z_i|\theta) -\int \dd D\, D^2\dd \ra \dd \dec
    \,\cos\dec\,\dens(X,Y,Z|\theta)\nonumber\\ & \ \ \int \dd J \dd (J-K_s)
      \,\denscmd(M_J,[J-K_s]_0|X,Y,Z)\,S(J,J-K_s,\ra,\dec)\,,\nonumber
\end{align}}
where in the second equation we have dropped terms that do not depend on
$\theta$.  As in \citet{Bovy16a} we simplify this expression by
defining the \emph{effective selection function} $\essf(\ra,\dec,D)$
defined by
{\footnotesize
\begin{align}\label{eq:essf}
  & \essf(\ra,\dec,D) \\ & \ \equiv \int \dd J \dd (J-K_s) \,\denscmd(M_J,[J-K_s]_0|X,Y,Z)\,S(J,J-K_s,\ra,\dec)\,,\nonumber
\end{align}}
where $5\log_{10}\left(D/10\pc\right) = J-M_J-A_J[X,Y,Z]$ and
$[J-K_s]_0 = J-K_s - E(J-K_s)[X,Y,Z]$. The ln likelihood then becomes
\begin{align}
  \ln \mathcal{L}(\theta) = & \sum_i \ln \dens(X_i,Y_i,Z_i|\theta)
  \\ & \ -\int \dd D\, D^2\dd \ra \dd \dec
    \,\cos\dec\,\dens(X,Y,Z|\theta)\,\essf(\ra,\dec,D)\,.\nonumber
\end{align}
Unlike the selection function $S(J,J-K_s,\ra,\dec)$ which depends on
the survey's operations only (which parts of the sky were observed,
for how long, \ldots), the effective selection function
$\essf(\ra,\dec,D)$ is a function of both the survey operations
\emph{and} the stellar population under investigation. Its usefulness
derives from the fact that it encapsulates all observational effects
due to selection and dust obscuration and turns the inference problem
into a purely geometric problem.  It directly gives the fraction of
stars in a given stellar population that are observed by the survey in
a given direction and at a given distance. Like the survey selection
function $S(J,J-K_s,\ra,\dec)$, $\essf(\ra,\dec,D)$ takes values
between zero (fully incomplete) and one (fully complete). Under the
assumption that the CMD $\denscmd(M_J,[J-K_s]_0|X,Y,Z)$ and the
three-dimensional extinction map $(A_J,E(J-K_s))[X,Y,Z]$ are known (or
at least fixed in the analysis), the effective selection function can
be computed \emph{once} for a given (survey, stellar population) pair.

To determine the best-fit parameters $\hat{\theta}$ of a parameterized
density law $\dens(X,Y,Z|\theta)$ one has to optimize the ln
likelihood given above. This ln likelihood can be marginalized
analytically over the overall amplitude of the density (the local
normalization if you will); this is discussed in \citet{Bovy16a} and
similar expressions would apply here if such a marginalization is
desired.

\subsection{Non-parametric binned density laws}\label{sec:method_binned}

Now suppose that one wants to determine the density $\dens(X,Y,Z)$ of
a stellar population in a set of non-overlapping bins in
$(X,Y,Z)$. The bins are given by a set $\{\Pi_k\}_k$ of rectangular
functions that evaluate to unity within the domain of the bin and zero
outside of it. The domain can have an arbitrary shape, but typically
this would be an interval in each of $X$, $Y$, and $Z$ or perhaps in
$R_{xy}$ and $Z$, where $R_xy = \sqrt{X^2+Y^2}$. We can then write the
density as
\begin{equation}
  \dens(X,Y,Z|\theta) = \sum_k n_k\,\Pi_k(X,Y,Z)\,,
\end{equation}
where $\theta\equiv\{n_k\}_k$ is a set of numbers that give the
density in each bin and that therefore parameterizes the density law.

{\allowdisplaybreaks
The ln likelihood then becomes
\begin{align}
 &  \ln \mathcal{L}(\{n_k\}_k) \nonumber\\ \ \ & = \sum_i \ln \sum_k n_k\,\Pi_k(X_i,Y_i,Z_i)\\
  & \qquad -\int \dd D\, D^2\dd \ra \dd \dec
    \,\cos\dec\,\sum_k n_k\,\Pi_k(X,Y,Z)\,\essf(\ra,\dec,D)\,,\nonumber
\end{align}
which, because the $\{\Pi_k\}_k$ are a set of non-overlapping bins,
can be simplified to
\begin{align}
& \ln \mathcal{L}(\{n_k\}_k) \nonumber\\ & \ \ = \sum_k \Big[ N_k \ln n_k \\ & \qquad - n_k \int
  \dd D\, D^2\dd \ra \dd \dec
  \,\cos\dec\,\Pi_k(X,Y,Z)\,\essf(\ra,\dec,D)\Big]\,,\nonumber
\end{align}
where $N_k$ is the number of points $i$ in the observed set that fall
within bin $k$. We can maximize this likelihood for each $n_k$
analytically and find best-fit $\hat{n}_k$
\begin{equation}
  \hat{n}_k = \frac{N_k}{\int \dd D\, D^2\dd \ra \dd \dec
    \,\cos\dec\,\Pi_k(X,Y,Z)\,\essf(\ra,\dec,D)}\,.
\end{equation}}

The denominator in this expression is known as the \emph{effective
  volume}. Using the same symbol $\Pi_k$ to denote the
three-dimensional integration region and using $x = (X,Y,Z)$ and
$(\ra,\dec,D)$ interchangeably because they are related through
coordinate transformation, this can be written as the following simple
expression
\begin{equation}\label{eq:bfdens}
  \hat{n}_k = \frac{N_k}{\int_{\Pi_k} \dd^3 x\,\essf(\ra,\dec,D)}\,.
\end{equation}
Thus, the effective volume corresponding to a given spatial region
$\Pi_k$ is the spatial integral of the effective selection function
over $\Pi_k$. We can then define the \emph{effective volume
  completeness} $\Xi(\Pi_k)$ of the spatial region $\Pi_k$ as
\begin{equation}
  \Xi(\Pi_k) = \frac{\int_{\Pi_k} \dd^3 x\,\essf(\ra,\dec,D)}{\int_{\Pi_k} \dd^3 x}\,,
\end{equation}
where $\int_{\Pi_k} \dd^3 x = V(\Pi_k)$ is simply the actual geometric
volume of $\Pi_k$. Because $\essf(\ra,\dec,D)$ is a function bounded
by zero and one, the effective volume completeness $\Xi(\Pi_k)$ also
takes values between zero and one. In terms of $\Xi(\Pi_k)$,
\equationname~(\ref{eq:bfdens}) becomes
\begin{equation}\label{eq:bfdens2}
  \hat{n}_k = \frac{N_k}{\Xi(\Pi_k)\,V(\Pi_k)} =
  \frac{1}{\Xi(\Pi_k)}\,\frac{N_k}{V(\Pi_k)}\,.
\end{equation}
This expression makes sense, because for a complete sample
$\essf(\ra,\dec,D) = 1$, such that $\Xi(\Pi_k) = 1$ and this
expression simplifies to the number divided by the volume of the bin,
the standard way to compute a number density.

From the second derivative of the ln likelihood, we find the
uncertainty on the $\hat{n}_k$
\begin{equation}
  \sigma_{\hat{n}_k} = \frac{\hat{n}_k}{\sqrt{N_k}}\,.
\end{equation}

\section{The \tgas\ completeness for different stellar populations}\label{sec:complete}

\begin{figure}
  \includegraphics[width=0.49\textwidth,clip=]{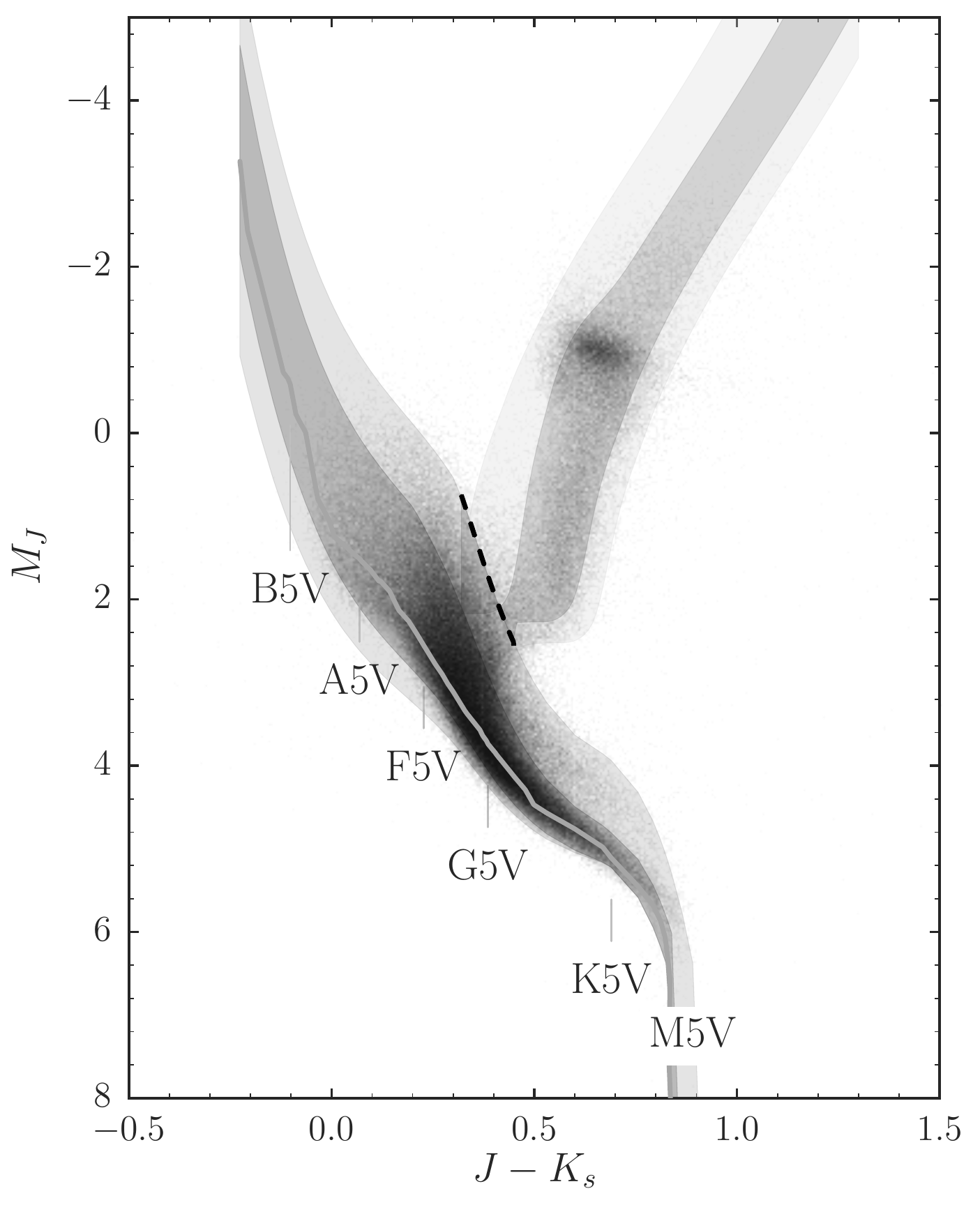}
  \caption{Near-infrared CMD for \tgas\ stars. The light gray line
    displays the mean dwarf stellar locus from \citet{Pecaut13a}. The
    lightly-shaded area shows how we select main-sequence and giant
    stars for the purpose of star counts; the darkly-shaded area
    displays the more stringent cuts that we use to determine the
    absolute magnitude distribution of different stellar types. The
    dashed line gives the assumed separation between the main sequence
    and the giant branch. Major stellar types along the main sequence
    are labeled.}\label{fig:cmd}
\end{figure}

From the discussion above, it is clear that for a given survey and a
given stellar population, the completeness functions of interest are
(i) the effective selection function $\essf(\ra,\dec,D)$ and (ii) the
effective volume completeness $\Xi(\Pi_k)$.  In this section, we
compute these functions for different stellar types along the main
sequence and along the giant branch using the \tgas\ selection
function determined in \appendixname~\ref{sec:tgascomplete}. These
functions will give a sense of how \tgas\ samples the extended solar
neighborhood for a given stellar type.

\subsection{Definitions of stellar types}\label{sec:stellartypes}

The near-infrared CMD for stars with well-determined parallaxes
\plx\ in \tgas\ is displayed in
\figurename~\ref{fig:cmd}. ``Well-determined parallaxes'' for the
purpose of this figure means (i) $\plx/\sigma_\plx > 10$ if $M_J < 0$,
(ii) $\plx/\sigma_\plx > 20$ for $M_J > 5$, and (iii)
$\plx/\sigma_\plx > 20-2\,(M_J-5)$ for $0 < M_J < 5$, with $M_J$
computed based on the \tgas\ \plx\ without correcting for
extinction. These cuts are chosen to have a relatively well-populated
upper main sequence and giant branch, because simple cuts on
$\plx/\sigma_\plx$ tend to select few intrinsically-bright, and thus
typically distant, stars. We use stars contained within the
darkly-shaded region as a sampling $(M_J,[J-K_s]_0)_j$ of the
intrinsic CMD $\denscmd(M_J,[J-K_s]_0)$, which we assume to be
independent of position. This is a reasonable assumption for the
$D\lesssim1\kpc$ probed by \tgas. The darkly-shaded region is defined
by a shifted and stretched version of the mean dwarf stellar locus
from \citet{Pecaut13a} and of the giant locus described below.

\begin{figure}
  \includegraphics[width=0.49\textwidth,clip=]{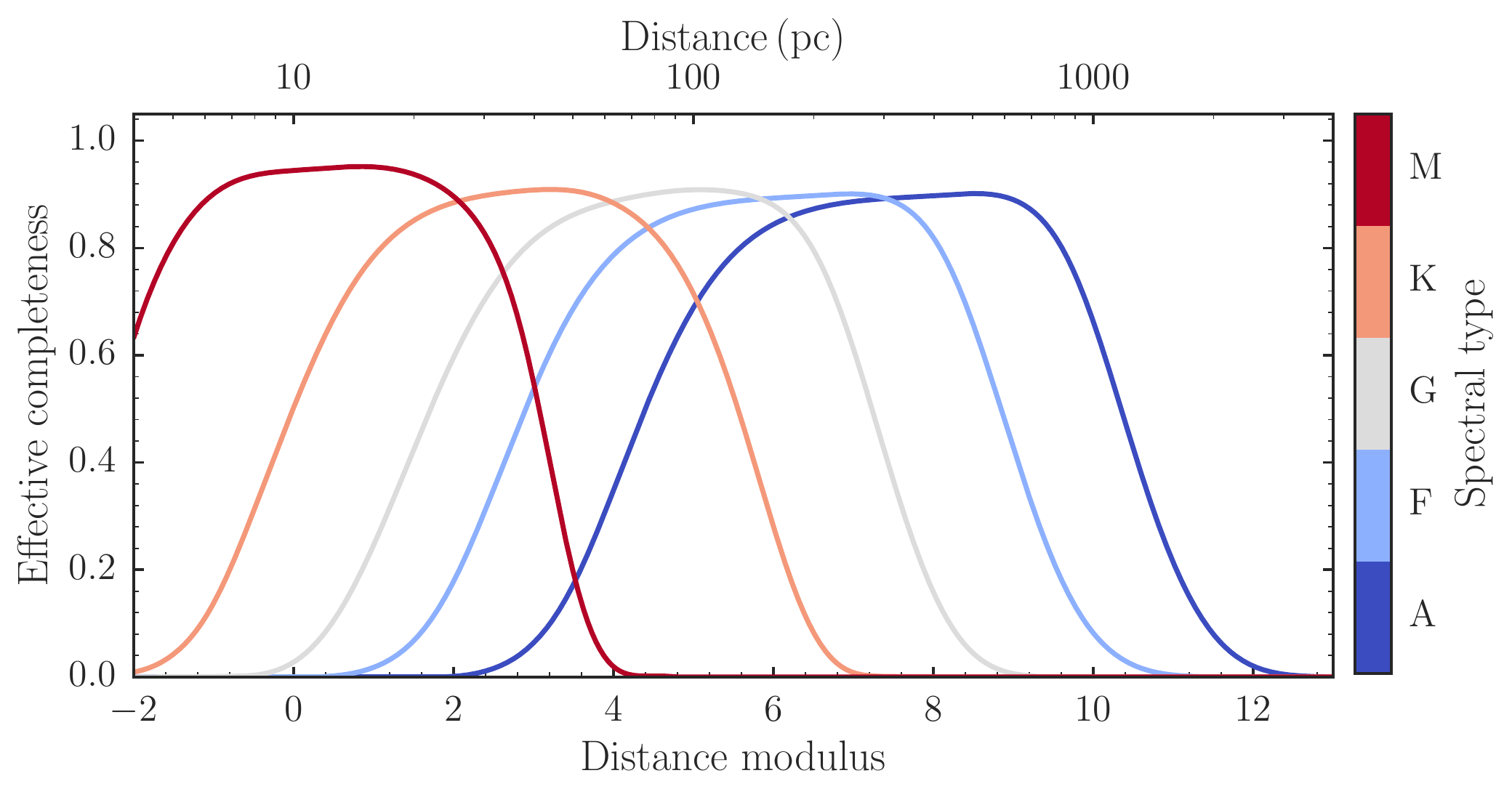}\\
  \includegraphics[width=0.49\textwidth,clip=]{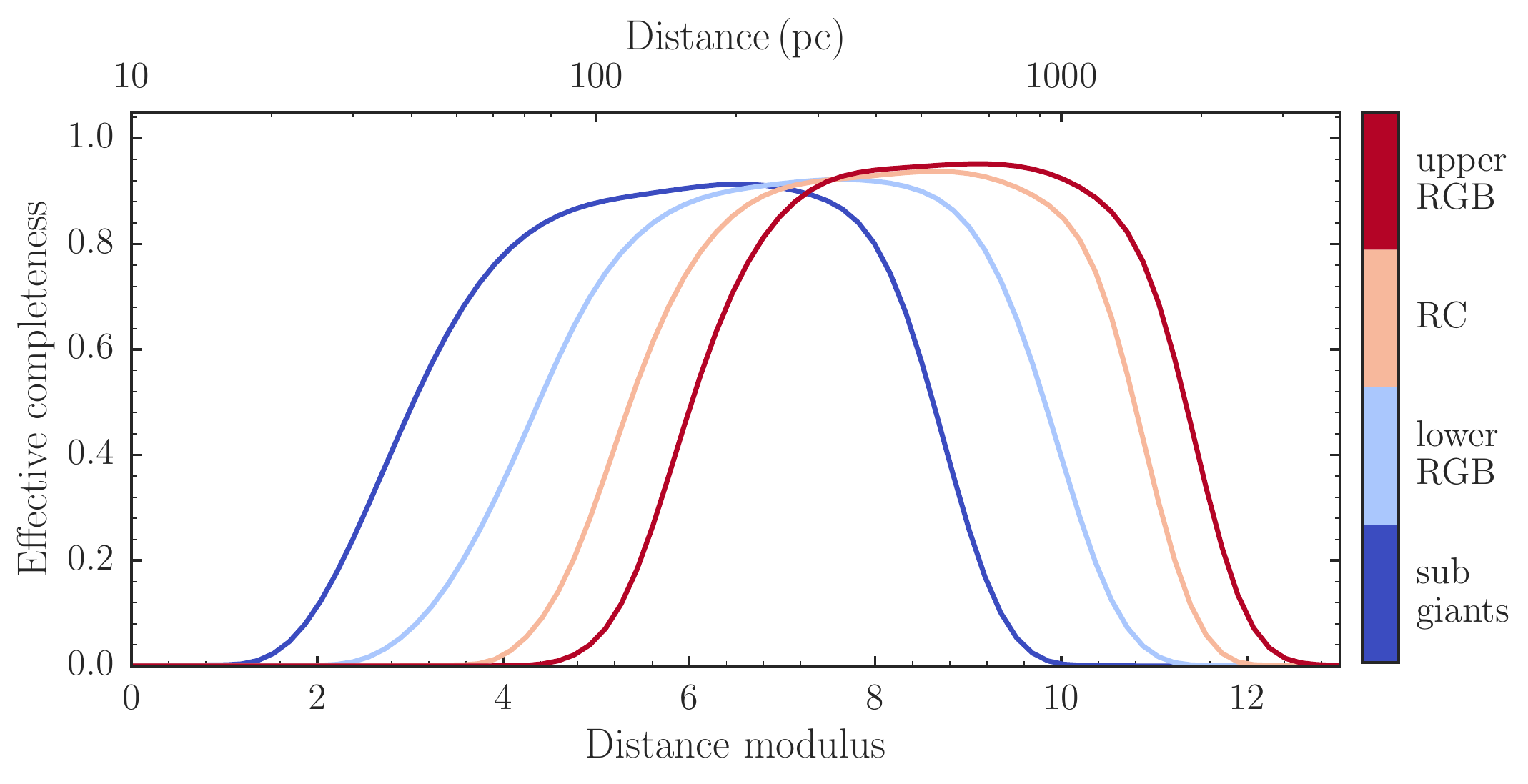}
  \caption{Effective selection function (or ``effective
    completeness'') for different stellar types in \tgas\ as a
    function of distance, assuming that extinction is
    negligible.}\label{fig:effsel_dist}
\end{figure}

Along the main sequence, we define stellar types everywhere in this
paper using the mean dwarf stellar locus from
\citet{Pecaut13a}\footnote{Specifically, we use version
  \texttt{2016.08.21} downloaded from
  \url{http://www.pas.rochester.edu/~emamajek/EEM_dwarf_UBVIJHK_colors_Teff.txt}~.},
displayed as a gray curve in \figurename~\ref{fig:cmd}. Thus, we
define A dwarfs as those stars along the main sequence with $J-K_s$
colors between the color of A0V and A9V stars according to the stellar
locus and similarly for F, G, and K stars. For M stars we only go as
red as the color of M5V dwarfs ($[J-K_s]_0 = 0.89$). In
\sectionname~\ref{sec:dens_ms} we further subdivide each stellar type
into subtypes A0V, A1V, etc. For those cases, the $J-K_s$ boundaries
are located halfway between the stellar subtype in question and the
adjoining subtypes. For example, for A1V the boundary goes from
halfway between A0V and A1V to halfway between A1V and A2V. The
sampling $(M_J,[J-K_s]_0)_j$ for a given stellar type then consists of
those stars in the correct color range that fall within the
darkly-shaded region of \figurename~\ref{fig:cmd}. We always limit
this sample to 1,000 stars for computational reasons.

We proceed similarly to define different types of giants. As a
fiducial giant locus, we use a solar metallicity ($Z = 0.017$) PARSEC
isochrone with an age of $10^{0.8}\Gyr$ \citep{Bressan12a}. For
simplicity, we define different types of giants using a simple cut on
$M_J$: (i) ``subgiants'' for $1 < M_J < 4$, (ii) ``lower red-giant
branch'' (RGB) for $-0.5 < M_J < 1$, (iii) ``red-clump'' (RC) for
$-1.5 < M_J < -0.5$, and (iv) ``upper RGB'' for $-4 < M_J < -1.5$. In
\sectionname~\ref{sec:dens_giants} below, we will further subdivide
these types into $\Delta M_J = 0.25\magunit$ bins. Similar to the
main-sequence stars above, a sampling $(M_J,[J-K_s]_0)_j$ for a given
giant type then consists of those stars in the correct $M_J$ range
that fall within the darkly-shaded region to the right of the dashed
line in \figurename~\ref{fig:cmd}. This selection of giants largely
avoids the asymptotic giant branch.

\subsection{Effective selection function for different stellar types in \tgas}\label{sec:effsel}

\begin{figure}
  \includegraphics[width=0.49\textwidth,clip=]{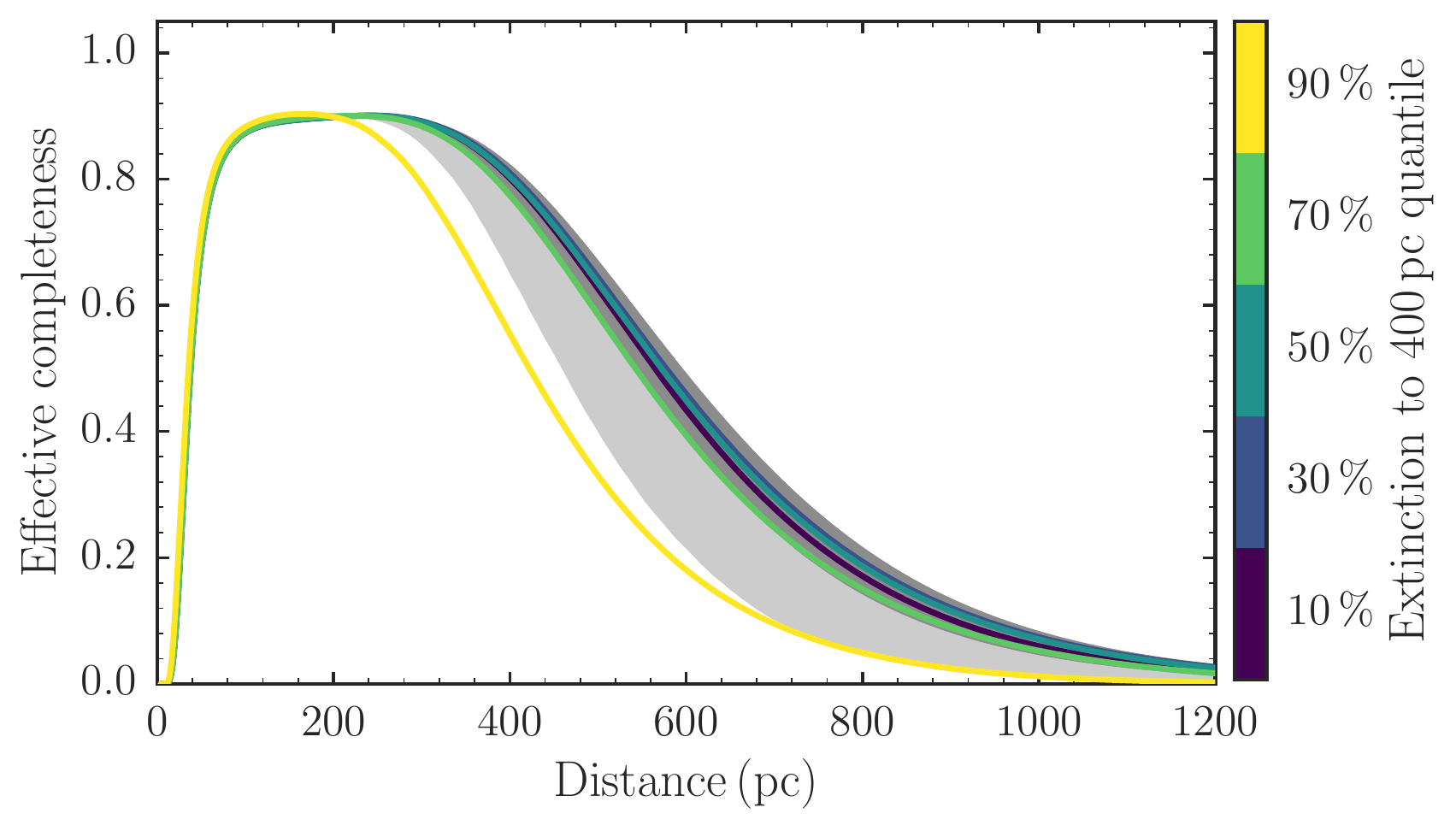}
  \caption{Effect of the three-dimensional dust extinction on the
    effective completeness. This figure displays the effective
    completeness of F dwarfs in \tgas\ as a function of distance in
    regions with different amounts of extinction. The colored curves
    show the effective completeness in $3.36\deg^{2}$ sky regions at
    five quantiles of the distribution of the mean extinction to
    $400\pc$. The dark and light gray bands displays the $68\,\%$ and
    $95\,\%$ lower limits of the completeness over the sky. The effect
    of extinction is small over the majority of the
    sky.}\label{fig:effsel_f_dist_ext}
\end{figure}

With the definitions of different stellar types from the previous
subsection, we compute the effective selection function
$\essf(\ra,\dec,D)$ for different stellar types. In particular, we use
the sampling $(M_J,[J-K_s]_0)_j$ from $\denscmd(M_J,[J-K_s]_0)$ for
each stellar type to approximate \equationname~(\ref{eq:essf}) as a
Monte Carlo integration
{\footnotesize \begin{align}\label{eq:essf_mc}
& \ \essf(\ra,\dec,D) \\ & \ \equiv \int \dd J \dd (J-K_s) \,\denscmd(M_J,[J-K_s]_0|X,Y,Z)\,S(J,J-K_s,\ra,\dec)\,,\nonumber
\\ & \approx \sum_j S(M_{J,j} + \mu + A_J,[J-K_s]_{0,j}+ E(J-K_s),\ra,\dec)\,,\nonumber
\end{align}}
where we substitute $(J,J-K_s) = (M_{J,j} + \mu + A_J,[J-K_s]_{0,j}+
E(J-K_s))$ in $S(J,J-K_s,\ra,\dec)$. In this equation, $\mu =
5\log_{10}\left(D/10\pc\right)$ is the distance modulus and we have
suppressed the dependence of the extinction map $(A_J,E(J-K_s))$ on
$(\ra,\dec,D)$. In the absence of extinction, the effective selection
function becomes
\begin{align}
\essf(\ra,\dec,D) & \approx \sum_jS(M_{J,j} + \mu,[J-K_s]_{0,j},\ra,\dec)\,.
\end{align}
Because our \tgas\ selection function $S(J,J-K_s,\ra,\dec)$ does not
depend on $(\ra,\dec)$ within the region of the sky over which it is
defined, this is a one-dimensional function of distance.

The effective selection function in the absence of extinction is shown
for different stellar types in
\figurename~\ref{fig:effsel_dist}. Along the main sequence, the
overall trend is that later types of stars can be seen out to a
smaller distance than earlier types of stars. This is because of two
reasons: not only are later-type dwarfs intrinsically fainter, but the
\tgas\ selection function also has a brighter faint-end cut-off for
redder, late-type stars (see \figurename~\ref{fig:comp_3jbins} in the
appendix). On the other hand, the \tgas\ selection function has a
slightly higher plateau value at intermediate magnitudes for redder
stars, which has the effect that the M-dwarf effective selection
function is slightly higher than for earlier types at its peak. The
lower distance limit results from the bright cut-off of the
\tgas\ selection function stemming from the exclusion of the brightest
($G \lesssim 6$) stars in \tgas. For A, F, and G dwarfs this causes a
hole around the location of the Sun for these stars extending up to
about $100\pc$ for A stars.

The effective selection function for different types of giants in
\figurename~\ref{fig:effsel_dist} behaves similarly to that of the
dwarfs. Intrinsically-brighter giants can be seen out to larger
distances, but not much larger distances, because the
intrinsically-brighter (in the near infrared) giants on the upper RGB
are also redder, where the \tgas\ selection function cuts off at
brighter magnitudes. For both early-type dwarfs and luminous giants the
effective selection function becomes small around $2\kpc$. The bright
cut-off of the \tgas\ selection function again causes a hole around
the location of the Sun of $\approx50\pc$ to $\approx200\pc$ for
giants depending on their luminosity.

The curves in \figurename~\ref{fig:effsel_dist} are for the case of
zero extinction. For the intrinsically-brighter stellar types that
extend out to $\gtrsim1\kpc$, extinction has important
effects. Because the extinction varies as a function of
three-dimensional position, the effective selection function computed
using \equationname~\eqref{eq:essf_mc} becomes a function of
$(\ra,\dec,D)$. We compute the extinction using the combined
extinction model from \citet{Bovy16a}, which merges the
three-dimensional extinction maps from \citet{Marshall06a},
\citet{Green15a}, and \citet{Drimmel03a}. These maps are combined in
that order in the case of overlap to create a full-sky
three-dimensional extinction map (see the appendix of
\citealt{Bovy16a} for full details). In what follows, we use this map
at its full resolution, which varies between $\approx4'$ to
$\approx20'$.

The effect of extinction on the zero-extinction curves in
\figurename~\ref{fig:effsel_dist} is displayed in
\figurename~\ref{fig:effsel_f_dist_ext}, focusing on F dwarfs. The
effect of extinction is mainly to lower the distance out to which a
stellar population can be observed. A secondary effect is that the
peak of the effective selection function can become higher, because of
the higher completeness to redder stars in \tgas, but this is a minor
effect. The narrow dark-gray region contains $68\,\%$ of the part of
the sky for which we have determined the \tgas\ selection function;
the light-gray region contains 95\,\%. To give another sense of this,
the colored lines are lines of sight (in $\nside = 32$ HEALPix pixels)
at the 10 through 90$^{\mathrm{th}}$ quantile of the distribution of
the mean extinction to $400\pc$. Thus, the effect of extinction is
overall small, except for the about 10\,\% of the sky with high
extinction, located near the Galactic plane.

\subsection{Effective volume completeness maps for \tgas}\label{sec:effvol}

\begin{figure}
  \includegraphics[width=0.49\textwidth,clip=]{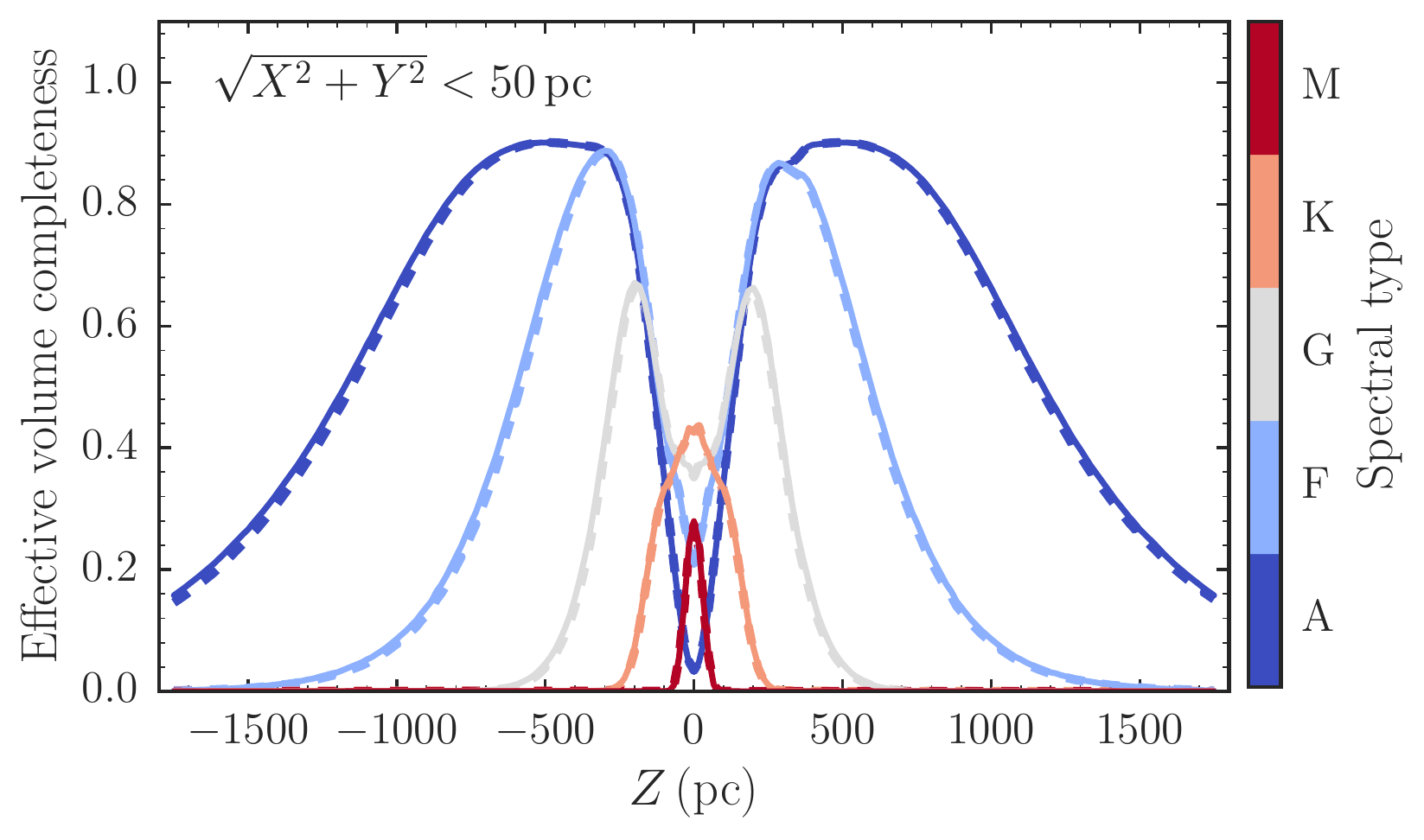}
  \includegraphics[width=0.49\textwidth,clip=]{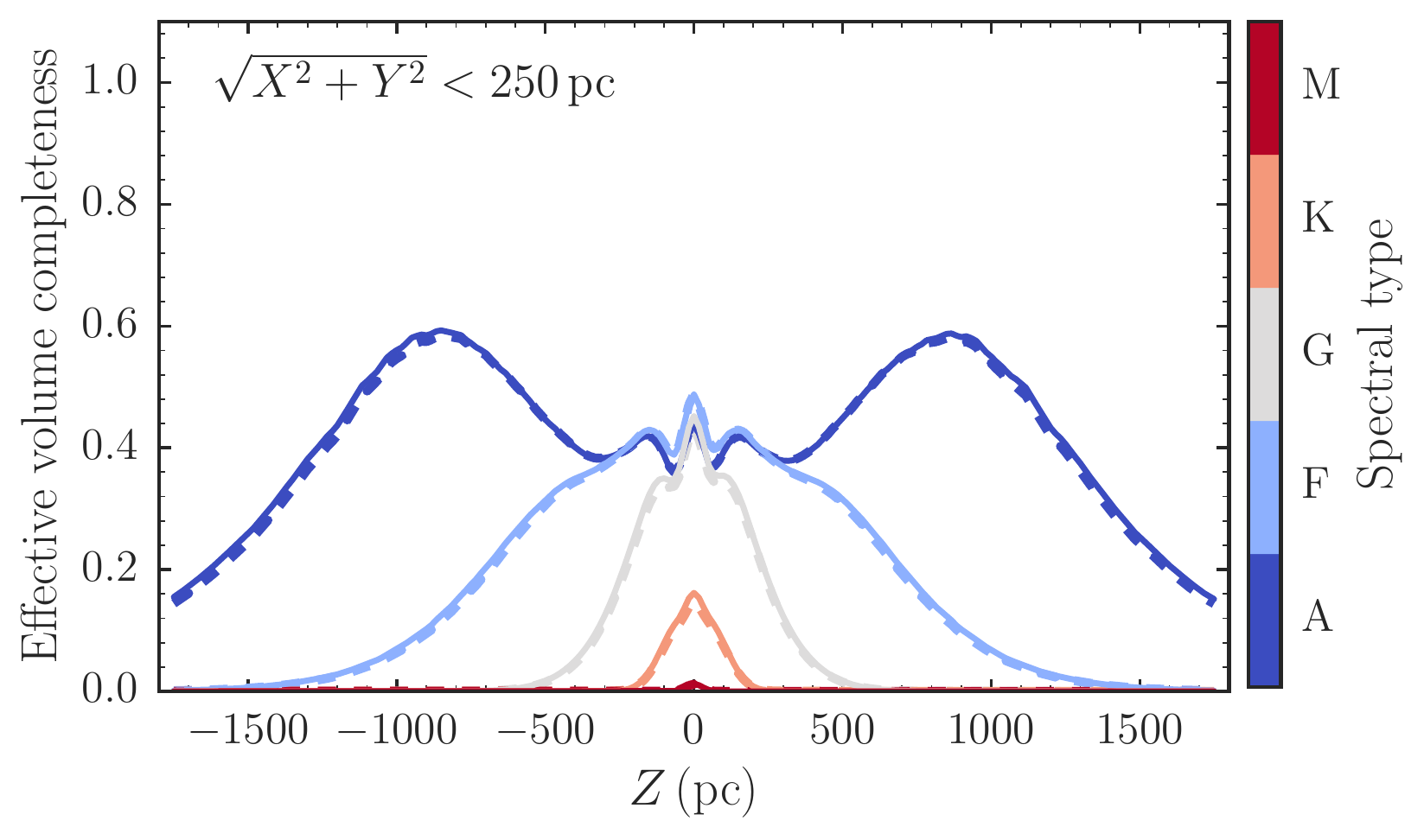}
  \caption{Effective volume completeness for different stellar types
    along the main sequence as a function of vertical distance from
    the Sun. The completeness is computed in cylinders centered on the
    Sun perpendicular to the $(X,Y)$ plane with a radius of
    $R_{xy}=50\pc$ and $R_{xy}=250\pc$ in the top and bottom panels,
    respectively, and in $Z$ slices with a width of $50\pc$. The solid
    curves assume zero dust extinction, while the dashed lines use a
    model for the three-dimensional dust distribution; the effects of
    extinction are small. \tgas's bright limit causes a hole in the
    narrow cylinder for early-type dwarfs. The effective completeness
    has small-scale structure that needs to be taken into account when
    using \tgas\ to determine the underlying stellar density profile
    of different stellar types.}\label{fig:effsel_cyl}
\end{figure}

Similar to the effective selection function $\essf(\ra,\dec,D)$, we
can compute the effective volume completeness $\Xi(\Pi_k)$ for
different kinds of volumes $\Pi_k$ and for different types of
stars. The effective volume completeness in cylindrical regions with
height $50\pc$ centered on $(X,Y) = (0,0)$ with the cylinder's axis
parallel to the $Z$ axis of the heliocentric coordinate frame is shown
in \figurename~\ref{fig:effsel_cyl} for different stellar types along
the main sequence. The top panel considers a narrow cylinder with a
radius of $R_{xy}=50\pc$, the bottom panel a broader cylinder with a
radius of $R_{xy}=250\pc$. Because the effective selection function of
giants in \figurename~\ref{fig:effsel_dist} closely resembles that of
earlier-type main-sequence stars, the effective volumes for giants are
similar to those of A, F, and G dwarfs and are not shown in
\figurename~\ref{fig:effsel_cyl}.

Because of the bright cut-off of \tgas, A, F, and G dwarfs have a hole
in their effective volume centered on the Sun (see above) for the
narrow cylinder. This hole is absent for the broader cylinder, because
it gets filled in by stars at $|Z| \approx 0$, but with $(X,Y) =
\mathcal{O}(100,100)\pc$. Because we only determined the
\tgas\ selection function over $48\,\%$ of the sky, the broader
cylinder has a complicated effective volume as a function of $Z$ for A
and F stars. The effective volume for K dwarfs is small beyond a few
$100\pc$ and M dwarfs are only sampled substantially by \tgas\ out to
about $50\pc$.

The dashed curves in \figurename~\ref{fig:effsel_cyl} represent the
effective volume when taking the full three-dimensional distribution
of interstellar extinction into account, while the solid curves assume
that there is no extinction. It is clear that extinction only has a
very minor, percent-level effect on the effective volume in vertical
slices near the Sun.

\begin{figure*}
  \includegraphics[width=0.99\textwidth,clip=]{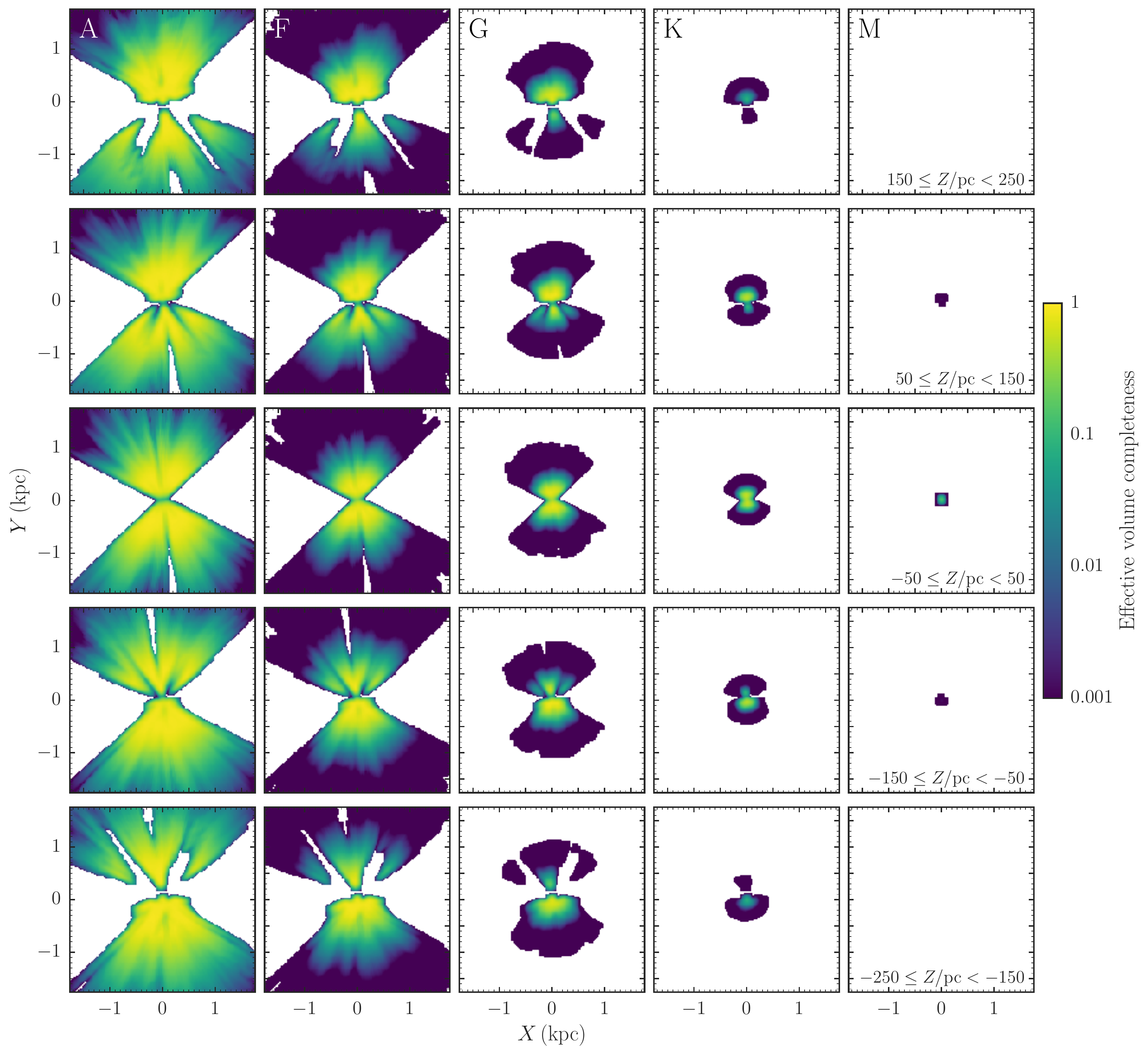}
  \caption{Effective volume completeness for different stellar types
    along the main sequence in $(100\pc)^3$ boxes. The volume
    completeness is displayed as a function of $(X,Y)$ in five $Z$
    slices. The incomplete, white parts for A and F stars are a result
    of the sky cut (mostly aligned with the ecliptic, see
    \appendixname~\ref{sec:tgascomplete}). The volume covered going
    from A to M dwarfs decreases rapidly, both because later type
    dwarfs are intrinsically fainter and because \tgas's completeness
    drops off at brighter magnitudes for red stars than for blue
    stars. }\label{fig:effsel_rectxy}
\end{figure*}

\begin{figure*}
  \includegraphics[width=0.99\textwidth,clip=]{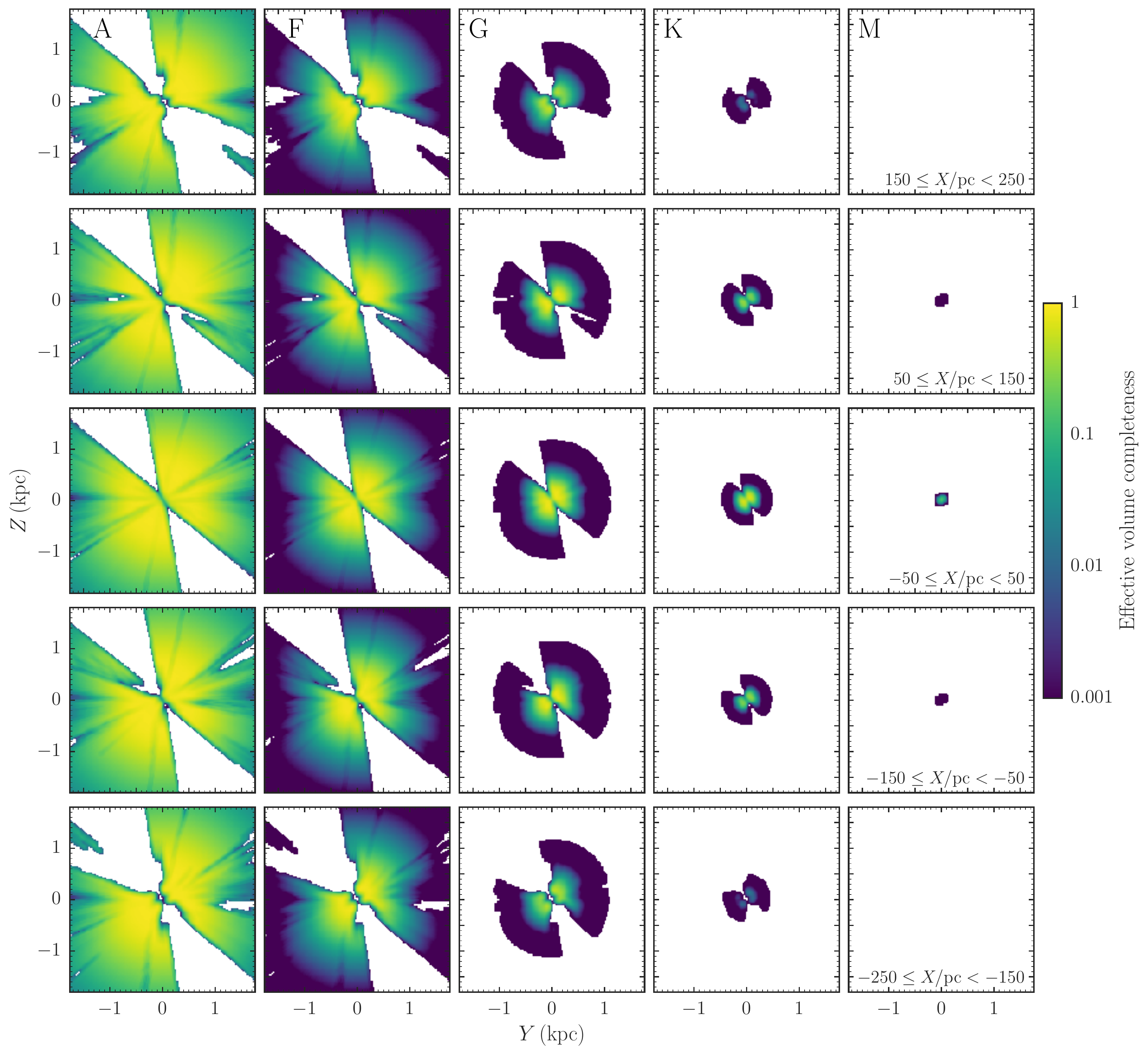}
  \caption{Like \figurename~\ref{fig:effsel_rectxy}, but for the
    volume completeness in the $(Y,Z)$ plane in five $X$
    slices. Because extinction is smaller looking out of the $Z=0$
    plane, the completeness remains high even at high $Z$ for A and F
    stars.}\label{fig:effsel_rectyz}
\end{figure*}

\begin{figure*}
  \includegraphics[width=0.86\textwidth,clip=]{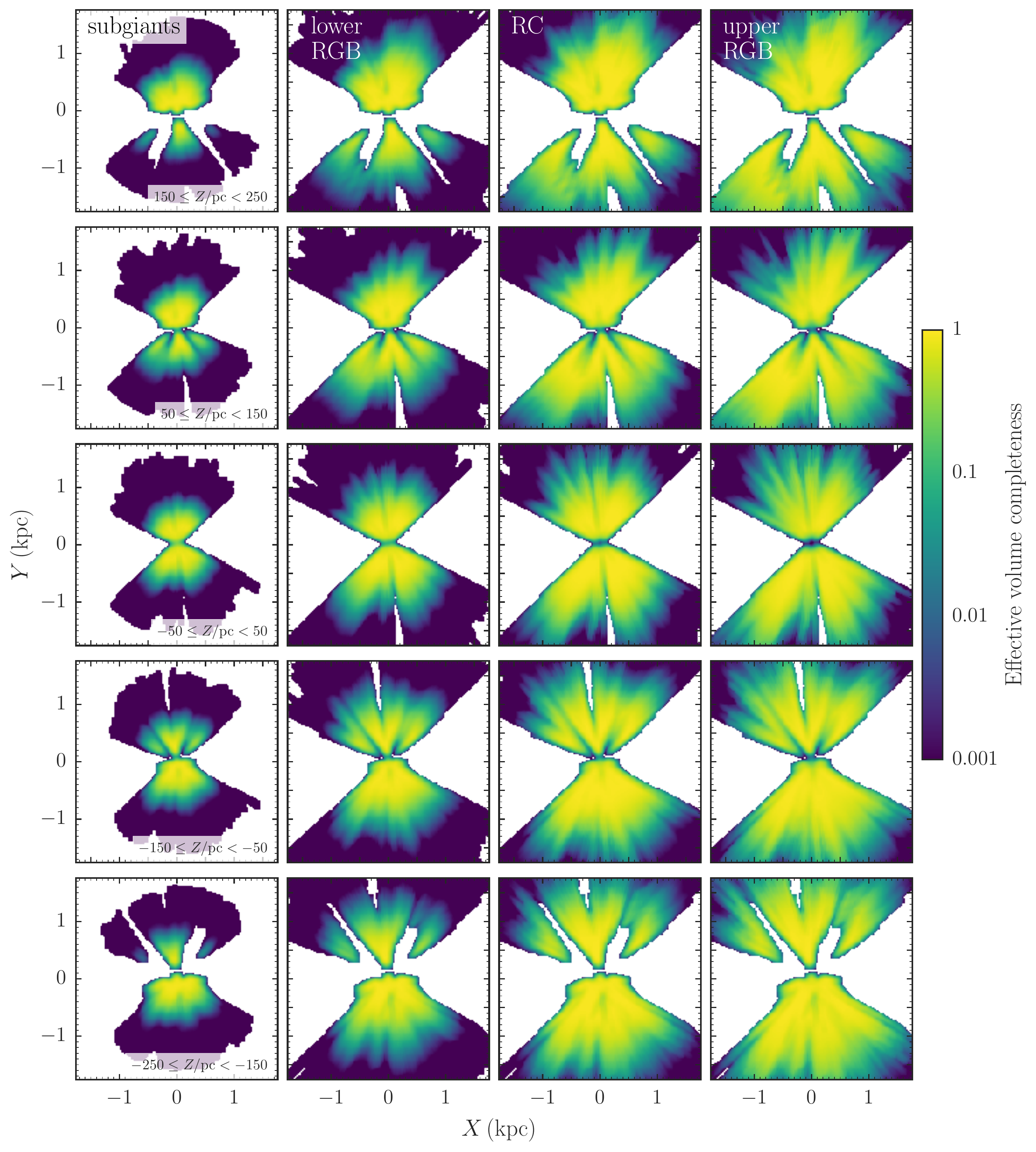}
  \caption{Like \figurename~\ref{fig:effsel_rectxy}, but for stars
    along the giant branch.}\label{fig:effsel_rectxy_giants}
\end{figure*}

\begin{figure*}
  \includegraphics[width=0.86\textwidth,clip=]{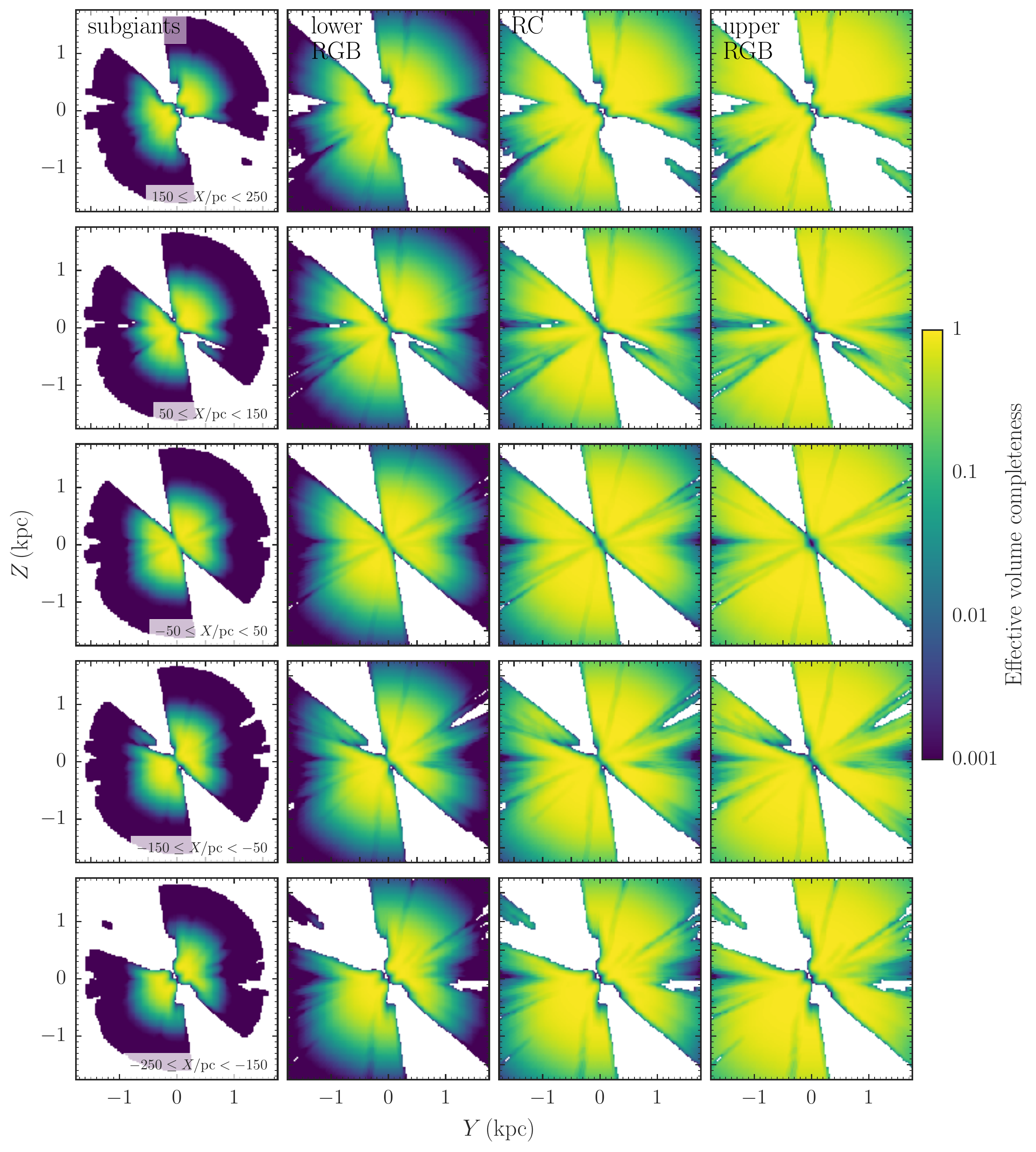}
  \caption{Like \figurename~\ref{fig:effsel_rectxy_giants}, but for
    the volume completeness in the $(Y,Z)$ plane in five $X$
    slices.}\label{fig:effsel_rectyz_giants}
\end{figure*}

To get a sense of how \tgas\ samples stars of different types over the
full three-dimensional volume around the Sun, we compute the effective
volume completeness $\Xi(\Pi_k)$ for cubic volumes in $(X,Y,Z)$ with a
volume of $(100\pc)^3$. This effective volume completeness is
displayed for different main-sequence stellar types in vertical slices
in \figurename~\ref{fig:effsel_rectxy} and in slices in $X$ in
\figurename~\ref{fig:effsel_rectyz}. The same for different types
along the giant branch is shown in \figurename
s~\ref{fig:effsel_rectxy_giants} and
\ref{fig:effsel_rectyz_giants}. These effective volume completenesses
take the full three-dimensional extinction map into account.

These maps of the \tgas\ volume completeness give a direct sense of
the volume probed by different stellar types, with early main-sequence
stars sampling a substantial fraction out to $\approx2\kpc$ out of the
plane and $\approx1\kpc$ in the plane and late-type dwarfs only
extending out to $\approx50\pc$. The intrinsically-bright giants ($M_J
\approx -2.5$) extend about as far as A dwarfs.

It is clear from these completeness maps that the \tgas\ completeness
is complex, because of the patchy nature of the `good'
\tgas\ footprint over which we determined the selection function and
because of the three-dimensional distribution of dust. Even if the
remaining 52\,\% of the sky were included in the selection function,
the volume completeness would remain complex, because the raw
\tgas\ completeness is lower and complicated in this 52\,\% of the
sky. These completeness maps need to be taken into account in any
investigation using \tgas\ for which the stellar density or the manner
in which a stellar population samples the local volume matters. Tools
to compute the effective completeness and the effective volume
completeness for a given stellar type are available in the
\texttt{gaia\_tools.select.tgasEffectiveSelect} class in the
\texttt{gaia\_tools} package available
at\\ \centerline{\url{https://github.com/jobovy/gaia_tools}~.}

\section{Stellar density laws for main-sequence stars}\label{sec:dens_ms}

We now combine the method for determining non-parametric, binned
stellar densities from \sectionname~\ref{sec:method_binned} with the
determination of the effective volume completeness from
\sectionname~\ref{sec:complete} to determine the stellar density and
its vertical dependence for different stellar types along the main
sequence. In \sectionname~\ref{sec:dens_ms_forward} we test the
effective volume completeness by directly comparing simple models for
the stellar density of various types of stars combined with the
effective volume completeness with the observed number counts in
\tgas. In \sectionname~\ref{sec:dens_ms_binned} we determine the
stellar density and its vertical dependence in narrow $Z$ bins and fit
these with simple analytic
profiles. \sectionname~\ref{sec:dens_ms_massfunc} focuses on the
measurement of the mid-plane stellar densities and its implications
for the present-day mass function, the star-formation history, and the
density of stellar remnants in the solar neighborhood. We postpone a
discussion of the Sun's offset from the mid-plane defined by different
types of stars to \sectionname~\ref{sec:zsun}, where we discuss the
measured offset for both main-sequence stars and giants.

In this and the subsequent section, we count as stars those
\tgas\ sources that fall within the lightly-shaded region shown in
\figurename~\ref{fig:cmd}, including the overlap with the
darkly-shaded region. This light-shaded region is a shifted and
stretched version of the mean dwarf stellar locus from
\citet{Pecaut13a} and of the giant locus described above, similar to
the dark-shaded region discussed in
\sectionname~\ref{sec:stellartypes}. It is designed, by hand, to
encompass the vast majority of stars that are plausibly part of the
dwarf and giant sequences, without including outliers. Because there
are only very few stars outside of the light-shaded area and all we
use are the number counts of stars, none of our results are
significantly affected by changes to this area. We estimate distances
simply as inverse parallaxes and only consider stars out to distances
of $444\pc = 0.2 / [0.45\mas]$. Because the parallax uncertainties in
the part of the sky that we consider are typically $<0.45\mas$ (see
\appendixname~\ref{sec:tgascomplete}), these stars typically have
relative parallax uncertainties better than 20\,\%, small enough that
the distance estimate from the inverse parallax is not strongly biased
\citep{BailerJones15a}. We only count stars up to $|Z| = 412.5\pc$ and
to a maximum $\sqrt{X^2+Y^2} < 250\pc$, so this distance cut only
affects these very furthest bins for the most luminous stars.

Because the effects of three-dimensional extinction are small (see
discussion in \sectionname\sectionname~\ref{sec:effsel} and
\ref{sec:effvol}), yet computationally laborious to compute, we ignore
extinction in this section. We have performed the analysis described
in this section taking into account the three-dimensional dependence
of extinction for the main spectral classes (all A dwarfs, all F
dwarfs, etc.) and find differences of only 1 to $2\,\%$ in the stellar
density distributions and inferred parameters of the vertical density
profile compared to assuming no extinction. These are a factor of a
few or more smaller than our statistical uncertainties.

\subsection{Forward modeling of the observed star counts}\label{sec:dens_ms_forward}

\begin{figure}
  \includegraphics[width=0.49\textwidth,clip=]{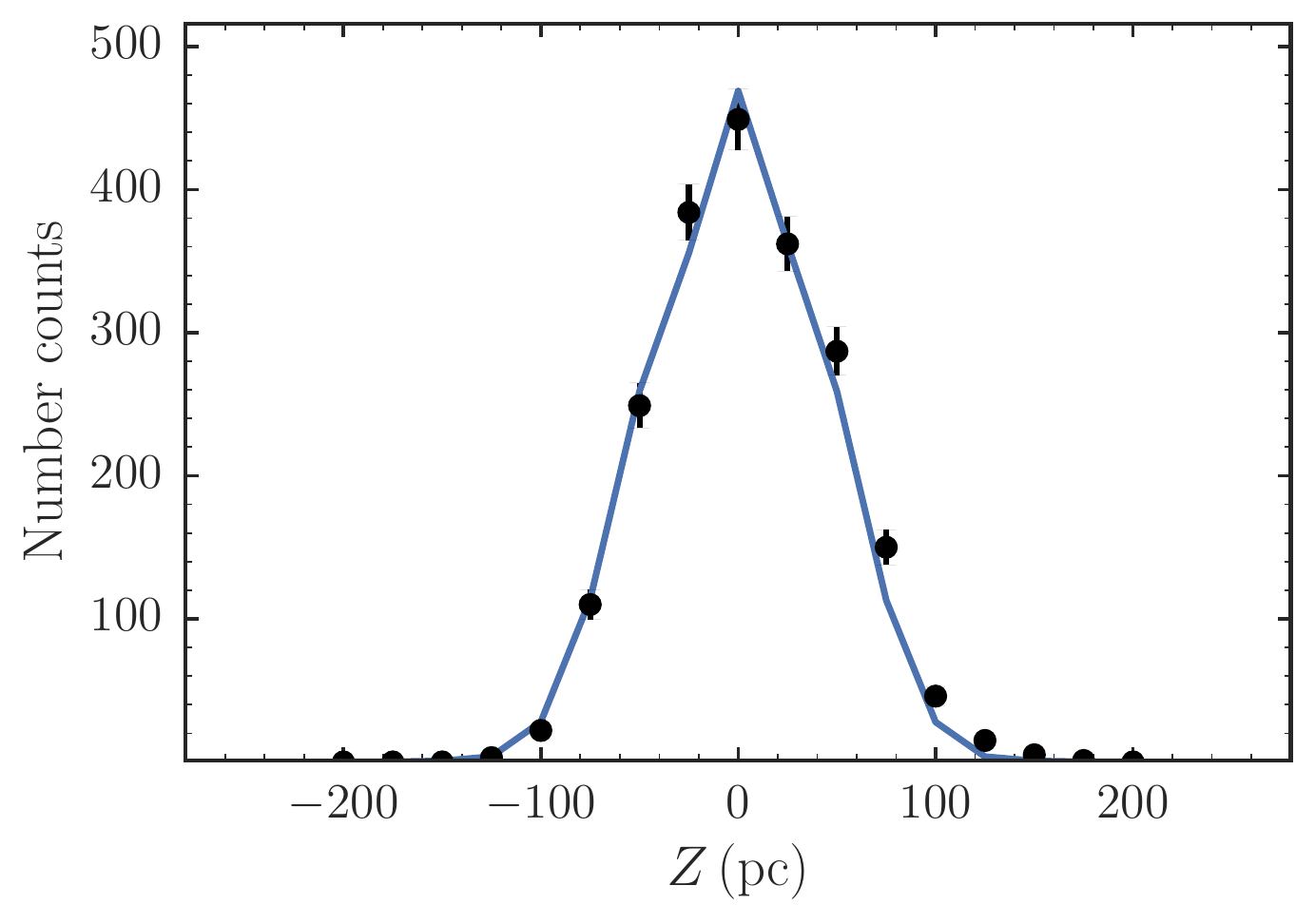}
  \caption{\tgas\ number counts of late K dwarfs (K5V through K9V, see
    \sectionname~\ref{sec:stellartypes}) in a cylinder with radius
    $R_{xy}=100\pc$ in $25\pc$ wide bins in $Z$. The blue curve is the
    effective volume completeness scaled to the number counts. Because
    the intrinsic density distribution of late K dwarfs is essentially
    constant within $-100\pc < Z < 100\pc$, the number counts should
    basically reflect the volume completeness, which is indeed what we
    find.}\label{fig:numcounts_latek}
\end{figure}

\begin{figure}
  \includegraphics[width=0.49\textwidth,clip=]{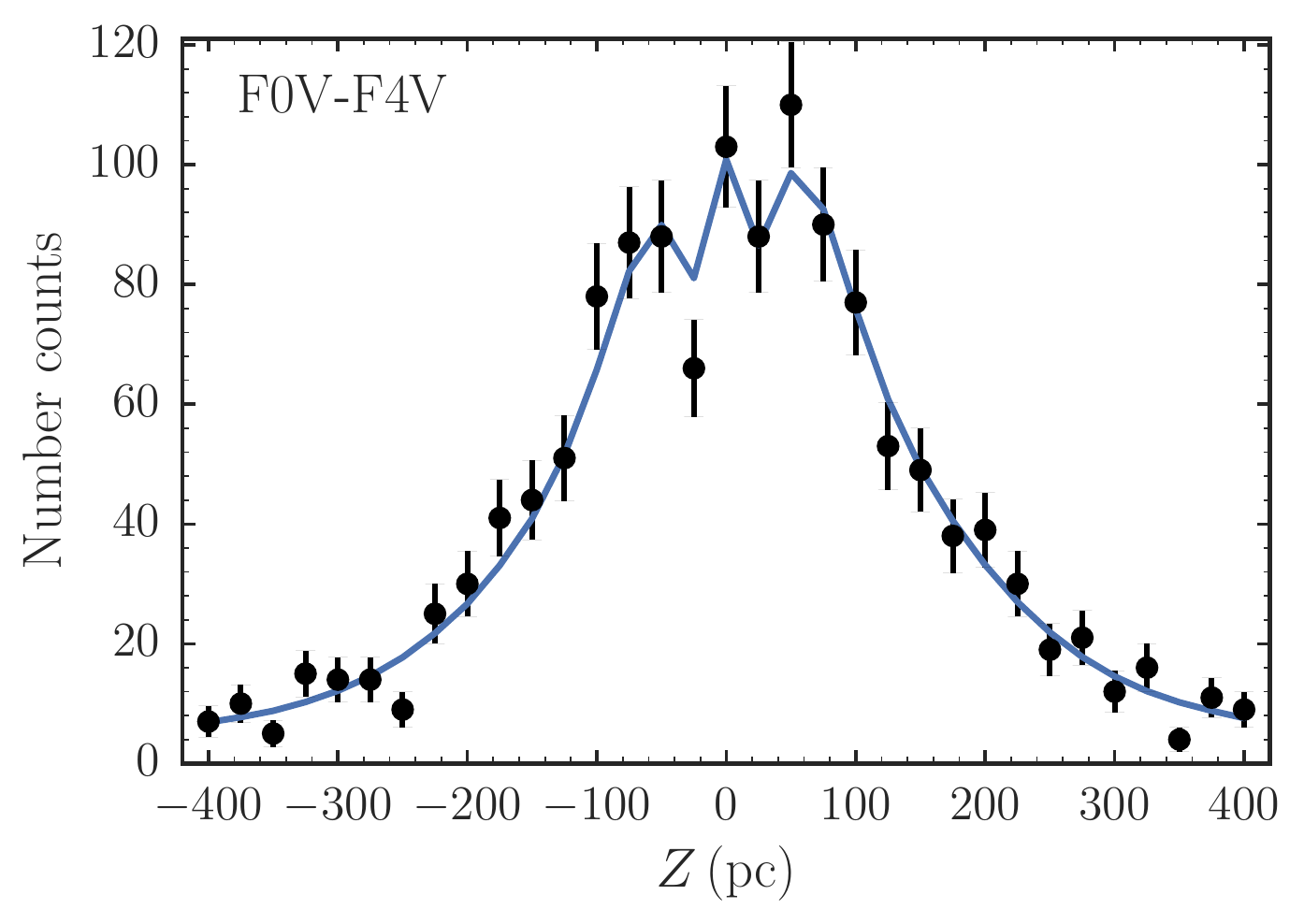}\\
  \includegraphics[width=0.49\textwidth,clip=]{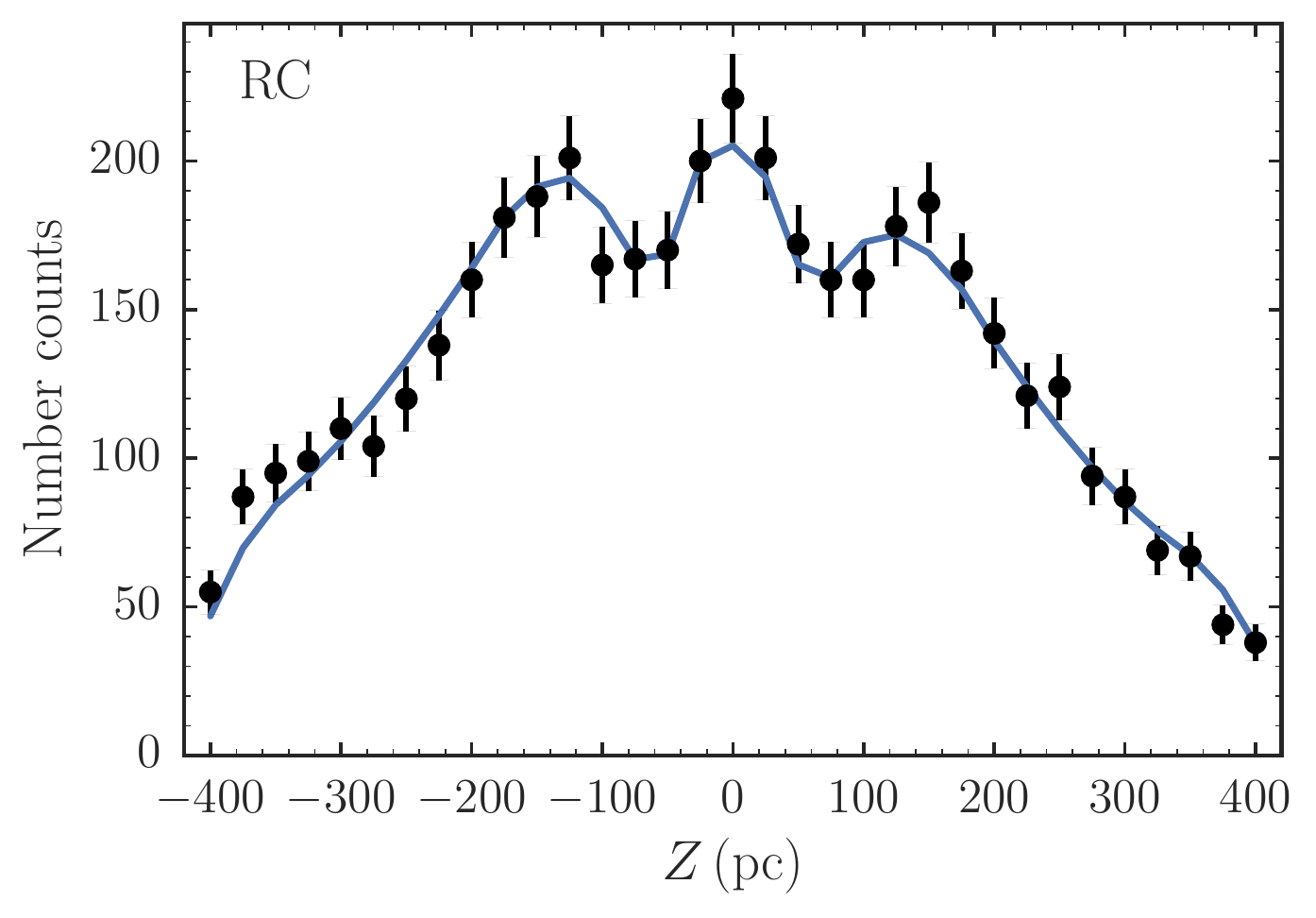}
  \caption{\tgas\ number counts of early F dwarfs (top) and of
    red-clump stars (bottom) in a cylinder with radius $R_{xy}=100\pc$
    and $250\pc$ for F dwarfs and red-clump giants respectively (see
    \sectionname~\ref{sec:stellartypes} for the definition of these
    subsamples). The blue curve displays the results from a fit of the
    underlying vertical stellar density with a $\sech^2$ profile
    multiplied with the effective volume.  The agreement between the
    number counts and the model is good.}\label{fig:numcounts_bright}
\end{figure}

Before discussing our determination of the underlying stellar density
profiles of main sequence stars, we test how well the observed star
counts for different stellar types are represented by the effective
volume completeness that we compute based on the raw \tgas\ selection
function from \appendixname~\ref{sec:tgascomplete} combined with the
model for the CMD of different stellar types from
\sectionname~\ref{sec:complete}.

The observed number counts of late K dwarfs (K5V through K9V) in a
$100\pc$ wide cylinder centered on the Sun in $25\pc$ wide bins in $Z$
are displayed in \figurename~\ref{fig:numcounts_latek}. From the
discussion above in \sectionname~\ref{sec:effvol}, it is clear that
late K dwarfs can only be observed at relatively high completeness out
to $|Z| \lesssim 100\pc$. From our prior understanding of the vertical
density profile of late K dwarfs, we know that the scale height of
these typically old stars is $\gg100\pc$ (\eg, \citealt{Juric08a}) and
the vertical density profile should thus be close to constant within
the observed \tgas\ volume. Therefore, we expect the observed number
counts to largely reflect the completeness of the survey rather than
the underlying stellar density profile. If the underlying density is
constant, then the counts in equal-volume bins are proportional to the
effective volume completeness: $N_k \propto \Xi(\Pi_k)$
(\equationname~\ref{eq:bfdens2}). The blue curve in
\figurename~\ref{fig:numcounts_latek} shows the vertical dependence of
the effective volume completeness, re-scaled to the median of the
three central points (to fix the proportionality constant). This
simple `model' matches the observed number counts well.

To compare the observed number counts to those predicted by the
effective volume completeness for stars that are more luminous or younger
than late K dwarfs, we need to model the underlying density
profiles. Below, we fit simple $\sech^2$ profiles to the binned
stellar density laws of different stellar
types. \figurename~\ref{fig:numcounts_bright} compares the observed
number counts for early F dwarfs (F0V to F4V) and for RC stars (discussed in
more detailed in \sectionname~\ref{sec:dens_giants} below) with
underlying stellar densities $\hat{n}_k$ fit as $\sech^2$ profiles and
multiplied by the effective volume ($N_k =
\hat{n}_k\times[\Xi(\Pi_k)\,V(\Pi_k)]$ from
\equationname~\ref{eq:bfdens2}). It is clear that these models fit the
observed number counts well: the vertical dependence and the dips near
$Z = 50\pc$ are the same in the model and the observed number
counts. The dips are caused by the geometry of the `good' part of the
sky.

We conclude that we can successfully model the observed number counts
of stars in \tgas\ using the effective volume completeness determined
in \sectionname~\ref{sec:complete}.

\subsection{Binned stellar densities along the main sequence}\label{sec:dens_ms_binned}

\begin{table*}
\caption{Properties, luminosity function, and mass function of stars along the main-sequence.}
\begin{tabular}{cccccr@{$\pm$}lr@{$\pm$}lr@{$\pm$}lr@{$\pm$}lr@{$\pm$}l}
\hline \hline
Spectral type & $J-K_s$ & $M_V$ & $M_J$ & Mass & \multicolumn{2}{c}{$n$} & \multicolumn{2}{c}{$\mathrm{d} n / \mathrm{d} M_V$} & \multicolumn{2}{c}{$\mathrm{d} n / \mathrm{d} M$}& \multicolumn{2}{c}{$\rho(>M)$} & \multicolumn{2}{c}{$z_d$}   \\ &  &  &  & $M_\odot$ & \multicolumn{2}{c}{$(10^{-4}\,\mathrm{pc}^{-3})$} & \multicolumn{2}{c}{$(10^{-4}\,\mathrm{pc}^{-3}\,\mathrm{mag}^{-1})$}  & \multicolumn{2}{c}{$(10^{-4}\,\mathrm{pc}^{-3}\,M_\odot^{-1})$} & \multicolumn{2}{c}{$(10^{-4}\,M_\odot\,\mathrm{pc}^{-3})$} & \multicolumn{2}{c}{(pc)}  \\\hline\hline
A0V & \!\!\!\!\!$-0.004$ & 1.34 & 1.25 & 2.15 & 0.38 & 0.04 & 1.4 & 0.2 & 2.2 & 0.3 & 0.8 & 0.1 & 49 & 6 \\
A1V & 0.007 & 1.48 & 1.32 & 2.05 & 0.38 & 0.04 & 2.0 & 0.2 & 3.0 & 0.3 & 1.6 & 0.2 & 51 & 6 \\
A2V & 0.024 & 1.55 & 1.35 & 2.00 & 0.41 & 0.04 & 3.9 & 0.3 & 5.5 & 0.5 & 2.4 & 0.2 & 51 & 6 \\
A3V & 0.032 & 1.76 & 1.47 & 1.90 & 0.69 & 0.04 & 4.9 & 0.3 & 9.1 & 0.6 & 3.7 & 0.3 & 46 & 3 \\
A4V & 0.059 & 1.84 & 1.51 & 1.85 & 0.75 & 0.05 & 5.2 & 0.3 & 10.1 & 0.6 & 5.1 & 0.4 & 52 & 3 \\
A5V & 0.069 & 1.89 & 1.54 & 1.83 & 0.35 & 0.04 & 5.4 & 0.6 & 10.1 & 1.1 & 5.8 & 0.5 & 37 & 6 \\
A6V & 0.074 & 2.07 & 1.64 & 1.76 & 0.60 & 0.04 & 5.2 & 0.4 & 13.4 & 0.9 & 6.8 & 0.6 & 53 & 4 \\
A7V & 0.095 & 2.29 & 1.78 & 1.67 & 1.01 & 0.05 & 5.0 & 0.3 & 12.6 & 0.7 & 8.5 & 0.6 & 55 & 3 \\
A8V & 0.117 & 2.30 & 1.78 & 1.67 & 0.63 & 0.04 & 5.5 & 0.3 & 14.0 & 0.8 & 9.6 & 0.7 & 52 & 4 \\
A9V & 0.121 & 2.51 & 1.90 & 1.59 & 0.65 & 0.03 & 5.9 & 0.3 & 16.1 & 0.9 & 10.6 & 0.8 & 56 & 4 \\
F0V & 0.143 & 2.79 & 2.13 & 1.50 & 1.19 & 0.06 & 4.8 & 0.2 & 14.0 & 0.7 & 12.4 & 0.8 & 64 & 3 \\
F1V & 0.166 & 2.99 & 2.26 & 1.44 & 1.41 & 0.06 & 5.9 & 0.2 & 18.9 & 0.8 & 14.4 & 0.9 & 67 & 3 \\
F2V & 0.190 & 3.08 & 2.32 & 1.43 & 1.23 & 0.05 & 8.5 & 0.4 & 35.1 & 1.5 & 16.2 & 1.0 & 64 & 7 \\
F3V & 0.198 & 3.23 & 2.42 & 1.39 & 0.88 & 0.04 & 7.4 & 0.3 & 35.3 & 1.7 & 17.4 & 1.1 & 75 & 4 \\
F4V & 0.211 & 3.40 & 2.55 & 1.33 & 1.48 & 0.05 & 9.3 & 0.3 & 29.6 & 1.1 & 19.4 & 1.1 & 62 & 7 \\
F5V & 0.227 & 3.70 & 2.78 & 1.25 & 2.54 & 0.07 & 10.8 & 0.3 & 36.3 & 1.0 & 22.5 & 1.2 & 80 & 3 \\
F6V & 0.256 & 3.87 & 2.90 & 1.21 & 3.19 & 0.08 & 13.6 & 0.3 & 53.2 & 1.3 & 26.4 & 1.3 & 89 & 4 \\
F7V & 0.273 & 4.01 & 3.01 & 1.18 & 2.41 & 0.07 & 15.5 & 0.4 & 68.8 & 1.9 & 29.2 & 1.4 & 97 & 5 \\
F8V & 0.286 & 4.15 & 3.11 & 1.14 & 2.54 & 0.06 & 18.1 & 0.5 & 72.5 & 1.8 & 32.1 & 1.5 & 104 & 6 \\
F9V & 0.300 & 4.45 & 3.34 & 1.08 & 4.46 & 0.09 & 20.3 & 0.4 & 89.3 & 1.8 & 36.9 & 1.6 & 118 & 8 \\
G0V & 0.329 & 4.50 & 3.38 & 1.07 & 4.10 & 0.11 & 23.4 & 0.6 & 117.0 & 3.1 & 41.3 & 1.7 & 108 & 11 \\
G1V & 0.335 & 4.79 & 3.57 & 1.02 & 4.21 & 0.09 & 24.7 & 0.5 & 140.2 & 3.1 & 45.6 & 1.8 & 177 & 17 \\
G2V & 0.366 & 4.86 & 3.64 & 1.00 & 4.76 & 0.12 & 26.4 & 0.7 & 135.9 & 3.5 & 50.4 & 1.9 & 123 & 20 \\
G3V & 0.373 & 4.94 & 3.70 & 0.99 & 2.05 & 0.07 & 27.4 & 0.9 & 136.9 & 4.6 & 52.4 & 2.0 & 133 & 39 \\
G4V & 0.382 & 4.98 & 3.74 & 0.98 & 1.49 & 0.07 & 24.8 & 1.1 & 148.9 & 6.7 & 53.9 & 2.0 &  \multicolumn{2}{c}{---} \\
G5V & 0.386 & 5.13 & 3.84 & 0.97 & 2.51 & 0.07 & 26.4 & 0.8 & 251.2 & 7.3 & 56.3 & 2.1 &  \multicolumn{2}{c}{---} \\
G6V & 0.403 & 5.18 & 3.88 & 0.96 & 2.59 & 0.08 & 25.9 & 0.8 & 259.1 & 7.6 & 58.8 & 2.2 &  \multicolumn{2}{c}{---} \\
G7V & 0.409 & 5.32 & 3.97 & 0.94 & 2.39 & 0.05 & 25.1 & 0.5 & 159.0 & 3.5 & 61.0 & 2.2 &  \multicolumn{2}{c}{---} \\
G8V & 0.424 & 5.55 & 4.14 & 0.90 & 4.67 & 0.14 & 25.2 & 0.8 & 155.5 & 4.7 & 65.2 & 2.4 &  \multicolumn{2}{c}{---} \\
G9V & 0.452 & 5.76 & 4.29 & 0.87 & 5.22 & 0.14 & 23.7 & 0.6 & 149.1 & 4.1 & 69.8 & 2.5 &  \multicolumn{2}{c}{---} \\
K0V & 0.478 & 5.91 & 4.47 & 0.85 & 4.01 & 0.12 & 22.3 & 0.7 & 160.4 & 4.7 & 73.2 & 2.6 &  \multicolumn{2}{c}{---} \\
K1V & 0.499 & 6.19 & 4.57 & 0.82 & 4.65 & 0.12 & 21.6 & 0.6 & 185.9 & 4.8 & 77.0 & 2.7 &  \multicolumn{2}{c}{---} \\
K2V & 0.531 & 6.57 & 4.76 & 0.78 & 8.89 & 0.20 & 26.9 & 0.6 & 254.1 & 5.7 & 83.9 & 2.8 &  \multicolumn{2}{c}{---} \\
K3V & 0.600 & 7.04 & 4.98 & 0.73 & 13.72 & 0.29 & 32.3 & 0.7 & 304.9 & 6.4 & 93.9 & 3.0 &  \multicolumn{2}{c}{---} \\
K4V & 0.669 & 7.25 & 5.11 & 0.72 & 10.87 & 0.35 & 32.0 & 1.0 & 362.4 & 11.7 & 101.8 & 3.3 &  \multicolumn{2}{c}{---} \\
\hline
\end{tabular}
\end{table*}

\begin{figure*}
  \includegraphics[width=0.99\textwidth,clip=]{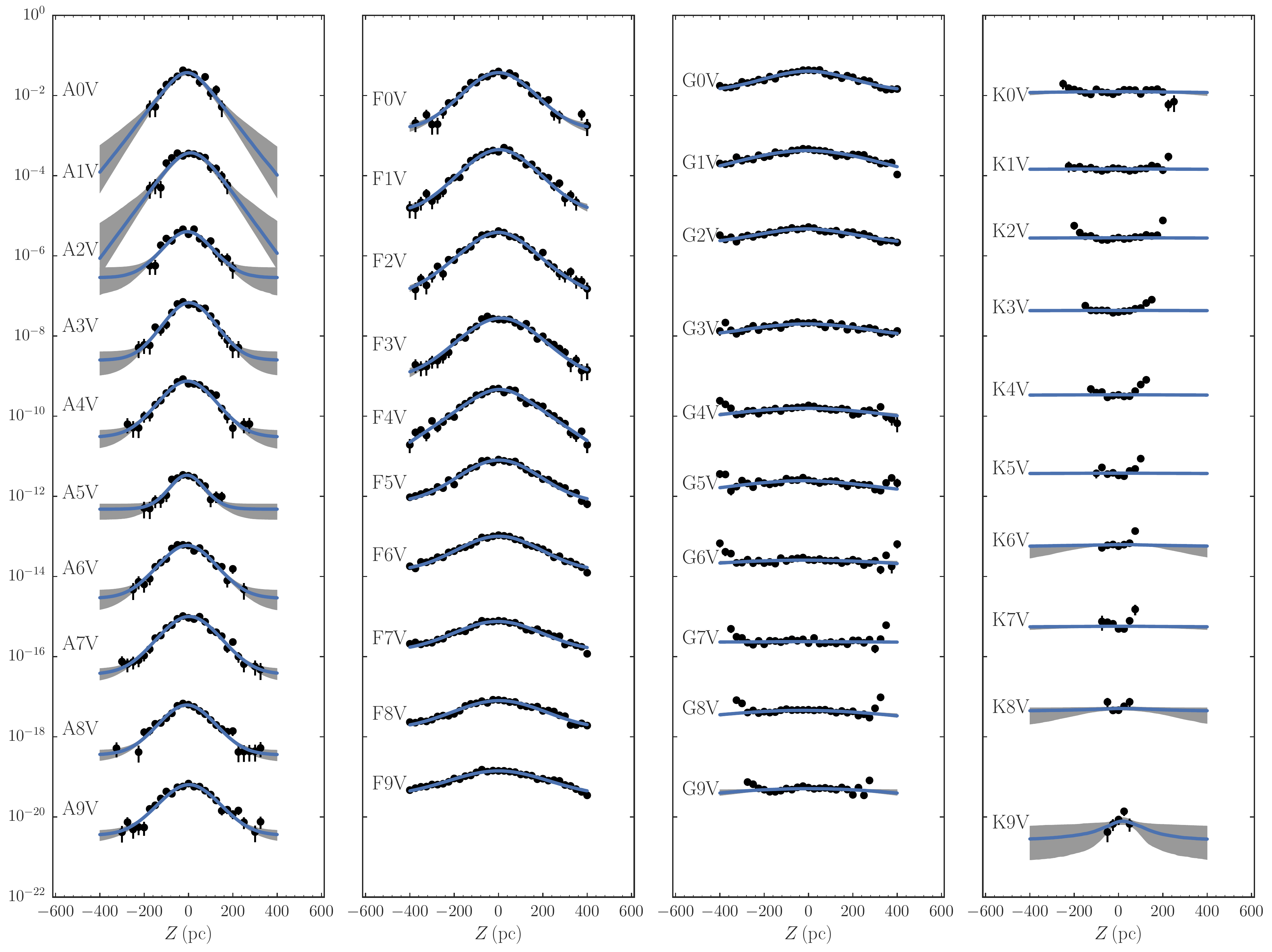}
  \caption{Vertical number density profiles of main-sequence stars of
    different stellar types. The black dots with uncertainties (mostly
    too small to see) are the data (\tgas\ number counts / effective
    volume), the blue curve is a $\sech^2$ fit and the gray band is
    the $68\,\%$ uncertainty range of the $\sech^2$ model. The
    profiles have been arbitrarily shifted in the $y$ direction. The
    vertical density flattens near the mid-plane for all stellar types
    and broadens as one moves from earlier to later stellar types. At
    later K dwarfs the density becomes difficult to measure because of
    \tgas's incompleteness.}\label{fig:densprofiles}
\end{figure*}

The binned vertical stellar density of stellar subtypes from A0V to
K9V are displayed in \figurename~\ref{fig:densprofiles}. To account
for the fact that luminous A dwarfs can be seen to much larger
distances than faint K dwarfs, these vertical densities are determined
in $25\pc$ wide ranges in $Z$ in a cylinder that extends out to
$R_{xy}=250\pc$ for A dwarfs, $200\pc$ for F dwarfs, $150\pc$ for G
dwarfs, and $100\pc$ for K dwarfs. We do not consider M dwarfs,
because the volume over which \tgas\ has relatively high completeness
to M dwarfs is very small. While we determine absolute stellar
densities, we shift the profiles in the $y$ direction in this figure
to better display the vertical dependence. For each stellar type, we
only show bins with (a) more than four stars, (b) effective volume
completeness larger than $3\,\%$ of the maximum effective volume
completeness for the stellar type in question, and (c) effective
volume completeness larger than 1 part in $10^5$, except for the $Z=0$
bin. These cuts are designed to weed out bins where our stellar
densities are highly noisy.

It is clear that the vertical profiles in
\figurename~\ref{fig:densprofiles} broaden significantly when going
from the earliest types of A dwarfs to later G and K dwarfs. By G5V,
\tgas\ does not extend far enough to see a substantial decrease in the
stellar density with vertical distance from the Sun. By K5V, the
volume over which \tgas\ has significant completeness has become so
small that it is difficult to measure the density at all.

We fit each stellar type's number density profile by minimizing the
$\chi^2$ deviation between the observed
$\hat{n}_k\pm\sigma_{\hat{n}_k}$ and a model composed of two $\sech^2$
profiles
\begin{equation}
  \dens(Z) =
  n\,\left[(1-\alpha)\,\sech\left(\frac{Z+\zsun}{2\,z_d}\right)^2
    +\alpha\,\sech\left(\frac{Z+\zsun}{2\,z_{d,2}}\right)^2\right]\,,
\end{equation}
where $n$ is the mid-plane density, $z_d$ is the scale height of the
main $\sech^2$ component, $\zsun$ is the Sun's offset from the
mid-plane, $\alpha$ is the fraction of the mid-plane density that is
part of the second $\sech^2$, and $z_{d,2}$ is the scale height of the
second $\sech^2$ component, constrained to be larger than $z_d$. This
$z_{d,2} \gg 1\kpc$ in almost all cases, such that the second
$\sech^2$ profile is essentially a constant density. The second
component typically has a small amplitude $\alpha$, especially for the
earlier types (for later types, it becomes degenerate with the first
component, because these types' densities are close to
constant). Thus, the measured number density profiles are
well-represented by $\sech^2$ profiles within $Z \lesssim4\,z_d$ or $Z
< 400\pc$ (if $4\,z_d > 400\pc$). The fit are shown as blue curves in
\figurename~\ref{fig:densprofiles}; the gray band displays the
$68\,\%$ confidence region obtained by MCMC sampling of the likelihood
defined as $\exp(-\chi^2/2)$.

The parameters for each stellar type together with their uncertainties
are given in \tablename~1. We list the mean color, absolute magnitude
in $V$ and $J$, and the mass of each type determined from the dwarf
locus from \citet{Pecaut13a}. We also give the derived quantities $\dd
n / \dd M_V$ (the luminosity function), $\dd n / \dd M$ (the mass
function), and the mid-plane mass density $\rho(>M)$ of stars with
masses down to that of the stellar type in question. For types later
than K4V, we are not able to determine reliable parameters because of
the lack of data (see \figurename~\ref{fig:densprofiles}); for types
later than G3V, we are unable to determine the scale height because
the distance out to which these stars can be seen in \tgas\ is too
small compared to their scale height.

\begin{figure}
  \includegraphics[width=0.49\textwidth,clip=]{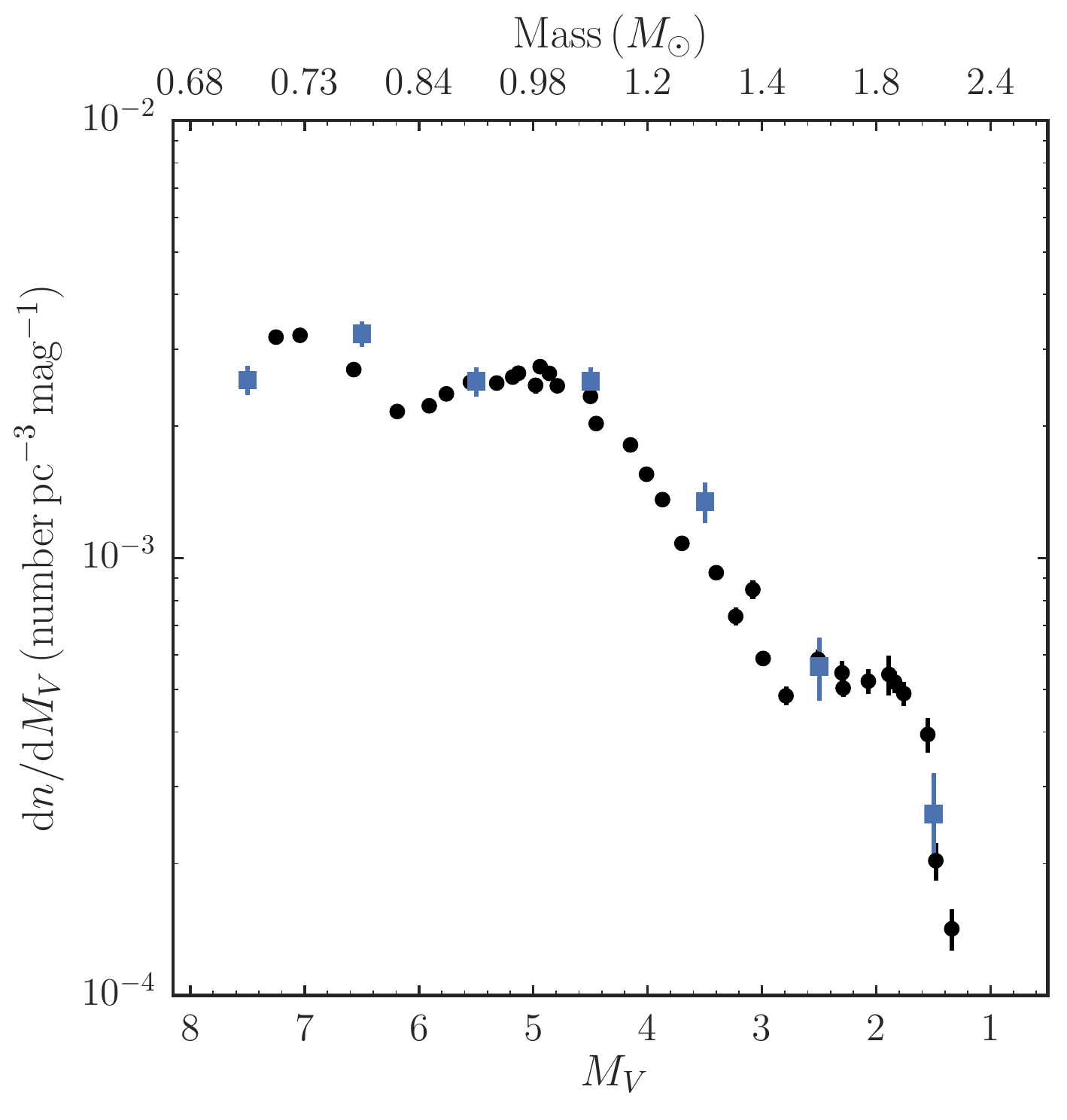}
  \caption{Luminosity function of main-sequence stars from A0V to
    K5V. Statistical uncertainties are typically smaller than the
    marker. The blue squares are measurements of the luminosity
    function from a volume complete sample within $25\pc$ of the Sun
    from \citet{Reid02a}.} \label{fig:lumfunc}
\end{figure}

The luminosity function $\dd n / \dd M_V$ is shown in
\figurename~\ref{fig:lumfunc}. We compare the luminosity function to
that determined by \citet{Reid02a} from a volume complete sample
within $25\pc$ from the Sun. The agreement between these two
determinations is good; importantly, they are essentially based on
non-overlapping volumes, because for G dwarfs and earlier types,
\tgas\ only starts observing stars at distances greater than about
$25\pc$. This good agreement is partly due to the fact that we find
that the Sun is at the mid-plane of the density of A and F stars (the
measured Solar offsets using different stellar types are displayed in
\figurename~\ref{fig:zsun} below and discussed in more detail together
with the equivalent measurements for giants in
\sectionname~\ref{sec:zsun}). If this were not the case, the $25\pc$
volume of \citet{Reid02a} would be offset from the mid-plane by half a
scale height. The new determination based on \tgas\ data has much
higher resolution in absolute magnitude than \citet{Reid02a}, which
has $\Delta M_V = 1$ compared to typical $\Delta M_V \approx 0.15$
here (of course, \citealt{Reid02a} extend down to $M_V = 20$, while we
are limited to $M_V \lesssim 7.5$).

\begin{figure}
  \includegraphics[width=0.49\textwidth,clip=]{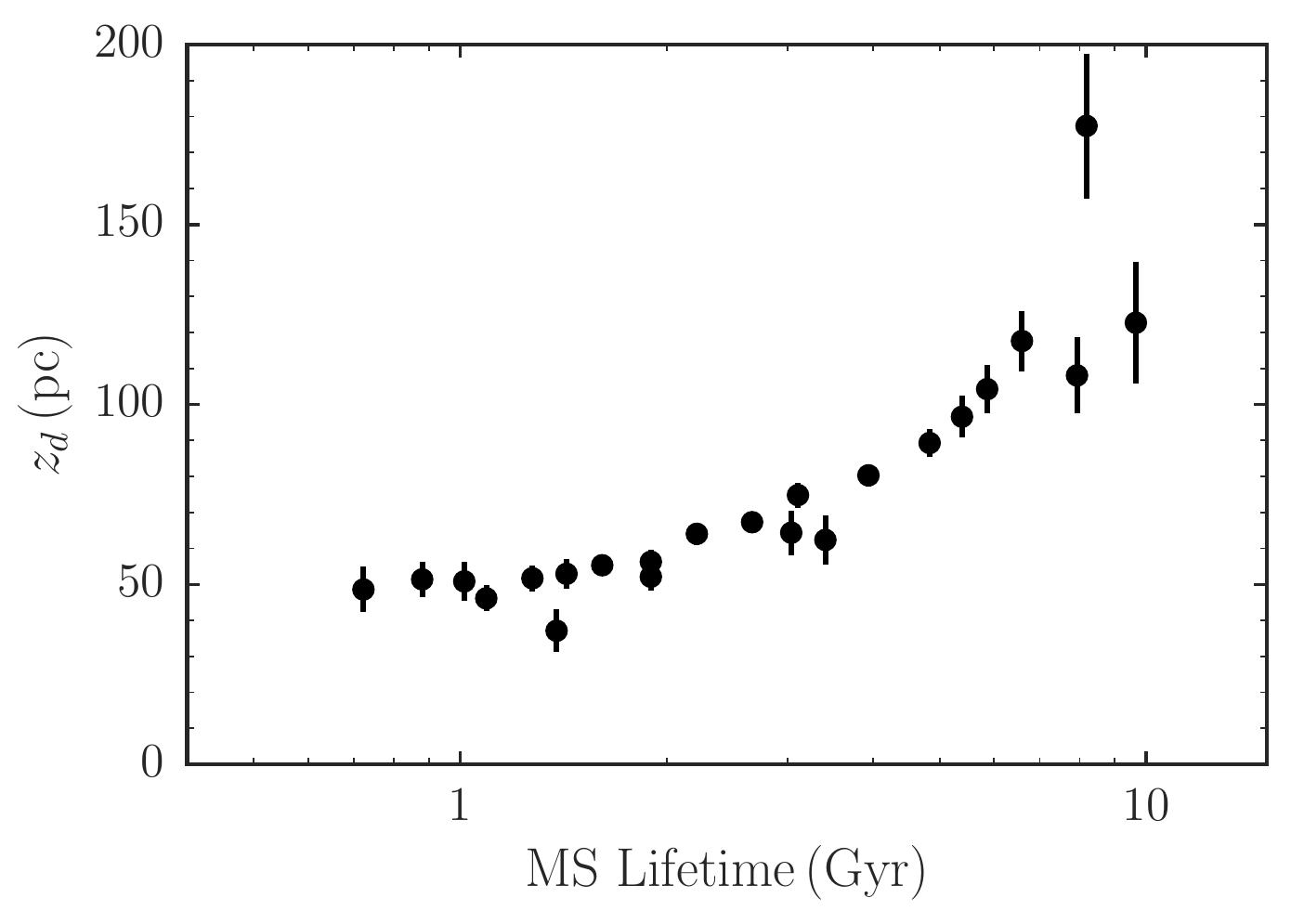}
  \caption{Scale height of the $\sech^2$ fits for stellar types along
    the main sequence (A, F, and early G-type dwarfs to G2V),
    displayed as a function of their main-sequence lifetime. The scale
    heights increase smoothly from $\approx50\pc$ for A stars to
    $\approx150\pc$ for early G dwarfs.}\label{fig:zd}
\end{figure}

The scale height for different stellar types is shown in
\figurename~\ref{fig:zd} as a function of
their main-sequence lifetime $\tau$, computed as in \citet{Reid02a}
\begin{equation}\label{eq:ms_lifetime}
  \log_{10} \tau(M) = 1.015-3.491\,\log_{10} M+0.8157\,(\log_{10} M)^2\,.
\end{equation}
Stars of a given stellar type will largely be younger than this
age. The scale height rises gradually from $\approx50\pc$ for A dwarfs
to $\approx100\pc$ for late F dwarfs and larger for older G
dwarfs. The scale height of young populations of stars is less than
half of that of the atomic and molecular hydrogen \citep{McKee15a},
but similar to that of young open clusters \citep[$\approx50$ to
  $75\pc$, \eg,][]{Bonatto06a,Buckner14a,Joshi16a} and of OB stars
\citep[$\approx50\pc$,][]{Reed00a}.

\subsection{The mass function and star-formation history of the solar neighborhood}\label{sec:dens_ms_massfunc}

\begin{figure}
  \includegraphics[width=0.49\textwidth,clip=]{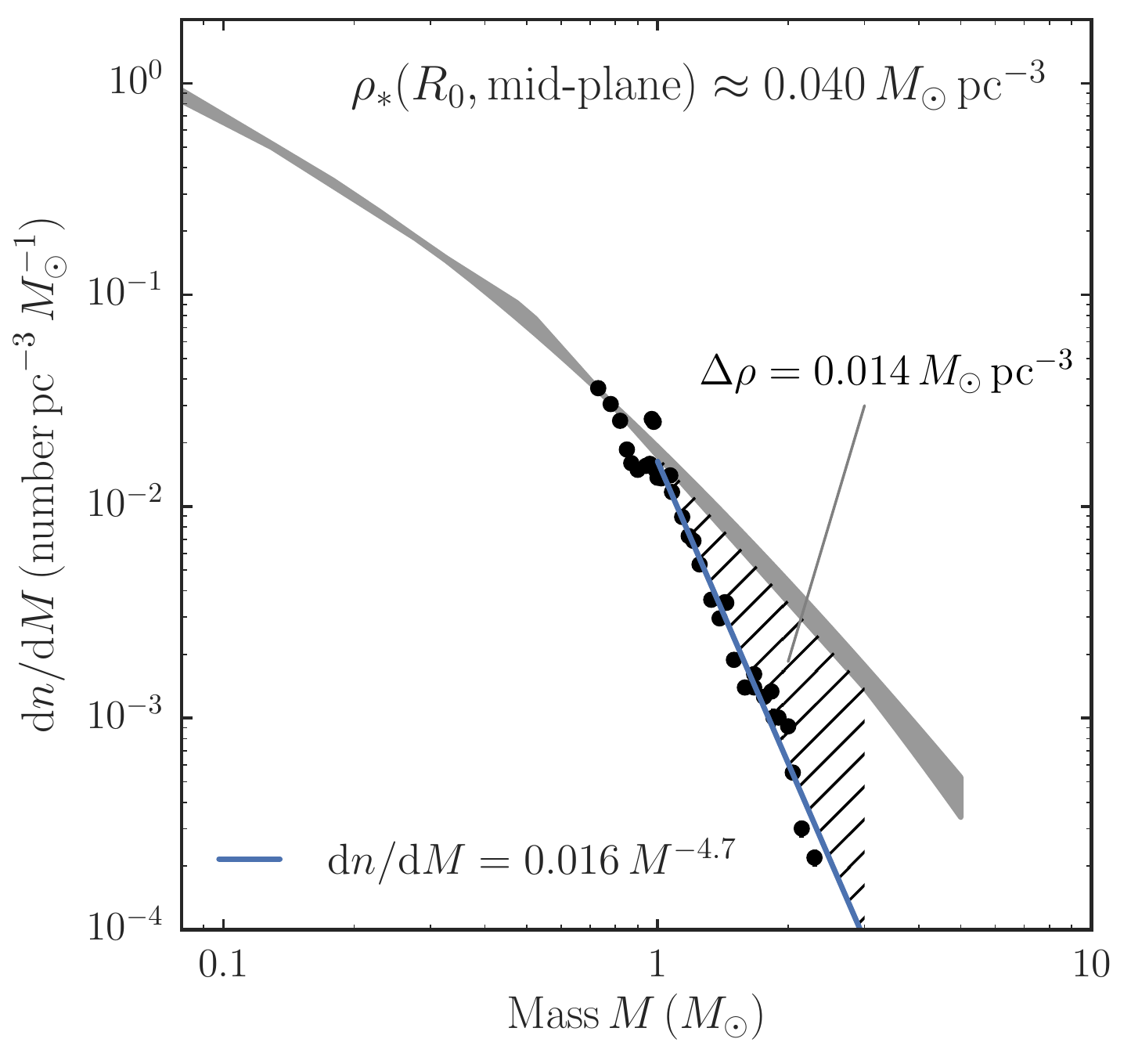}
  \caption{Mass function of main-sequence stars from A0V to K5V. The
    gray band shows the range spanned by models of the IMF from
    \citet{Chabrier01a} (lognormal and exponential) and from
    \citet{Kroupa01a} when anchored to the measured density of
    long-lived stars ($0.7\msun \leq M \leq 0.8\msun$). The high-mass
    end ($M > 1\msun$) is fit with the power-law model given by the
    blue line. Extrapolating below K5V using the IMF model, the total
    mid-plane stellar density is
    $0.040\pm0.002\msun\pc^{-3}$. Compared to the IMF, the amount of
    mass in the shaded region is missing due to stellar
    evolution.}\label{fig:massfunction}
\end{figure}

Using a relation for the mass as a function of stellar type along the
main sequence, we can translate the luminosity function of
\figurename~\ref{fig:lumfunc} into a mass function $\dd n / \dd M$. We
use masses for stellar types derived from the dwarf mean stellar locus
from \citet{Pecaut13a}. The resulting mass function is shown in
\figurename~\ref{fig:massfunction}. This function gives the mass
density at the mid-plane in units of numbers of stars per solar mass
and cubic pc. This is the present-day mass function. For long-lived
stars this should reflect the IMF, while for short-lived stars the
present-day mass function results from the combination of the IMF,
stellar evolution, and kinematic heating. Our latest-type stars for
which we can measure reliable mid-plane densities (K2V through K4V)
are long-lived enough to trace all mass formed over the history of the
disk \citep[\eg,][]{Bressan12a} and thus reflect the IMF for these
types. Therefore, we anchor IMF models to the observed $\dd n / \dd M$
of these populations. In particular, we employ the lognormal and
exponential IMF models from \citet{Chabrier01a} and the broken-power
law IMF model from \citet{Kroupa01a}. The range spanned by these three
IMFs when anchored to the K dwarfs (using the median proportionality
constant) is shown as the gray band in
\figurename~\ref{fig:massfunction}.

We fit the high-mass end of the mass function using a power-law and
find
\begin{equation}
\frac{\dd n}{\dd M} = 0.016\,\left(\frac{M}{\msun}\right)^{-4.7}\,\msun^{-1}\pc^{-3}\,,\quad M > 1\msun\,.
\end{equation}
This can be compared to the mass function obtained by
\citet{Scalo86a}, who finds $\dd n / \dd M =
0.019\,(M/\msun)^{-5.4}\,\msun^{-1}\pc^{-3}$; these differ by less
than $25\,\%$ within the relevant mass range. Combining the power-law
high-mass mass function at $M > 1\msun$ with the IMF fit to the K
dwarfs at $M \leq 1\msun$, we can compute the total mid-plane density
of main-sequence stars. This is
\begin{equation}\label{eq:rhoms}
\rho^{\mathrm{MS}}_* = 0.040\pm0.002\msun\pc^{-3}\,,
\end{equation}
where the uncertainty includes a contribution from the uncertainty in
the raw \tgas\ selection function for K dwarfs. As shown in
\tablename~1, only $0.01\msun\pc^{-3}$ of this is directly determined
from our stellar density measurements and the rest is an extrapolation
down to $M = 0.08\msun$ (assumed here to be the smallest mass above
which core-hydrogen-burning occurs). This is in good agreement with
previous determinations, \eg, the measurement of $\rho^{\mathrm{MS}}_*
= 0.036\pm0.004\msun\pc^{-3}$ from \citet{McKee15a} based in large
part on the measurements from \citet{Reid02a}.

Above $1\msun$, the difference between the present-day mass function
and the IMF is $\approx0.014\msun\pc^{-3}$ between $1 < M/\msun <
3$. If we extrapolate to $M = 8\msun$, this difference is
$\approx0.027\msun\pc^{-3}$ between $1 < M/\msun < 8$. Assuming that
all of the stars in this initial mass range turn into white dwarfs,
the corresponding mass density of white dwarfs would be
$0.0035\msun\pc^{-3}$ using the initial-mass to final-mass relation of
\citet{Kalirai08a} (as we will see below, the mass contained in giant
populations is negligible in this context). This is a lower limit,
because this estimate does not take into account that the older stars
have more extended vertical profiles. Accounting for a typical factor
of two difference in $z_d$ between old and young populations, the
predicted density of white dwarfs becomes $0.0065\msun\pc^{-3}$, in
good agreement with the value of $0.0056\pm0.0010\msun\pc^{-3}$ from
\citet{McKee15a}.

\begin{figure}
  \includegraphics[width=0.49\textwidth,clip=]{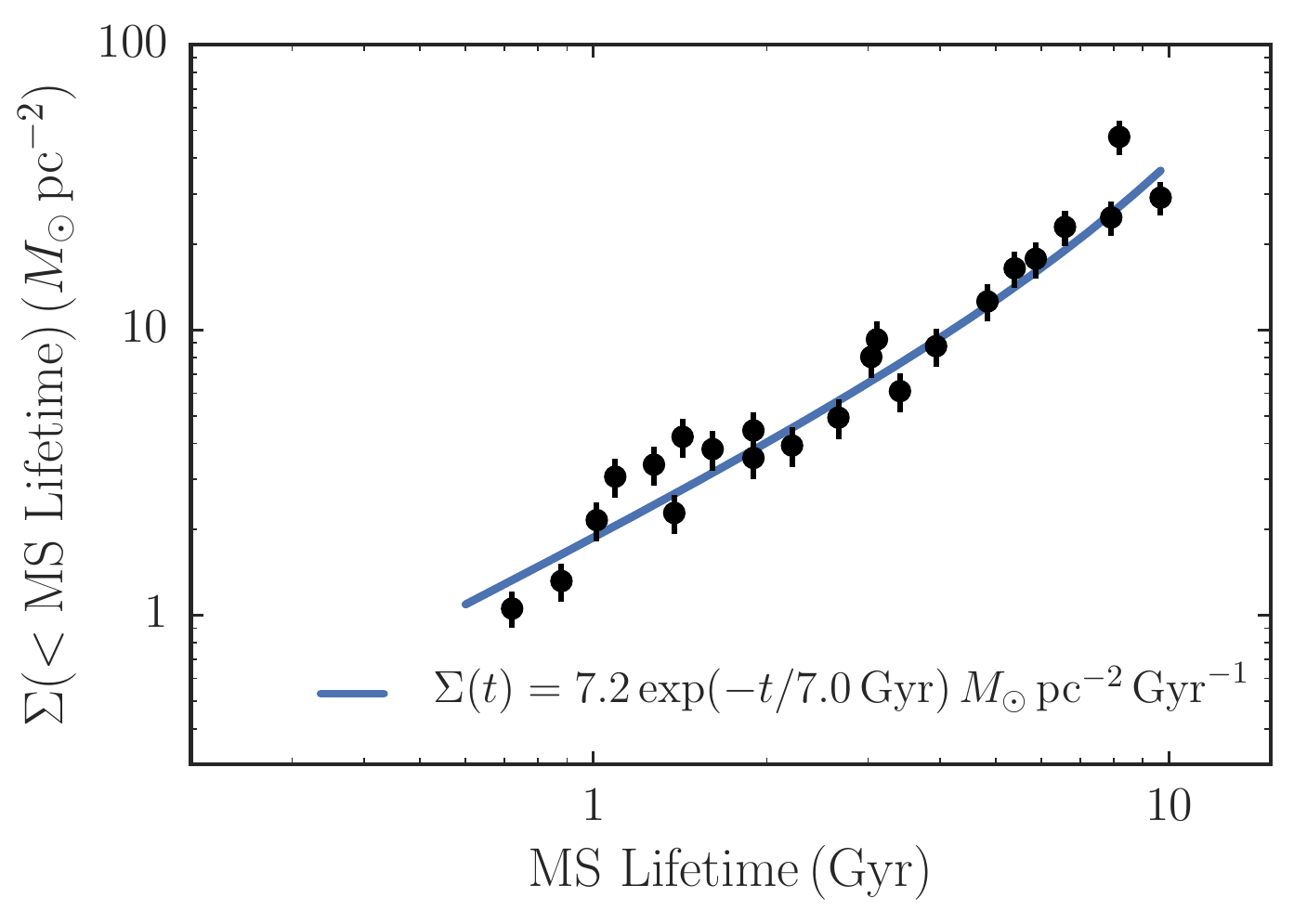}
  \caption{Star-formation history of the solar neighborhood. This
    figure shows the cumulative total surface density of stars formed
    as a function of lookback time. These surface densities are
    obtained by extrapolating the observed column densities of
    different stellar types to full stellar populations using IMF
    models under the assumption that each stellar type traces all
    stars formed up to its main-sequence lifetime. The blue curve is a
    fit of an exponentially-declining star-formation rate to these
    data; uncertainties in the fit parameters are about $15\,\%$ and
    almost exactly anti-correlated. The exponentially-declining
    star-formation rate provides a good fit to the data. The total
    amount of mass formed into stars is
    $38.5\pm2.5\msun\pc^{-2}$.}\label{fig:sfh}
\end{figure}

When corrected for the effect of vertical heating, we can translate
the present-day mass function into a cumulative star-formation
history. To do this, we integrate the observed number volume density
corresponding to the dominant $\sech^2$ component over $Z$ and obtain
the observed number surface density for all stellar types. Similar to
the volume density above, we can represent the high-mass end as a
power-law:
\begin{equation}\label{eq:dNdM}
\frac{\dd N}{\dd M} = 9.3\,\left(\frac{M}{\msun}\right)^{-6.5}\,\msun^{-1}\pc\
^{-2}\,,\quad M > 1\msun\,.
\end{equation}
We turn the observed number surface density into an estimate of the
total stellar mass formed going backwards in time from the present day
up to the main-sequence lifetime of the stellar type, by running an
IMF through the surface density measurement of each stellar type and
integrating over all masses. That is, the current surface density $\dd
N_i / \dd M (\tau_i)$ for a stellar type $i$ with main-sequence
lifetime $\tau_i$ corresponds to a total stellar mass
$\Sigma_*(\tau_i)$ up to $\tau_i$ in the past if we anchor an IMF to
$\dd N_i / \dd M(\tau_i)$ and integrate over all masses. The (reverse)
cumulative star-formation history of the solar neighborhood thus
determined is displayed in \figurename~\ref{fig:sfh}, where the
uncertainties are due to the uncertainty in the IMF (marginalizing
over the three IMF models that we consider). We see that about
$2\msun\pc^{-2}$ was formed in the last Gyr and that the increase
towards higher ages is faster than linear, indicating that the
star-formation rate has decreased in time.

We fit a model for an exponentially-declining (or increasing)
star-formation rate to these measurements and find that the
star-formation rate declines in time and is given by
\begin{equation}\label{eq:sfh}
  \Sigma_{\mathrm{SFR}}(t) = 7.2\pm1.0\,\exp(-t/7\pm1\,\mathrm{Gyr})\,
  M_\odot\,\mathrm{pc}^{-2}\,\mathrm{Gyr}^{-1}\,,
\end{equation}
where $t$ is time measured from $10\Gyr$ ago.  The total amount of
stellar mass formed in the last 10 Gyr is $\Sigma_{\mathrm{form}} =
38.5\pm2.5\msun\pc^{-2}$. The declining star-formation rate found here
is consistent with the analysis of \citet{Aumer09a}, who find the
star-formation rate to be $2--7$ times lower now than at the start of
star formation in the disk.

From this measurement of the total mass formed, we can estimate the
mass that is presently still contained in visible stars. Based on
maximum masses as a function of age from PARSEC \citep{Bressan12a}
isochrones and a \citet{Kroupa01a} IMF, we find that the ratio of
current-to-formed stellar mass is
\begin{align}
  \frac{\Sigma_*(\tau)}{\Sigma_{\mathrm{form}}} & = -0.0313\,x^2-0.182\,x+0.75\,,\\ & \qquad \mathrm{with}\ x = \log_{10}\left(\tau / \mathrm{Gyr}\right)\quad (\mathrm{solar\ metallicity})\,.\nonumber
\end{align}
At approximately half-solar metallicity, this fraction is only about a
percent lower; for other IMFs (\eg, those from \citealt{Chabrier01a}),
this fraction can be different up to $\approx5\,\%$. For the
exponentially-declining star-formation history, this gives $\Sigma_* =
23.0\pm1.5\msun\pc^{-2}$ or about $60\,\%$ of the formed mass; this
agrees, largely by construction, with directly counting all of the
mass above $1\msun$ using \equationname~\eqref{eq:dNdM} and combining
it with mass below $1\msun$ obtained in a similar manner as for
$\rho^{\mathrm{MS}}_*$ above: $\Sigma_* =
22.8\pm1.0\msun\pc^{-2}$. For a flat star-formation history, the ratio
would be $\approx63\,\%$; for a steeply-declining star-formation
history with $e$-folding time of $1\Gyr$, the ratio would be
$55\,\%$. Considering that we only include stellar mass contained in
simple $\sech^2$ components within $400\pc$ from the mid-plane (what
would traditionally be called the ``thin'' disk), this estimate is in
good agreement with previous determinations of the surface-density of
the thinner component of the disk \citep{Flynn06a,Bovy12a,McKee15a}.

\begin{figure*}
  \includegraphics[width=0.99\textwidth,clip=]{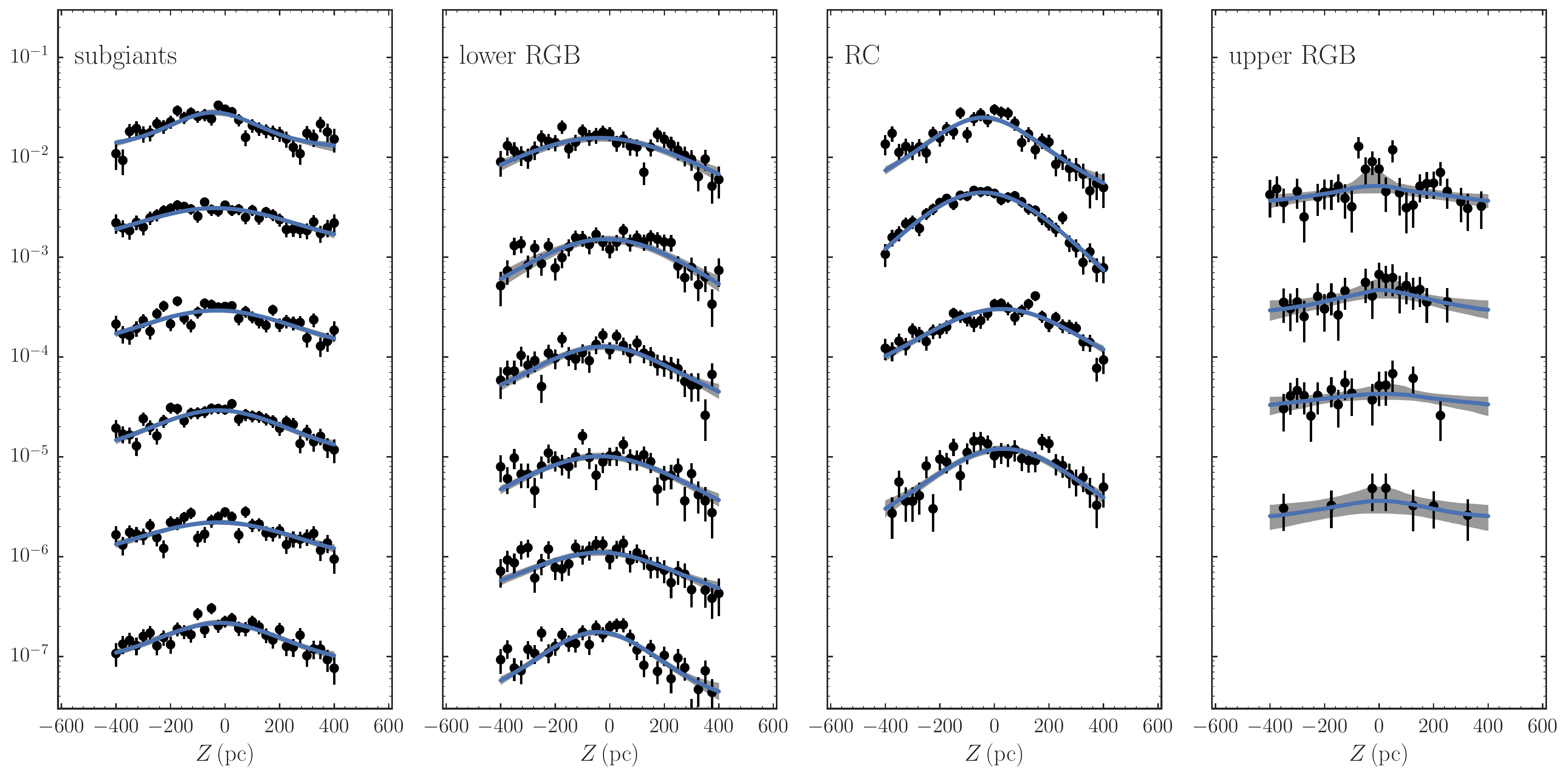}
  \caption{Vertical number density profiles of stars along the giant
    branch. Giants are separated into $\Delta M_J = 0.25\magunit$
    bins. The black points with uncertainties are the data
    (\tgas\ number counts / effective volume), the blue curve is a
    $\sech^2$ fit, and the gray band is the $68\,\%$ uncertainty range
    of the $\sech^2$ model. The profiles have been arbitrarily shifted
    in the $y$ direction. As for main-sequence stars, we see that the
    vertical density clearly flattens near the mid-plane for almost
    all types of giants.}\label{fig:densprofiles_giants}
\end{figure*}

Similarly, we determine the ratio of the current mass in white dwarfs
versus the total stellar mass formed using the initial--final mass
relation from \citet{Kalirai08a} and find that
\begin{align}
  \frac{\Sigma_{\mathrm{WD}}(\tau)}{\Sigma_{\mathrm{form}}} & = 0.0222\,x^2+0.0553\,x+0.04\,,\\ & \qquad \mathrm{with}\ x = \log_{10}\left(\tau / \mathrm{Gyr}\right)\quad (\mathrm{solar\ metallicity})\,.\nonumber
\end{align}
At lower metallicity, this ratio is about $0.5\,\%$ higher. For the
exponentially-declining star-formation history, this gives
$\Sigma_{\mathrm{WD}} = 3.6\pm0.5\msun\pc^{-2}$ or about $10\,\%$ of
the formed mass (and about $16\,\%$ of the current stellar mass). For
a flat star-formation history, the ratio would be $\approx8.4\,\%$
(and about $13\,\%$ of the current stellar mass); for a star-formation
history with $e$-folding time of $1\Gyr$, the ratio would be
$11.5\,\%$ (and about $21\,\%$ of the current stellar
mass). Accounting for the about $7\msun\pc^{-2}$ currently in the
thicker component of the disk \citep{Bovy12a} that we have ignored and
that likely has a star-formation history more sharply peaked in the
past, we estimate a total disk $\Sigma_{\mathrm{WD}} =
5\pm1\msun\pc^{-2}$, again in good agreement with \citet{McKee15a}'s
$\Sigma_{\mathrm{WD}} =4.9\pm0.8\msun\pc^{-2}$.

All in all, the mass distribution of the solar neighborhood and more
broadly the solar cylinder determined from the \tgas\ star counts is
in excellent agreement with previous studies and in many instances
substantially more precise.

\begin{figure*}
  \includegraphics[width=0.99\textwidth,clip=]{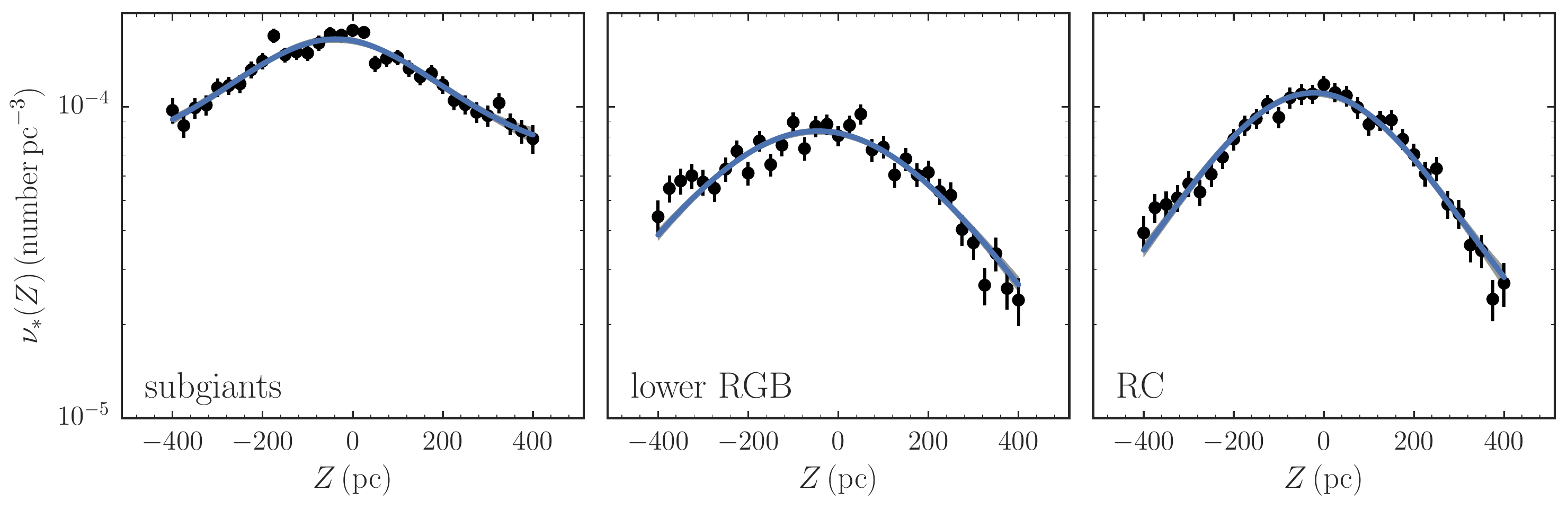}
  \caption{Vertical number density profiles of giants in broad bins in
    $M_J$. All stars for a given giant type in
    \figurename~\ref{fig:densprofiles_giants} are combined to give a
    higher quality measurement of their density profiles. The fitted
    $\sech^2$ profiles have scale heights of $\approx150\pc$. The
    Sun's offset from the mid-plane defined by giants is $\zsun =
    29\pm4\pc$, disagreeing with the Sun's offset from the mid-plane
    defined by A and F stars ($\zsun = -0.9\pm0.9\pc$;
    \figurename~\ref{fig:zsun}).}\label{fig:densprofiles_giants_coarse}
\end{figure*}

\section{Stellar density laws for giant stars}\label{sec:dens_giants}

In the previous section, we discussed the stellar density profiles of
different stellar types along the main sequence in detail. We do the
same in this section for stars along the giant branch. Similar to the
case of the main sequence above, we count as stars those objects that
fall within the lightly-shaded region indicated in
\figurename~\ref{fig:cmd}, including the darkly-shaded region. We
again ignore dust extinction, because its effect on the effective
volume completeness is negligible within the volume that we consider
for the purpose of determining vertical density profiles.

\subsection{Binned stellar densities along the giant branch}

\figurename~\ref{fig:densprofiles_giants} shows the vertical stellar
density profiles for giants in $\Delta M_J = 0.25$ bins, determined
from the number counts in a cylinder with a radius of $250\pc$ in
$\Delta Z = 25\pc$ bins. Comparing these to each other and to the
density profiles of the later stellar types along the main sequence in
\figurename~\ref{fig:densprofiles}, we find that these profiles are
all very similar. The giant profiles are significantly more noisy than
the dwarf profiles, because of the relative paucity of giants in the
\tgas\ sample (each $\Delta M_J = 0.25$ bin typically only has about
1,000 stars). The measured profiles are well represented by single
$\sech^2$ profiles: we find that the amplitude $\alpha$ of a second
component when included in the fit is always so small that it is
practically insignificant.

To obtain density profiles with smaller uncertainties, we also
determine the vertical density of all stars belong to each major type
of giant, because from \figurename~\ref{fig:densprofiles_giants} it
appears that they all have the same vertical profile. This is shown
for the subgiants, lower RGB, and RC in
\figurename~\ref{fig:densprofiles_giants_coarse}; there are too few
stars on the upper RGB to obtain a high precision measurement of the
vertical profile. These high-precision profiles demonstrate that the
$\sech^2$ fit is indeed a good representation of the giants' density
profiles: the profiles clearly display the exponential decline at $|Z|
> z_d$ and the flattening near $Z = 0$. The scale heights of these
$\sech^2$ fits are $\approx150\pc$ (subgiants: $130\pm20\pc$, lower
RGB: $100\pm33\pc$, RC: $154\pm10\pc$).

\subsection{The luminosity function of giants}

\begin{figure}
  \includegraphics[width=0.49\textwidth,clip=]{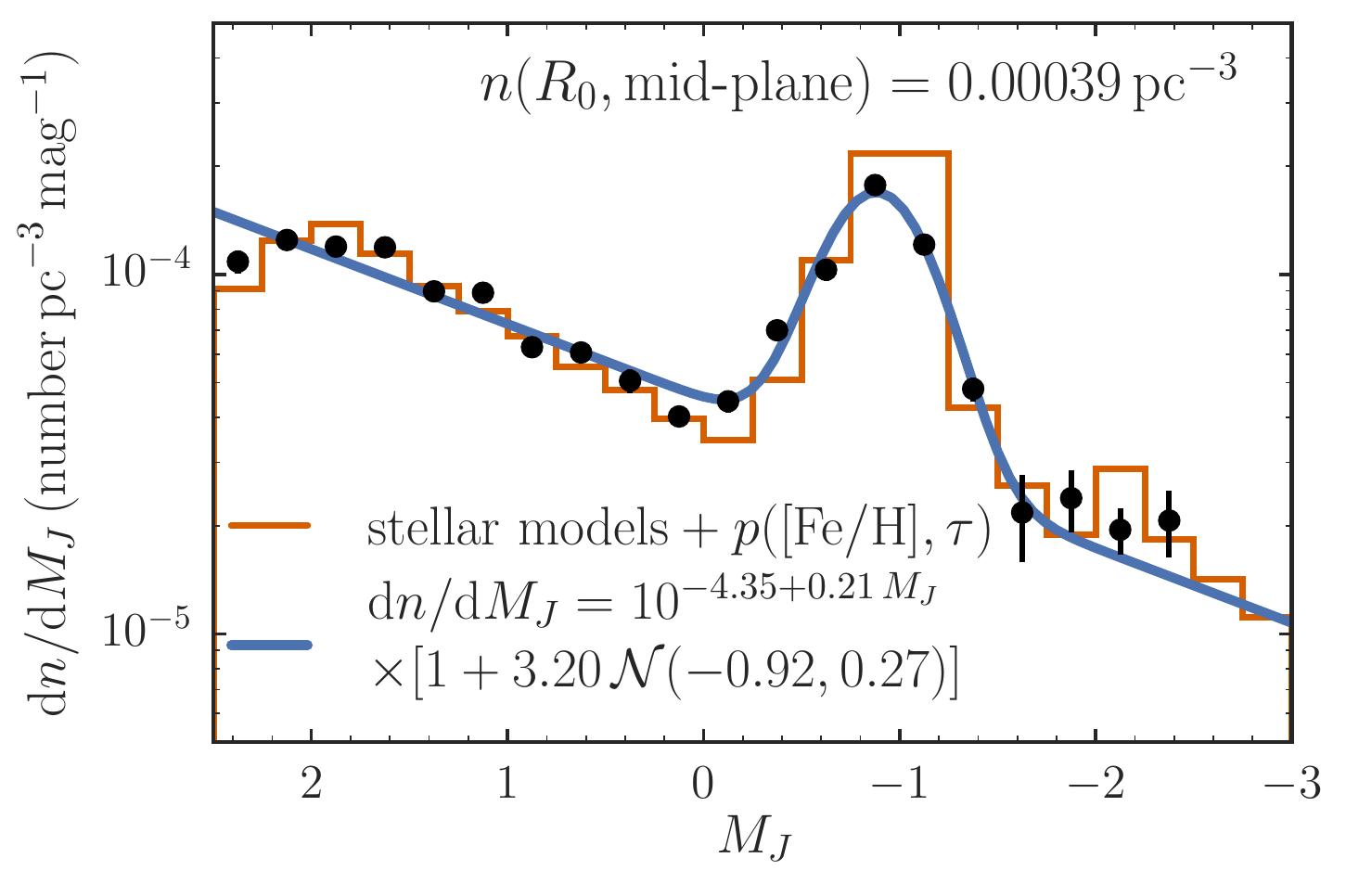}
  \caption{Luminosity function of giants as a function of
    $M_J$. Statistical uncertainties are typically smaller than the
    marker. The luminosity function is well-represented by an
    exponential with a Gaussian red-clump bump superimposed. The total
    mid-plane density of giants is $0.00039\pm0.00001\pc^{-3}$, which
    corresponds to a total mass density of
    $0.00046\pm0.00005\msun\pc^{-3}$. The orange line shows the
    prediction from PARSEC stellar models combined with a
    solar-neighborhood metallicity distribution and a uniform age
    distribution; the overall amplitude of the models is fixed to give
    our measured total mid-plane density of $0.01\msun\pc^{-3}$ in $M
    > 0.72\msun$ stars and is thus not a free parameter. The agreement
    between the observed and predicted giant luminosity function in
    both amplitude and shape is striking.}\label{fig:lf_giants}
\end{figure}

From the $\sech^2$ fits to the density profiles in
\figurename~\ref{fig:densprofiles_giants} we determine the mid-plane
number densities of giants. The luminosity function obtained from
these is displayed in \figurename~\ref{fig:lf_giants} and tabulated in
\tablename~2. This luminosity function has an overall exponential
decline toward the upper RGB, with a bump at the location of the
RC. The luminosity function is well fit by an exponential function of
$M_J$ (or, equivalently, a power-law of luminosity), with a Gaussian
bump superimposed
\begin{equation}
  \frac{\dd n}{\dd M_J} = 10^{-4.35+0.21\,M_J}\times\left[1+3.20\,\mathcal{N}(-0.92,0.27)\right]\,,
\end{equation}
in number\,$\mathrm{pc}^{-3}\,\mathrm{mag}^{-1}$ and where
$\mathcal{N}(\mu,\sigma)$ is the (normalized) Gaussian probability
distribution with mean $\mu$ and standard deviation $\sigma$. 

The orange curve in \figurename~\ref{fig:lf_giants} shows the
predicted luminosity function from the PARSEC \citep{Bressan12a}
isochrones weighted using a lognormal \citet{Chabrier01a} IMF, a
solar-neighborhood-like metallicity distribution function from
\citet{Casagrande11a}, and a uniform age distribution, using the same
cuts to define the giant branch as used for the data. The amplitude of
this predicted luminosity function is set such that the total stellar
mass represented by $M > 0.72\msun$ stars in the model is
$0.01\msun\pc^{-3}$, the total mid-plane density that we directly
measured from number counts along the main sequence in
\sectionname~\ref{sec:dens_ms_massfunc} above. The amplitude is thus
not fit to the observed luminosity function of giants. It is clear
that the agreement between the predicted and the observed luminosity
functions for giants is excellent. The overall amplitude, overall
decline toward more luminous giants, and the RC bump are all in good
agreement.

The total mid-plane density of giants from integrating over the giant
luminosity function between $-4 < M_J < 2.5$ is
\begin{equation}\label{eq:ngiants}
n = 0.00039\pm0.00001\,\mathrm{giants}\pc^{-3}\,.
\end{equation}
\citet{Jahreiss97a} find
$n=0.00049\pm0.00009\,\mathrm{giants}\pc^{-3}$ from a complete sample
of giants within $25\pc$. This agrees well with the more precise value
found here.

To turn this measurement of the mid-plane number density of giants
into an estimate of the mid-plane mass density of giants requires the
average stellar mass of giants of different luminosities. We determine
the stellar mass of giants of a given luminosity $M_J$ using PARSEC
\citep{Bressan12a} isochrones, selecting giants as stars with surface
gravities $\log g < 3.75$ and $J-K_s > 0.4$ and marginalizing over a
flat age distribution and the solar-neighborhood-like metallicity
distribution function from \citet{Casagrande11a}. The mass function
thus obtained is well fit by
\begin{equation}
  \frac{\dd \rho}{\dd M_J} = 10^{-4.25+0.17\,M_J}\times\left[1+2.77\,\mathcal{N}(-0.93,0.28)\right]\,,
\end{equation}
in $\msun\pc^{-3}\,\mathrm{mag}^{-1}$. The total mid-plane
mass density in giants is
\begin{equation}\label{eq:rhogiants}
\rho^{\mathrm{giants}}_* = 0.00046\pm0.00005\,\msun\pc^{-3}\,,
\end{equation}
where the uncertainty is dominated by an (estimated) uncertainty of
stellar mass along the giant branch. This is close to the value of
$\rho^{\mathrm{giants}}_* = 0.00060\pm0.00012\msun\pc^{-2}$ from
\citet{Flynn06a}.

\begin{table}
\caption{Luminosity function of giants.}
\begin{tabular}{cr@{$\pm$}lr@{$\pm$}lr@{$\pm$}l}
\hline \hline
$M_J$ & \multicolumn{2}{c}{$n$} & \multicolumn{2}{c}{$\mathrm{d} n / \mathrm{d} M_J$} & \multicolumn{2}{c}{$n(<M_J)$} \\ & \multicolumn{2}{c}{$(10^{-4}\,\mathrm{pc}^{-3})$} & \multicolumn{2}{c}{$(10^{-4}\,\mathrm{pc}^{-3}\,\mathrm{mag}^{-1})$} & \multicolumn{2}{c}{$(10^{-4}\,\mathrm{pc}^{-3})$} \\\hline\hline
$-2.375$ & 0.05 & 0.01 & 0.21 & 0.05 & 0.05 & 0.01 \\
$-2.125$ & 0.05 & 0.01 & 0.19 & 0.03 & 0.10 & 0.01 \\
$-1.875$ & 0.06 & 0.01 & 0.24 & 0.05 & 0.16 & 0.02 \\
$-1.625$ & 0.05 & 0.02 & 0.22 & 0.07 & 0.21 & 0.03 \\
$-1.375$ & 0.12 & 0.01 & 0.48 & 0.04 & 0.33 & 0.03 \\
$-1.125$ & 0.30 & 0.02 & 1.21 & 0.07 & 0.64 & 0.03 \\
$-0.875$ & 0.44 & 0.02 & 1.77 & 0.07 & 1.08 & 0.04 \\
$-0.625$ & 0.26 & 0.02 & 1.03 & 0.07 & 1.34 & 0.04 \\
$-0.375$ & 0.17 & 0.01 & 0.70 & 0.05 & 1.51 & 0.04 \\
$-0.125$ & 0.11 & 0.01 & 0.44 & 0.03 & 1.62 & 0.04 \\
$\phantom{-}0.125$ & 0.10 & 0.01 & 0.40 & 0.03 & 1.72 & 0.04 \\
$\phantom{-}0.375$ & 0.13 & 0.01 & 0.51 & 0.04 & 1.85 & 0.04 \\
$\phantom{-}0.625$ & 0.15 & 0.01 & 0.61 & 0.05 & 2.00 & 0.05 \\
$\phantom{-}0.875$ & 0.16 & 0.02 & 0.63 & 0.07 & 2.16 & 0.05 \\
$\phantom{-}1.125$ & 0.22 & 0.01 & 0.89 & 0.05 & 2.38 & 0.05 \\
$\phantom{-}1.375$ & 0.22 & 0.01 & 0.90 & 0.04 & 2.61 & 0.05 \\
$\phantom{-}1.625$ & 0.30 & 0.01 & 1.19 & 0.05 & 2.90 & 0.05 \\
$\phantom{-}1.875$ & 0.30 & 0.01 & 1.20 & 0.06 & 3.20 & 0.06 \\
$\phantom{-}2.125$ & 0.31 & 0.01 & 1.25 & 0.06 & 3.51 & 0.06 \\
$\phantom{-}2.375$ & 0.27 & 0.02 & 1.09 & 0.07 & 3.78 & 0.06 \\
\hline
\end{tabular}
\end{table}

\section{The Sun's height above the mid-plane}\label{sec:zsun}

\begin{figure}
  \includegraphics[width=0.49\textwidth,clip=]{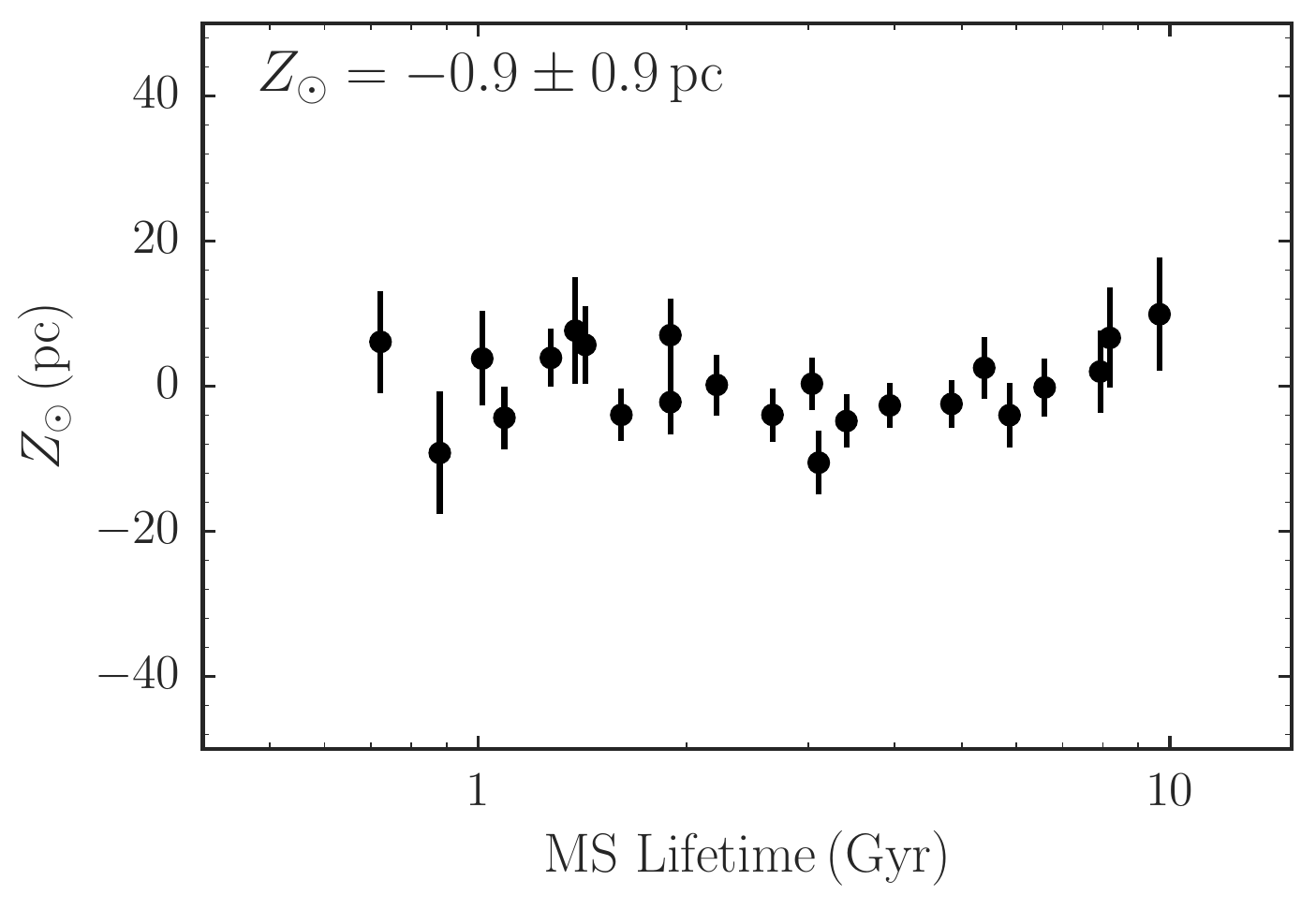}
  \caption{Solar offset from the mid-plane defined by different
    stellar types. The Sun is consistent with being at the mid-plane
    defined by A and F dwarfs.}\label{fig:zsun}
\end{figure}

The value of the Sun's offset from the mid-plane defined by each
spectral subtype from A0V to G3V is displayed in
\figurename~\ref{fig:zsun}. For later-type dwarfs we are unable to
determine $\zsun$, because the stellar density for these stars is
almost entirely flat within the observed $Z$ range (see
\figurename~\ref{fig:densprofiles}). Remarkably, the Sun is consistent
with being at the mid-plane defined by each spectral type, with a
combined measurement of $\zsun = -0.9\pm0.9\pc$. For A and F dwarfs,
\zsun\ is determined from the vertical distribution over $\approx4$
scale heights in both directions away from the mid-plane, leading to
the high precision in the measured \zsun.

We measure the Sun's offset from the mid-plane defined by the three
major, different types of giants (subgiants, lower RGB, and RC) using
the density profiles shown in
\figurename~\ref{fig:densprofiles_giants_coarse}. The measured
\zsun\ for these three populations are: $\zsun = 42\pm7\pc$
(subgiants), $\zsun > 45\pc\,(1\sigma)$ (lower RGB), and $\zsun =
22\pm6\pc$ (RC). For the lower RGB we only measure a lower limit with
an a priori upper limit of $\zsun < 50\pc$. This inferred value is
strongly affected by the density profile at $|Z| > 200\pc$ for lower
RGB stars, which is somewhat asymmetric. Only considering lower RGB
stars below $200\pc$, we find $\zsun = 25\pm13\pc$. Combining this
value with the measurements using subgiants and RC stars, we find
$\zsun = 29\pm4\pc$ with respect to the mid-plane defined by
giants. If we only include giants at $|Z| < 200\pc$ for all three
types, we find $\zsun = 35\pm7\pc$.

We have investigated whether the inferred \zsun\ depends on the
assumed extinction model. The basic fits shown in
\figurename~\ref{fig:zsun} ignore extinction in the calculation of the
effective volume completeness, because it has only a minor effect. We
have also determined the vertical stellar profiles of entire spectral
types---all stars of type AV, all stars of type FV, etc.---taking into
account the three-dimensional extinction model discussed in
\sectionname~\ref{sec:effsel}. The inferred \zsun\ only differ by
$\approx0.1\pc$ from those determined without taking extinction into
account for A and F dwarfs and by only $1\pc$ for G dwarfs.

As discussed in \appendixname~\ref{sec:tgascomplete}, our model for
the raw \tgas\ selection function is not perfect at the bright
end. This may affect the results derived in this section, especially
for the intrinsically-bright A and F dwarfs and for red giants, many
of which are located near the bright-end cut-off of \tgas. Therefore,
we have re-determined the stellar densities, removing all stars with
$J_G < 8$ (where $J_G$ is defined in
\appendixname~\ref{sec:tgascomplete}), which is where our model for
the raw selection function performs poorly. The resulting density
profiles are very similar to those derived from the full data set,
although they are much noisier for the earliest A dwarfs. The combined
value of the Sun's height above the mid-plane is $\zsun =
-0.3\pm1.1\pc$, consistent with the value determined from the full
data set. The scale height and mid-plane densities are also the
same. For giants, we measure $\zsun = 36\pm5\pc$ when removing the
bright stars and the bright-end behavior of the \tgas\ selection can
therefore not explain the difference that we find between the
\zsun\ determined from A/F dwarfs and from giants.

We also investigate whether the wide cylinders (radii of $250\pc$ and
$200\pc$) used to determine the vertical density profiles of A and F
dwarfs affect the Solar offset from the mid-plane defined by these
populations. We determine the density profiles for all types of
main-sequence stars for cylinders with width $150\pc$. We find density
profiles and $\sech^2$ fits that are almost the same as those based on
the wider cylinders for A and F dwarfs. In particular, the Sun's
height above the mid-plane defined by A and F dwarfs is $\zsun =
-0.7\pm1.3\pc$, fully consistent with the value determined from the
wider cylinders.

Finally, we determine the Solar offset only considering density
measurements at $|Z| > 100\pc$. We do this because the effective
volume completeness for A and F stars has a mildly-peaked structure
centered on $Z = 0$, due to the bright-end \tgas\ cut-off and the sky
cut (see \figurename~\ref{fig:effsel_cyl}). If the effective volume
completeness is even slightly incorrect, this could induce an apparent
peak in the stellar density profile, which is shallow near the
plane. At $|Z| > 100\pc$, the effective volume for A and F stars is
almost constant with $Z$. In this case we find the Sun's height as
defined by A and F dwarfs to be $\zsun = -2.5\pm1.2\pc$, consistent
with the result from the full data set.

Therefore, the \tgas\ data strongly prefer a value of $\zsun$ when
measured using A and F dwarfs that is consistent with zero within a
few pc. However, while we have performed a number of tests, described
in the previous paragraphs, of the impact of the \tgas\ selection
function on the measurement of $\zsun$, our \tgas\ selection function
is only an empirical model derived using a number of assumptions,
including that the selection function does not vary with position on
the sky (within the `good' part of the sky). These assumptions are
unlikely to be correct in detail and may affect the $\zsun$
measurement in ways that are not covered by the tests discussed
above. The next \Gaia\ data release will not depend on the
\tyctwo\ catalog and will likely have a simpler selection function
than \tgas. This will allow for a more robust determination of
\zsun\ using various stellar types.

Our results for \zsun\ with respect to A/F dwarfs and with respect to
giants are at odds with each other. The measured \zsun\ with respect
to A/F dwarfs also appears to be at odds with previous determinations,
which find a canonical value of $\zsun = 15$ to $25\pc$
\citep[\eg,][]{Binney97a,Chen01a,Juric08a} and agree with the Solar
offset we determine using giants. However, these literature
determinations are based on old stellar populations, rather than the
young populations traced by A and F stars here. A value of $\zsun =
25\pc$ corresponds to one half of a scale height for the A dwarfs and
can be clearly ruled out by the \tgas\ star counts. For younger
tracers (young open clusters and OB stars), $\zsun$ is typically found
to be smaller, with $\zsun = 6$ to $15\pc$
\citep[\eg,][]{Bonatto06a,Joshi07a,Joshi16a}, closer to what we
determine here. Moreover, measurements of the vertical distribution of
atomic \citep{Malhotra95a} and molecular gas \citep{Nakanishi06a} in
the Milky Way find typical fluctuations in the location of their
mid-plane of tens of pc. These are most likely inherited by
newly-formed stars and could persist for a few dynamical times for the
younger stars traced by A and F dwarfs.

\section{Discussion}\label{sec:discussion}

\subsection{Computational complexity}

In this paper, we have determined stellar densities using the
$N=\mathcal{O}(10^6)$ stars in the \tgas\ catalog. The full
\Gaia\ data set will eventually consist of $\mathcal{O}(10^9)$ stars
and it is therefore urgent to ask how the method used in this paper
scales to this much larger number of stars. When determining stellar
densities using the formalism in \sectionname~\ref{sec:method_binned},
the main computational complexity arises from the calculation of the
effective volume in the denominator in
\equationname~\eqref{eq:bfdens}. The numerator in that expression can
be efficiently computed for all volume bins simultaneously (e.g.,
using a histogram function) using $\mathcal{O}(N)$ operations and is
fast even for $\mathcal{O}(10^9)$ stars. The calculation of the
effective volume does not scale directly with $N$, because it does not
involve the data, only the selection function. However, because future
data releases will cover a larger region of the Galaxy, the numerical
integration of the effective volume will extend over a larger area and
take longer to compute.

In the limit of vanishing dust extinction, computing the effective
volume is easy, because the distance integral can be factored out from
the integral on the sky. This is the case as long as the raw selection
function does not depend on sky position, except for a purely
geometric sky cut, like in our model for the \tgas\ selection
function. While the selection function will certainly depend on sky
position in small parts of the sky with high stellar density (where
\Gaia\ cannot catalog all sources due to telemetry limitations), over
most of the sky future \Gaia\ selection functions will likely not
depend on position on the sky. However, extinction cannot be avoided
when determining stellar densities close to the mid-plane and this
complicates the effective-volume integral. \figurename
s~\ref{fig:effsel_cyl}, \ref{fig:effsel_rectxy},
\ref{fig:effsel_rectyz}, \ref{fig:effsel_rectxy_giants}, and
\ref{fig:effsel_rectyz_giants} take the full three-dimensional
dependence of dust extinction into account and took altogether a few
days to compute using dozens of processors. Because the effective
volume does not directly scale with the number of data points, as long
as the extinction model is kept fixed (or only a small number of
alternative models are explored), this computational time will not
increase much in future data releases.

\subsection{Other uses of the effective completeness for \tgas}

We have used the effective completeness and the effective volume
completeness for \tgas\ from \sectionname~\ref{sec:complete} only for
measuring the vertical density profiles of different stellar
types. However, the effective completeness should be taken into
account in any use of the \tgas\ data that depends on how the
underlying stellar density distribution is catalogued in \tgas\ and
this, in principle, includes essentially all use of the
\tgas\ parallaxes as distance indicators. Because \Gaia\ measures
parallaxes rather than distances, any distance inferred from the
\tgas\ parallaxes requires a distance prior
\citep{BailerJones15a}. This prior should take the effective
completeness of \tgas\ into account. For example, to get the best
inferred distances using the method of \citet{Astraatmadja16a}, their
priors should be multiplied by the effective selection function, which
they currently assume is uniform down to a magnitude limit (and they
fully ignore the bright cut-off in \tgas; \citealt{Astraatmadja16b}).

Similarly, because the effective completeness of \tgas\ is a strong
function of stellar type (see \figurename~\ref{fig:effsel_dist}), any
representation of the CMD from \tgas\ data \citep[\eg, in][]{GaiaDR1}
has strong selection biases and densities within the color--magnitude
plane do not reflect true, underlying densities, but rather the
density of objects \emph{contained in \tgas}. This has strong
implications for any analysis that models the observed CMD and uses it
to, for instance, improve distances inferred from the
\tgas\ parallaxes without attempting to correct for \tgas's selection
biases (\eg, \citealt{Leistedt17a}, L. Anderson et al. 2017, in
prep.). Such models of the CMD cannot be applied to data other than
the \tgas\ data themselves, because other data will have a different
effective completeness and thus have different biases in the
color--magnitude plane.

More broadly, many analyses of the \tgas\ data that use more than a
small volume can be affected by the selection function, even if they
do not directly depend on the number counts or underlying stellar
densities. For example, analyses of the kinematics---typically assumed
to be free of selection biases---of different stellar populations
based on \tgas\ may depend on the selection function if they average
over large volumes and gradients across this volume are important. For
example, suppose one examines the kinematics of stars along the main
sequence for a volume with $\sqrt{X^2+Y^2} < 500\pc$ and $-100\pc < Z
< 100\pc$. Then the most luminous main-sequence stars will sample this
volume almost uniformly, while the faintest main-sequence stars will
only cover the nearest few tens of pc and the measured kinematic
properties will effectively cover widely different volumes. If
Galactic gradients are important, the effective completeness needs to
be taken into account when comparing different stellar
populations. The completeness maps displayed in \figurename
s~\ref{fig:effsel_rectxy}, \ref{fig:effsel_rectyz},
\ref{fig:effsel_rectxy_giants}, and \ref{fig:effsel_rectyz_giants} can
act as a guide to determine how important selection effects may be.

\subsection{Astrophysical implications}

We have obtained a number of new constraints on the
stellar-populations structure of the local Milky-Way disk from the
analysis of the vertical densities of main-sequence and giant stars in
\tgas. One of the most directly measured quantities is the
distribution of stellar mass in the mid-plane along the main sequence
covering $0.7\msun \lesssim M \lesssim 2.2\msun$. While the
\tgas\ data represent only a small fraction of the full, final
\Gaia\ catalog, this range extends both to high enough $M$ such that
there is little mass in stellar populations with higher $M$ and to low
enough $M$ that we reach the part of the main sequence where stars of
any age have not evolved into giants yet. Thus, assuming an IMF model
for $M \lesssim 0.7\msun$, we obtain a census of all stellar mass in
the solar neighborhood, an important ingredient in the baryonic mass
budget in the solar neighborhood \citep{McKee15a}. Because---beside
filling in the $G \lesssim 6$ part of the magnitude distribution and
the part of the sky that we have ignored in the selection
function---improvements in the sampling of the \Gaia\ catalog will
primarily come from going to fainter magnitudes and thus larger
distances, the census described in this paper is practically the
definitive census of $M \gtrsim 1\msun$ stars in the solar
neighborhood. However, future data releases will allow the low-mass
IMF to be determined from the \Gaia\ data and thus improve the census
by directly including the contribution of low-mass stars.

We have used the measurements of the intrinsic stellar number counts
at $M > 1\msun$, after translating them to surface densities to
account for vertical heating, to determine the star-formation history
of the solar neighborhood. Our measurement in
\figurename~\ref{fig:sfh} is a direct measurement of the
reverse-cumulative star-formation history and thus presents an
important new constraint on models for the evolution of the Milky Way
disk. However, to make this measurement we had to assume that the
slope of the high-mass IMF is constant over the history of the disk
and that it is known. Our results are degenerate with the slope of the
high-mass IMF. For example, if we assume that the high-mass IMF tracks
the present-day mass function that we measure, the points in
\figurename~\ref{fig:sfh} would scatter around a constant value and we
would infer that stars only formed during the last few hundred
Myr. Using the kinematics of stars along the main sequence in addition
to their number densities can help break this degeneracy
\citep{Binney00a} and using the number densities along the giant
branch would provide additional constraints (see
\figurename~\ref{fig:lf_giants}).

The star-formation history that we measure is corrected for vertical
heating. However, stars may also heat and migrate in the radial
direction \citep{Sellwood02a} and the locally-measured star-formation
history in that case is essentially that of annuli in the disk
convolved with the migration history \citep[\eg,][]{Schoenrich09a}. If
migration is extremely efficient and the disk is fully mixed, the
locally-measured star-formation history would simply be the
star-formation history of the entire disk. Conversely, if migration is
inefficient, the locally-measured star-formation history reflects the
local evolution. The truth most probably lies somewhere in the middle
and it is thus difficult to directly relate the locally-measured
star-formation history to the evolution of the solar circle.

\subsection{Future work}

Future \Gaia\ data releases will allow the stellar census performed
here to be extended in many ways. Star counts down to $G \approx 20.7$
will allow all stars down to $M = 0.08\msun$ to be mapped within
$\approx25\pc$ and in rapidly-increasing volumes for higher masses
(\eg, $\approx190\pc$ for $M=0.15\msun$, $\approx500\pc$ for
$M=0.3\msun$). Thus the completeness maps shown for \tgas\ in
\figurename s~\ref{fig:effsel_rectxy}, \ref{fig:effsel_rectyz},
\ref{fig:effsel_rectxy_giants}, and \ref{fig:effsel_rectyz_giants}
will cover much larger volumes of the Galaxy beyond the closest
kpc. For all but the faintest stars, the vertical and radial density
law within the Galaxy can be determined to high precision, which will
provide an unprecedented view of the distribution of mass and light in
a large disk galaxy.

To get the most out of the \Gaia\ data, various improvements to the
analysis presented in the paper will have to be made. First of all,
because no all-sky, complete stellar catalog currently exists down to
$G \approx 20.7$, we will not be able to determine the catalog
completeness by comparing to an external, complete catalog, as we did
for \tgas\ using the 2MASS catalog in
\appendixname~\ref{sec:tgascomplete}. Ultimately, the completeness of
the \Gaia\ catalog should be determined from the \Gaia\ data and
instrument themselves, by considering how likely it is that sources
with a given set of properties (sky location, brightness, color,
\ldots) appear in the final catalog. From the way in which
\Gaia\ scans the sky it is expected that, but for the most crowded
regions of the sky, the final \Gaia\ catalog will be 100\,\% complete
down to some faint limit. For making the most complete stellar census
based on \Gaia\ data, determining the completeness around the faint
limit will be of great importance, as small magnitude differences at
the faint end represent large numbers of stars and large volumes. For
example, by determining the completeness of \tgas\ using 2MASS rather
than the \tyctwo\ catalog, we were able to extend the magnitude range
over which the completeness is understood by $\approx1.5\magunit$ or a
factor of two in distance. For \Gaia\ DR2, we can still use 2MASS down
to $J\approx15$ to determine the completeness of the catalog, an
improvement of $\approx3\magunit$, or a factor of four in distance,
over \tgas.

A second necessary improvement concerns the methodology used to infer
the underlying stellar density. We have presented the statistical
framework for inferring the underlying density from an incomplete
survey in \sectionname~\ref{sec:method}. In the current application of
this general methodology, we have assumed that only the underlying
stellar density is unknown, while the survey selection function, the
density in the CMD $\denscmd$, and the three-dimensional extinction
map are presumed to be known. While the survey selection function will
hopefully be well known enough from the data processing itself (see
previous paragraph), the density in the CMD and the three-dimensional
extinction map are far from well known. Determining them is among the
other main scientific goals of the \Gaia\ mission
\citep{GaiaMission}. Because we have focused in this paper on the
nearest few hundred pc and have used near-infrared photometry, the
three-dimensional extinction has only a minor effect on the observed
number counts. For the CMD, we were able to approximate the underlying
density by focusing on the high-precision parallaxes. 

Ultimately, to gain a full, empirical understanding of the stellar
distribution, CMD, and three-dimensional extinction throughout the
Milky Way, all three should be obtained through a single analysis
using the methodology described in \sectionname~\ref{sec:method}. The
likelihood given in the first line of \equationname~\eqref{eq:lnl} is
general and could include parameters of flexible models for the CMD
and the extinction map (we have only assumed that these are known in
the second line of \equationname~\ref{eq:lnl}). So far, studies of the
CMD or the three-dimensional extinction map have assumed that the
other ingredients are perfectly known. For example, studies of the CMD
using \tgas\ data assume that the density distribution and extinction
are known (\eg, \citealt{Leistedt17a}, L. Anderson et al. 2017, in
prep.). Similarly, the sophisticated determination of the
three-dimensional extinction by \citet{Green15a} assumes that the
stellar locus (the CMD) and the density distribution of the Milky Way
are known \citep{Green14a}. With more \Gaia\ data, especially with the
BP/RP photometry, it should be possible to constrain all three
ingredients simultaneously.

\section{Conclusion}\label{sec:conclusion}

We have conducted the first detailed stellar inventory of the solar
neighborhood using the \Gaia\ DR1 \tgas\ data. To do this, we have
performed a detailed analysis of the raw \tgas\ selection function and
have successfully derived the \tgas\ completeness over 48\,\% of the
sky with `good' \tgas\ observations. Using this raw completeness, we
have determined the effective completeness in distance and volume of
different stellar populations in \tgas, taking into account their
intrinsic distribution in color and absolute magnitude as well as the
three-dimensional dependence of dust extinction. Maps of the
completeness of \tgas\ for different stellar tracers are given in
\figurename s~\ref{fig:effsel_rectxy}, \ref{fig:effsel_rectyz},
\ref{fig:effsel_rectxy_giants}, and \ref{fig:effsel_rectyz_giants} and
these should provide a useful guide for many studies making use of
\tgas\ data.

Using our determination of the completeness of \tgas, we have measured
the intrinsic stellar density distribution for different stellar types
along the main sequence and along the giant branch. This results in a
detailed new inventory of the stellar mass distribution in the solar
neighborhood given in \tablename s~1 and 2. We have determined the
luminosity function along the main sequence for $7.25 > M_V > 1.34$
(\figurename~\ref{fig:lumfunc}), the present-day mass function for
stars with masses $M \gtrsim0.72\msun$
(\figurename~\ref{fig:massfunction}), and total mid-plane density in
stars (\equationname~\ref{eq:rhoms}). From these, we have determined
the implications for the mass in white dwarfs and we have measured the
star-formation history of the solar neighborhood
(\figurename~\ref{fig:sfh} and \equationname~\ref{eq:sfh}). We have
further determined the luminosity function of stars along the giant
branch (\figurename~\ref{fig:lf_giants}) and the total number and mass
density of giants in the mid-plane (\equationname s~\ref{eq:ngiants}
and \ref{eq:rhogiants}).

We have also measured the vertical density profiles of different types
of dwarfs and giants. For all stellar types, we clearly see that the
vertical density profiles flatten at $|Z| \lesssim 100\pc$ and all
profiles are well represented as $\sech^2$ profiles within a few scale
heights ($Z \lesssim 4\,z_d$). The scale height of these profiles
increases smoothly when going from the earliest A-type dwarfs ($z_d
\approx 50\pc$) to late K-type dwarfs ($z_d \approx 150\pc$); giants
have similar profiles as late dwarfs. Surprisingly, we find that the
Sun is at the mid-plane defined by A-type and F-type dwarfs ($\zsun =
-0.9\pm0.9\pc$), in tension with previous measurements. However, we
are unable to identify any systematic in our analysis that would
produce this result. With respect to older stars on the main sequence
and on the giant branch, we find that the Sun is offset from the
mid-plane by $\zsun = 29\pm4\pc$, in good agreement with previous
measurements.

The new stellar inventory made possible by \Gaia\ DR1 is in good
agreement with previous studies, but substantially more precise for
the stars that we directly observe ($M \gtrsim0.72\msun$). The
detailed determination of the completeness of \tgas\ opens up many
avenues of investigation using \tgas\ data that depend on how
\tgas\ samples the underlying stellar distribution. The methodology
for determining the completeness can also be straightforwardly extended
three magnitudes fainter for \Gaia\ DR2 and the tools developed here
will thus remain useful. As described in
\appendixname~\ref{sec:tgascomplete} and
\sectionname~\ref{sec:complete}, we have made code available that
allows the raw and effective selection function for \tgas\ to be
evaluated. All of the code used to perform the analysis presented in
this paper is available at
\\ \centerline{\url{https://github.com/jobovy/tgas-completeness}~,}
and can serve as an example of how to use the selection function.


{\bf Acknowledgments} I thank Dustin Lang for providing the
TGAS-matched 2MASS data and Chris Flynn and Wilma Trick for helpful
comments. JB received support from the Natural Sciences and
Engineering Research Council of Canada. JB also received partial
support from an Alfred P. Sloan Fellowship and from the Simons
Foundation.

This work has made use of data from the European Space Agency (ESA)
mission {\it Gaia} (\mbox{\url{http://www.cosmos.esa.int/gaia}}),
processed by the {\it Gaia} Data Processing and Analysis Consortium
(DPAC,
\mbox{\url{http://www.cosmos.esa.int/web/gaia/dpac/consortium}}). Funding
for the DPAC has been provided by national institutions, in particular
the institutions participating in the {\it Gaia} Multilateral
Agreement. This publication makes use of data products from the Two
Micron All Sky Survey, which is a joint project of the University of
Massachusetts and the Infrared Processing and Analysis
Center/California Institute of Technology, funded by the National
Aeronautics and Space Administration and the National Science
Foundation.

Some of the results in this paper have been derived using the
\texttt{HEALPix} \citep{Gorksi05a}, \texttt{astropy} \citep{astropy},
and \texttt{emcee} \citep{Foreman13a} software packages.

\appendix
\onecolumn
\begingroup
\centering
\section{The completeness of \protect{\tgas} in color--magnitude--sky-position}\label{sec:tgascomplete}
\endgroup

\begin{figure*}
  \includegraphics[width=0.49\textwidth,clip=]{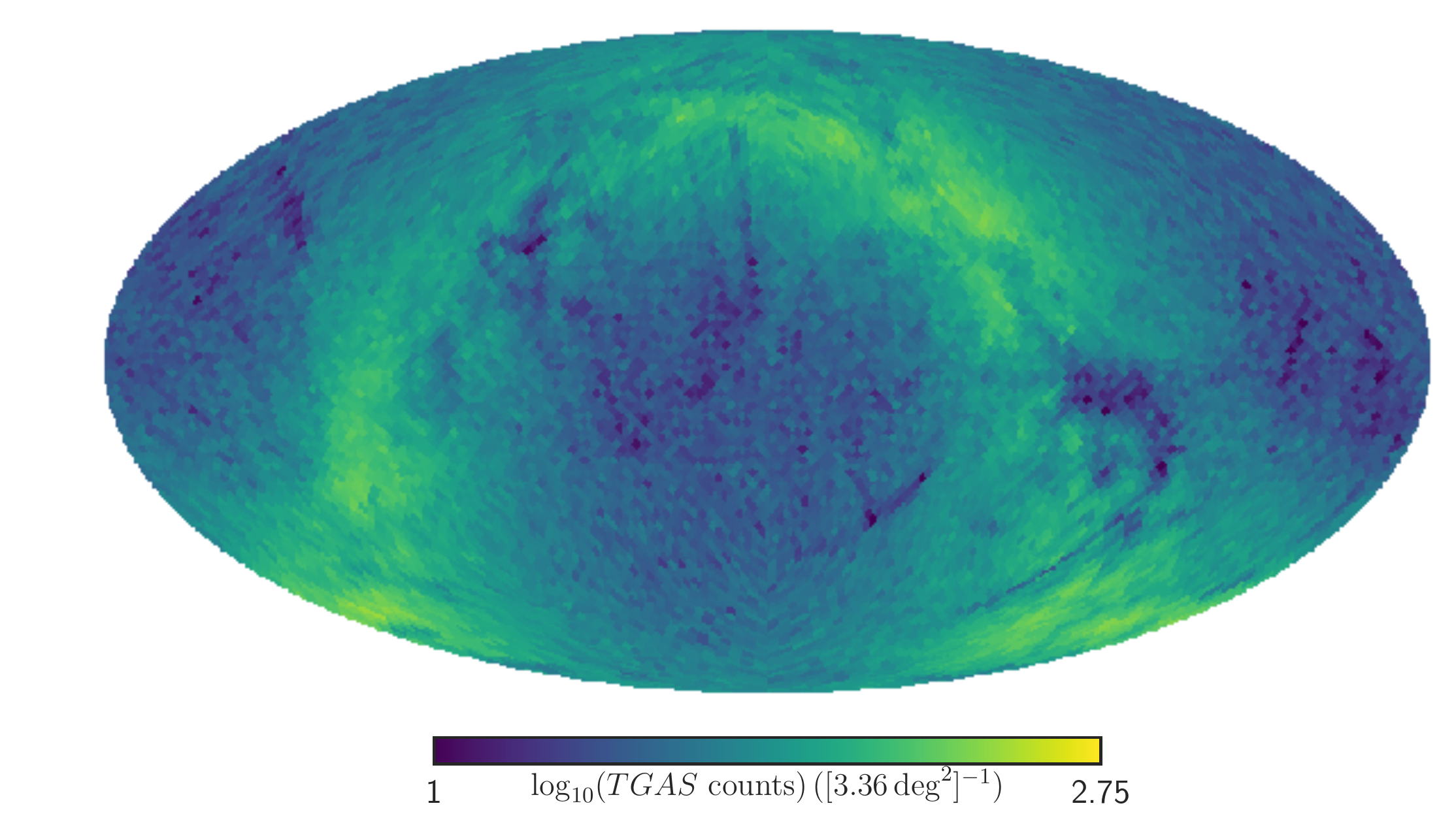}
  \includegraphics[width=0.49\textwidth,clip=]{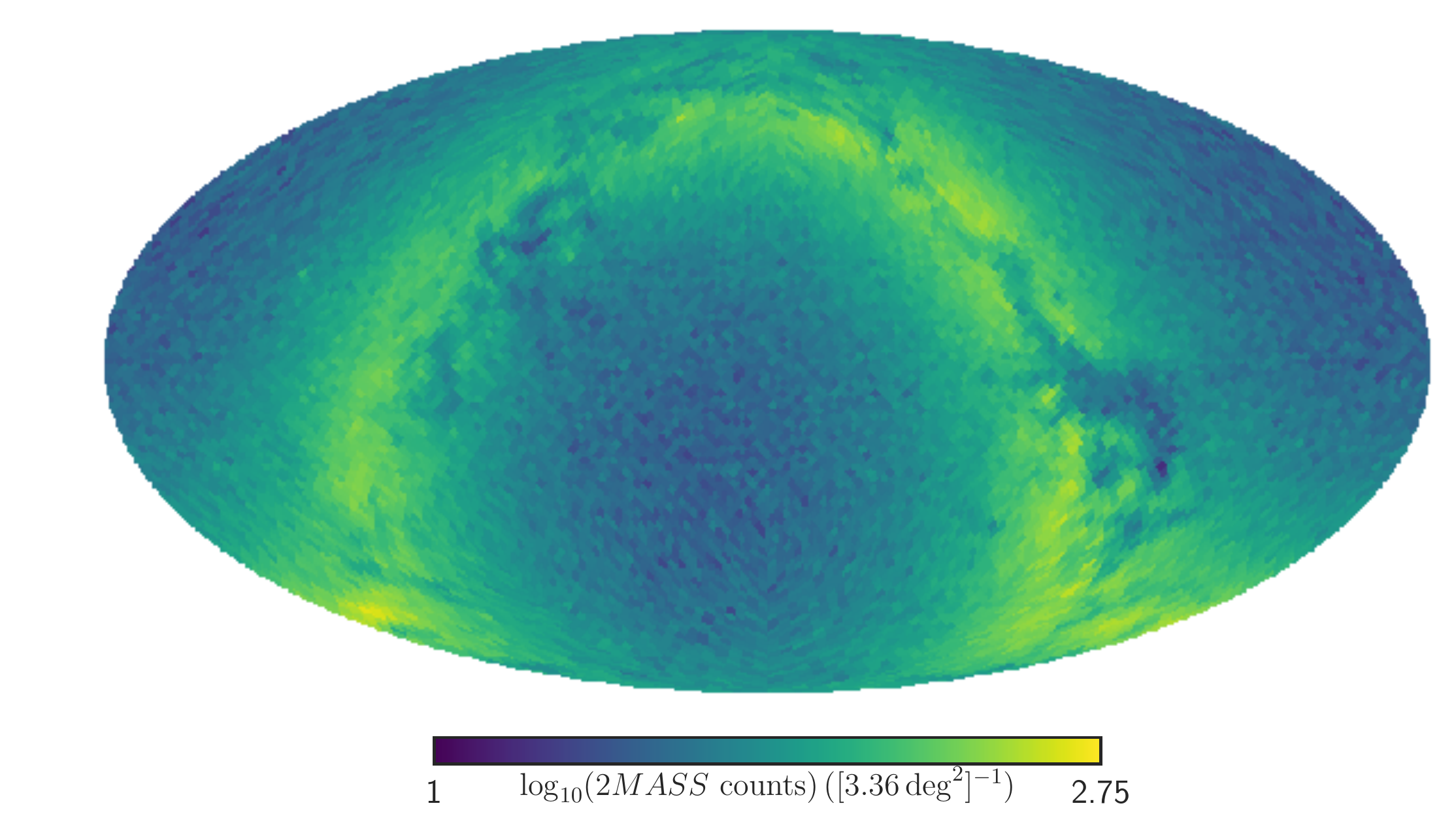}\\
  \caption{\tgas\ and 2MASS number counts for stars with $0 < J-K_s <
    0.8$ and $6 < J < 10$. While 2MASS is complete in this range,
    \tgas\ has artifacts that are mainly due to its scanning
    pattern. We define the overall completeness in $3.36\deg^2$
    HEALPix pixels ($N_{\mathrm{side}} = 2^5$) in the following
    figures to be the ratio of these \tgas\ counts to the 2MASS
    counts.}\label{fig:sky_counts}
\end{figure*}

\Gaia\ DR1 consists of two astrometric solutions
\citep{GaiaDR1,Lindegren16a}: the primary \tgas\ data set containing
positions, parallaxes, and proper motions for a subset of the
\tyctwo\ catalog \citep{Hog00a} and the secondary data set with
approximate positions for stars brighter than $G \approx 20.7$. The
primary solution consists of 2,057,050 stars out of 2,539,913 stars in
\tyctwo. The \tyctwo\ catalog is 99\,\% complete down to $V\approx11$
and its completeness drops quickly at fainter magnitudes. However,
\tgas\ does not share this simple completeness limit, as many of the
$\approx20\,\%$ of missing stars are at much brighter magnitudes and
the very brightest stars ($G \lesssim 6$) are missing, because the
specialized observing mode that they require is not yet sufficiently
calibrated to produce reliable results. Due to the inhomogeneity of
the scanning law, the completeness also strongly varies over the
sky. In this Appendix, we investigate the completeness of \tgas.

\begingroup
\centering
\subsection{Overall completeness}\label{sec:overall}
\endgroup

Almost all of the \tyctwo\ stars were considered in the primary
astrometric solution and the main reasons that stars failed to be
included in \tgas\ are: (i) they are too bright ($G \lesssim 6$) or
(ii) the quality of their astrometric solution is too low (as
evidenced by a parallax uncertainty larger than 1 mas or a position
uncertainty larger than 20 mas; \citealt{Lindegren16a}). While the
hope is that the astrometric uncertainties will eventually be
dominated by photon noise and, therefore, apparent magnitude, the
quality of the astrometry in \Gaia\ DR1 is primarily set by
observational limitations and systematic uncertainties: the lack of
high numbers of observations in certain parts of the sky, limitations
in the current attitude model, and the simplistic modeling of the
point-spread function. The completeness is therefore also in large
part a function of broad observational properties (such as the number
of astrometric transits) and color, but to a lesser degree of apparent
magnitude (above the faint limit of \tyctwo).

\begin{figure*}
  \includegraphics[width=0.49\textwidth,clip=]{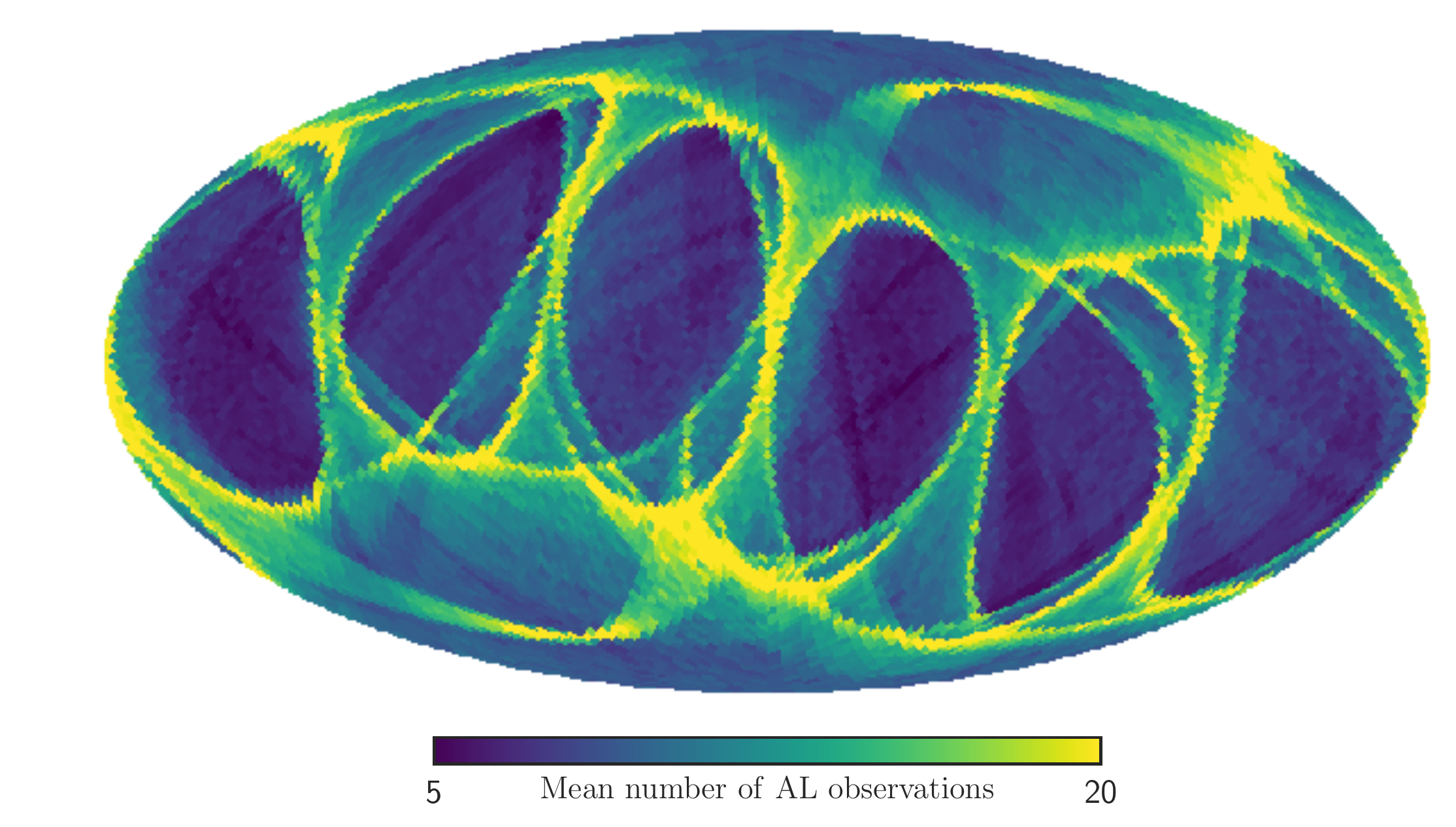}
  \includegraphics[width=0.49\textwidth,clip=]{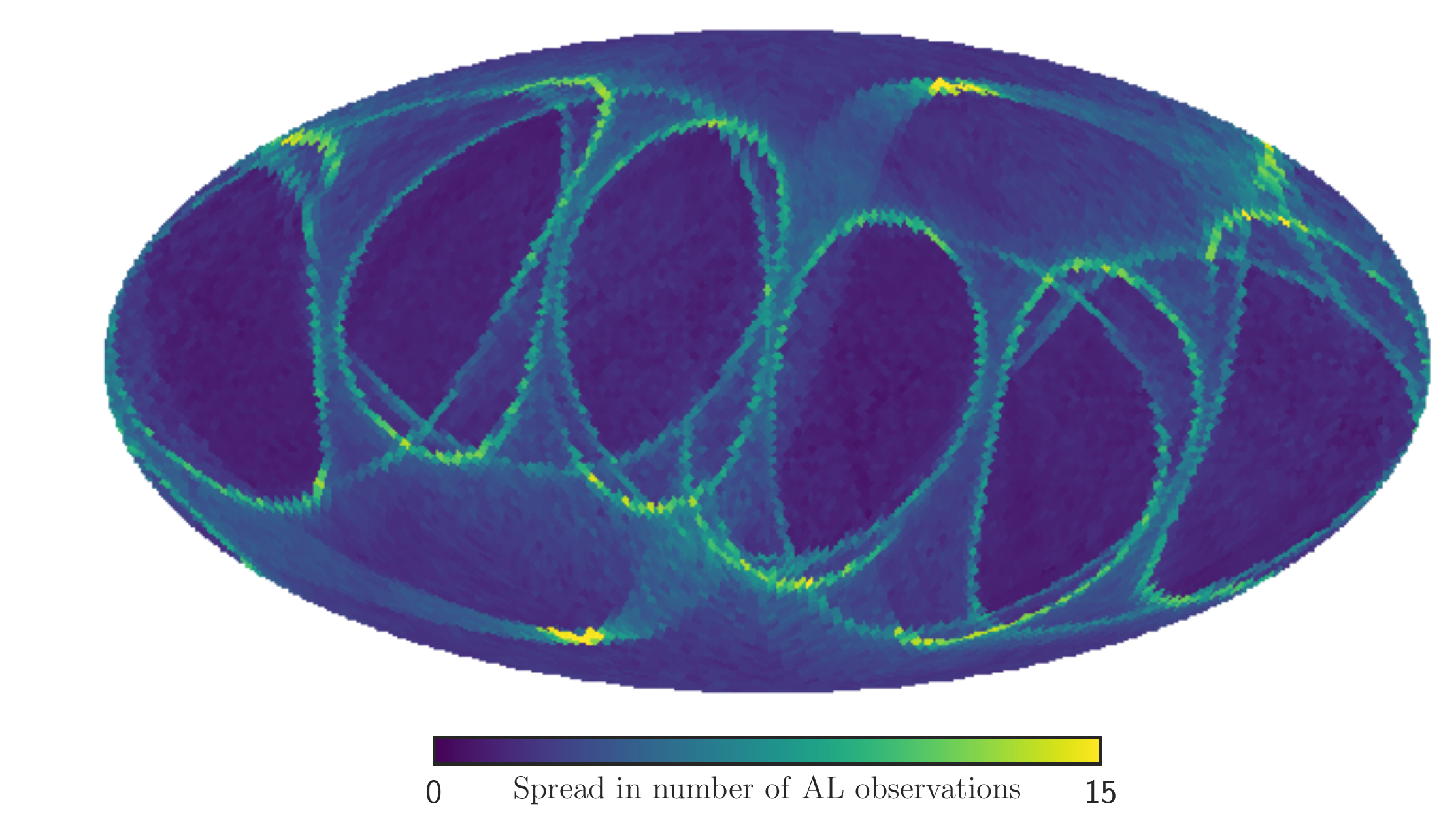}\\
  \includegraphics[width=0.49\textwidth,clip=]{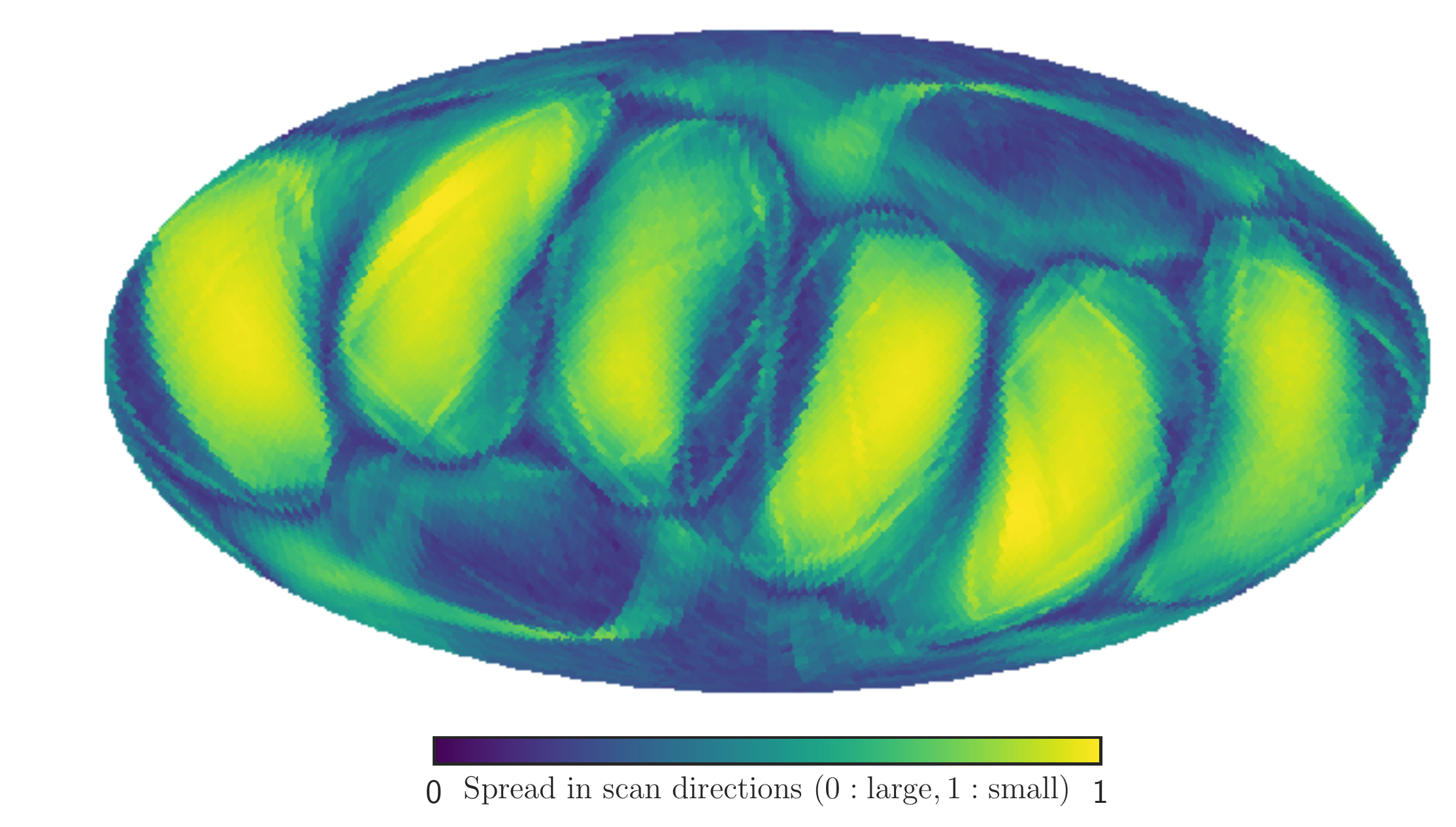}
  \includegraphics[width=0.49\textwidth,clip=]{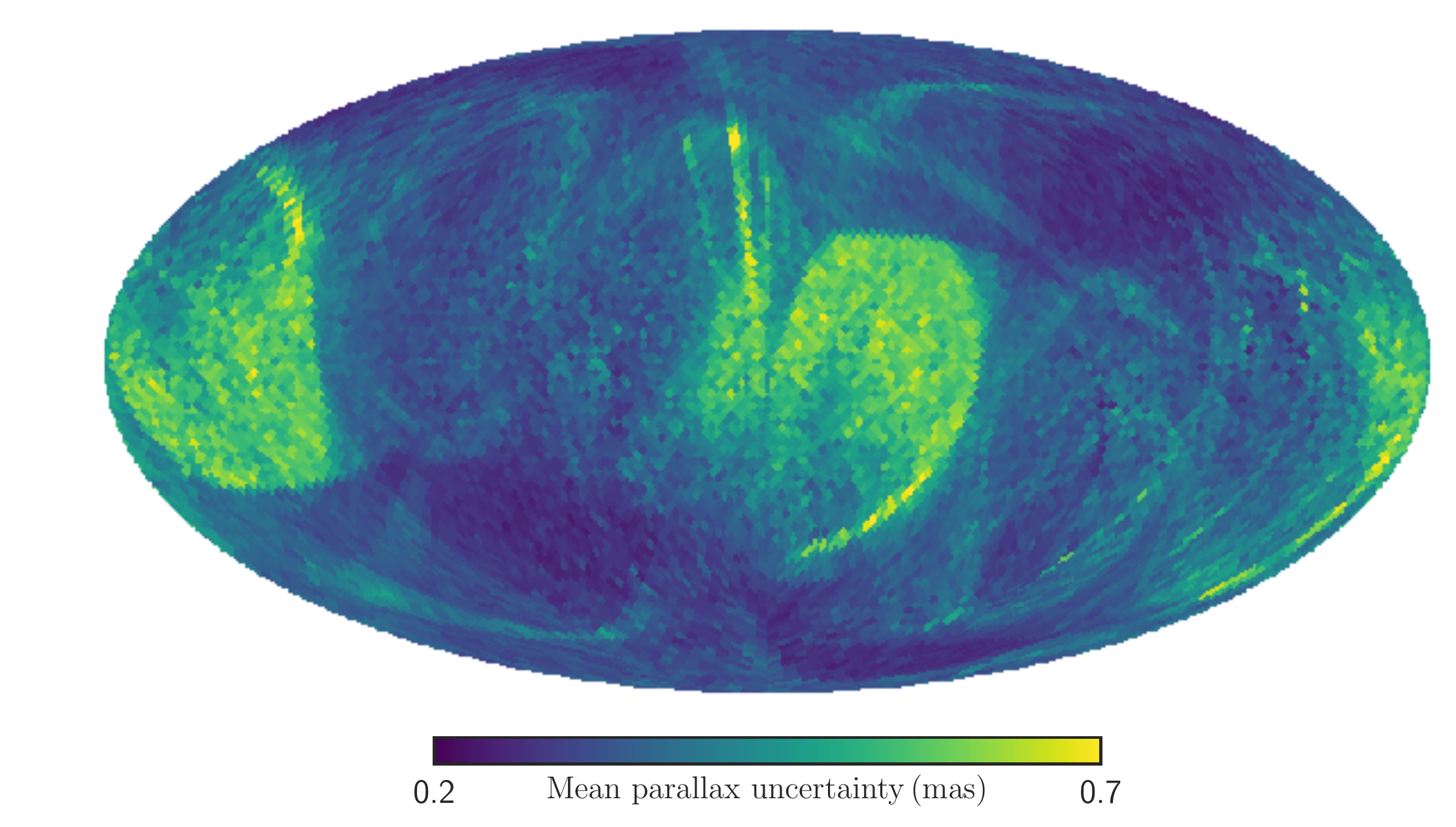}
  \caption{Properties of the \tgas\ scanning law and catalog relevant
    for the determination of the \tgas\ selection function. The panels
    show the following in $3.36\deg^2$ ($N_{\mathrm{side}} = 2^5$)
    HEALPix pixels: (a) the mean number of along-scan (AL)
    observations (divided by 9 to account for the 9 astrometric CCDs,
    such that 1 observation corresponds to 1 focal-plane crossing)
    [top left], (b) the standard deviation of the number of AL
    observations [top right], (c) the spread in the scan directions of
    different transits (as traced by
    \texttt{scan\_direction\_strength\_k4}; a small value indicates a
    large spread in the scan directions, which improves the
    astrometric solution) [bottom left], and (d) the mean parallax
    uncertainty [bottom right]. Due to the short period of
    observations covered by \Gaia\ DR1, a large area around the
    ecliptic currently has a small number of transits, with
    accordingly a small spread in the scan directions necessary for
    good astrometry. Areas with high numbers of transits are so narrow
    that the spread in the number of transits even in $3.36\deg^2$
    pixels is large and the astrometric quality likely also varies
    strongly within these pixels. These effects lead to an increase in
    the typical parallax uncertainty and to decreased
    completeness.}\label{fig:sky_props}
\end{figure*}

Because the data processing for the secondary solution, which contains
sky positions and broad band $G$ magnitudes down to $G \approx 20.7$,
is still in a preliminary state, we cannot assume that the secondary
solution is a complete sample that we can use to assess the
completeness of \tgas. We could use \tyctwo\ as the reference catalog,
but this would limit any use of the \tgas\ selection function to $V
\lesssim11$, the 99\,\% completeness limit of \tyctwo\ (without
modeling the completeness of \tyctwo\ itself). As we will see below,
\tgas\ is 50\,\% complete in at least half of the sky down to $V
\approx 12$ and about 20\,\% complete down to $V\approx 12.5$,
allowing us to extend \tgas\ coverage about a factor of two in
distance and eight in volume beyond the nominal \tyctwo\ limit.

We therefore use the Two Micron All Sky Survey
(2MASS; \citealt{Skrutskie06a}) Point Source Catalog, which is
$>99\,\%$ complete down to $J = 15.8$ and $K = 14.3$ over almost the
entire sky. As we will see below, this is at least two magnitudes
fainter than the \tgas\ completeness limit even in the best parts of
the sky and therefore suffices for our purposes. We select unique
($\texttt{use\_src} = ''\negmedspace\negmedspace 1''$), reliable point
sources in 2MASS that either have signal-to-noise ratio greater than
10 in $J$ ($\texttt{ph\_qual}=''\negmedspace\negmedspace A''$) or are
brighter than the detector saturation limit in either the ``Read 2 $-$
Read 1'' or in the ``Read 1'' exposures and similarly in $K_s$.

We assess the overall completeness by comparing the number of point
sources with $6 < J < 10$ and $0 < J-K_s < 0.8$ in \tgas\ and
2MASS. The number counts on the sky in the two catalogs are shown in
\figurename~\ref{fig:sky_counts}. Here and in what follows we bin the
sky using HEALPix\footnote{See \url{http://healpix.sourceforge.net}~.}
level $\nside = 2^5 = 32$. At this level, the sky is divided into
12,288 equal-area pixels that have an approximate size of $1.8^\circ$
and an approximate area of $3.36\deg^2$. This size was chosen as a
compromise between large pixels for good Poisson statistics on number
counts in \tgas\ and 2MASS and small pixels that resolve the
small-scale structure in the \Gaia\ DR1 scanning law. These pixels are
small enough to clearly display the structure in the scanning law (see
\figurename~\ref{fig:sky_props} below). By comparing the 2MASS counts
to the \tgas\ counts in this relatively bright magnitude range, it is
clear that while overall the counts are similar, the \tgas\ counts
have significant features that are absent in 2MASS.

We extract some of the main properties of the \Gaia\ DR1 scanning law
and data processing directly from the
\tgas\ catalog. \figurename~\ref{fig:sky_props} displays the mean
number of astrometric transits, the spread in this number, the mean
spread in the scan directions (the direction along which stars transit
the CCDs, using the catalog entry
\texttt{scan\_direction\_strength\_k4}), and the mean parallax
uncertainty as a function of position on the sky. The mean number of
observations clearly shows the imprint of the scanning law, with very
few transits near the ecliptic, many near the ecliptic poles, and
narrow ridges of abundant transits in ellipsoidal regions around the
ecliptic. The spread in the number of transits is small, except in the
ellipsoidal regions, which in reality are narrower than our pixel
size. Below, we will remove the small part of the sky with high spread
in the number of observations, because our sky pixelization is
inadequate there. Over the majority of the sky our pixelization
captures the properties of the observations well.

The lower-left panel in \figurename~\ref{fig:sky_props} displays a
measure of how well distributed the scan directions are in different
parts of the sky. Because \Gaia\ essentially performs one-dimensional
scans, the astrometric accuracy is higher when different transits scan
through a given field at a large variety of angles. As expected, the
spread in fields near the ecliptic is small, because these fields have
had few transits and thus cannot have a large spread in scan
directions. However, the regions near the ecliptic poles also have a
relatively small spread in scan directions, even though they have had
many observations. As is clear from the lower-right panel with the
mean parallax uncertainty, the regions near the ecliptic have
relatively large astrometric uncertainties because of this small
spread in scan directions.

\begin{figure*}
  \includegraphics[width=0.99\textwidth,clip=]{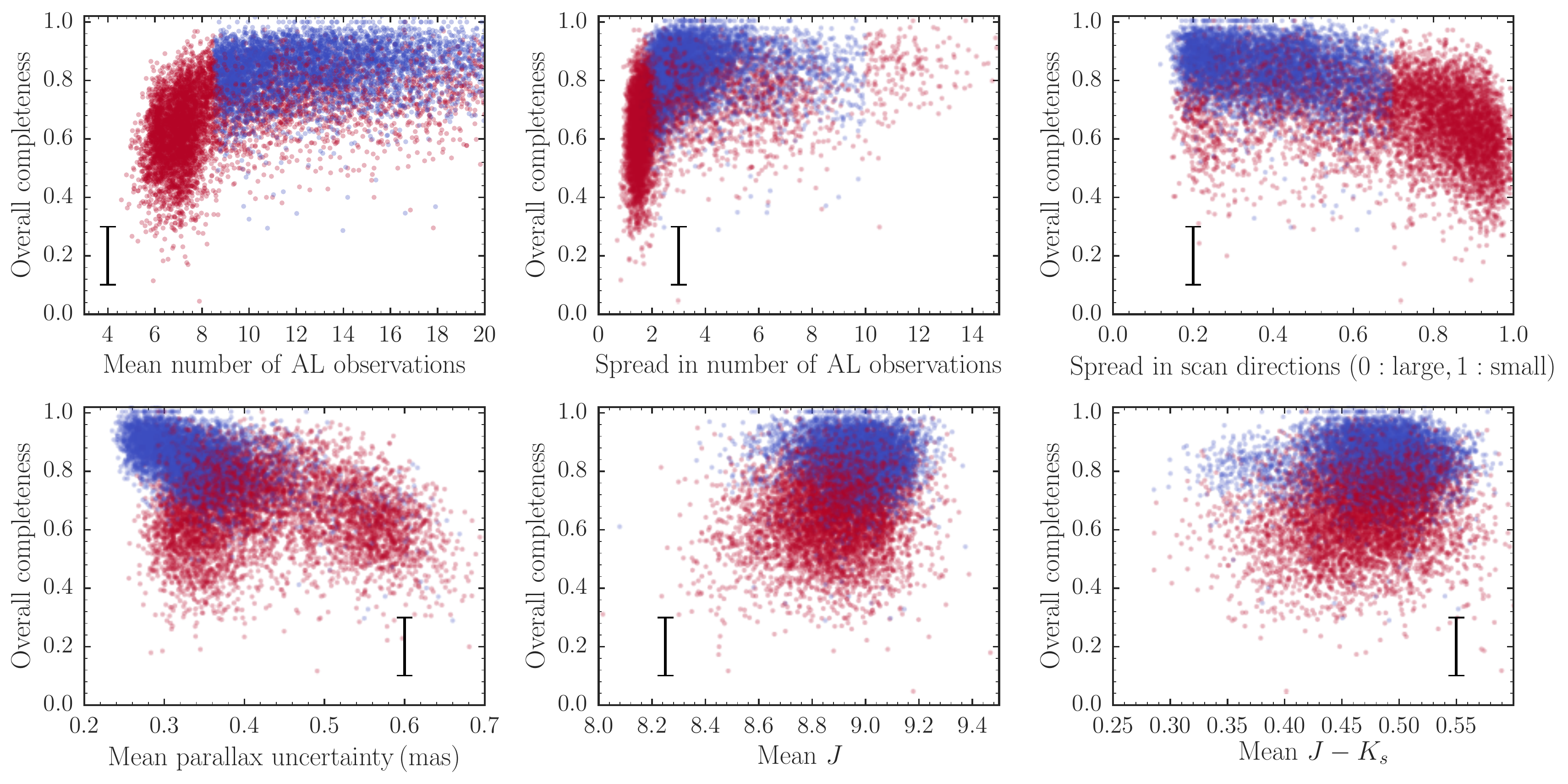}
  \caption{Overall completeness in $3.36\deg^2$ sky pixels as a
    function of properties of the observations in these pixels (see
    \figurename~\ref{fig:sky_props}; additionally we display the mean
    $J$-band magnitude and the mean $J-K_s$ color). The errorbar gives
    the typical uncertainty in the completeness. Low numbers of
    transits or a small spread in the scan directions are associated
    with a lower overall completeness. The blue points are sky
    locations that satisfy our \tgas\ observational quality cuts that
    select regions of high completeness, the red points cover the
    remaining part of the sky. The typical magnitude and color of the
    locations satisfying our quality cuts are similar to those of the
    rest of the sky.}\label{fig:props}
\end{figure*}

Comparing \figurename s~\ref{fig:sky_counts} and \ref{fig:sky_props}
one can see that many of the artifacts in
\figurename~\ref{fig:sky_counts} are aligned with features in the
scanning law. In many cases, these features coincide with the narrow
ridges of the ellipsoidal rings around the ecliptic, which have large
numbers of transits and one would, thus, naively expect to have high
completeness.

We compute the overall completeness as the ratio of the \tgas\ counts
in the range $6 < J < 10$ and $0 < J-K_s < 0.8$ to those in
2MASS. These ranges were chosen to contain enough stars in \tgas\ to
allow a precise measurement (that is, not too hobbled by Poisson
noise) of the completeness as a function of position on the sky using
the above pixelization. The overall completeness is shown as a
function of the properties of the scanning law and of the mean $J$ and
$J-K_s$ color in \figurename~\ref{fig:props}. It is clear that a low
number of transits is most strongly associated with low overall
completeness (top-left panel). The overall completeness quickly drops
for regions with less than 10 transits. The overall completeness is
also low when the spread in the scan directions is small (top-right
panel). As discussed above, some of the fields with a large number of
transits, but also a large spread in the number of transits have low
completeness.

Based on these considerations of the \Gaia\ DR1 scanning law, we
define a `good', or well observed, part of the sky as those pixels
that satisfy the following cuts:\\$\bullet$ Mean number of AL
observations $\geq 8.5$;\\$\bullet$ Spread in number of AL
observations $\leq 10$;\\$\bullet$ Spread in scan directions $\leq
0.7$;\\$\bullet$ $|\mathrm{ecliptic\ latitude}| \geq 20^\circ$.\\ The
last cut serves only to remove about $11\,\%$ of the sky that would
otherwise remain as small, isolated islands around the ecliptic. These
cuts select $48\,\%$ of the sky. Of the remaining $52\,\%$, $7\,\%$ is
removed because it has a low number of transits (but otherwise good
observations), $5\,\%$ because it has a small spread in scan
directions, and $28\,\%$ because it has both a low number of transits
and a small spread in scan directions. About $1\,\%$ of the sky is
excluded because it has a large spread in the number of transits (the
ellipsoidal ridges in \figurename~\ref{fig:sky_props}).

\begin{figure*}
  \includegraphics[width=0.69\textwidth,clip=]{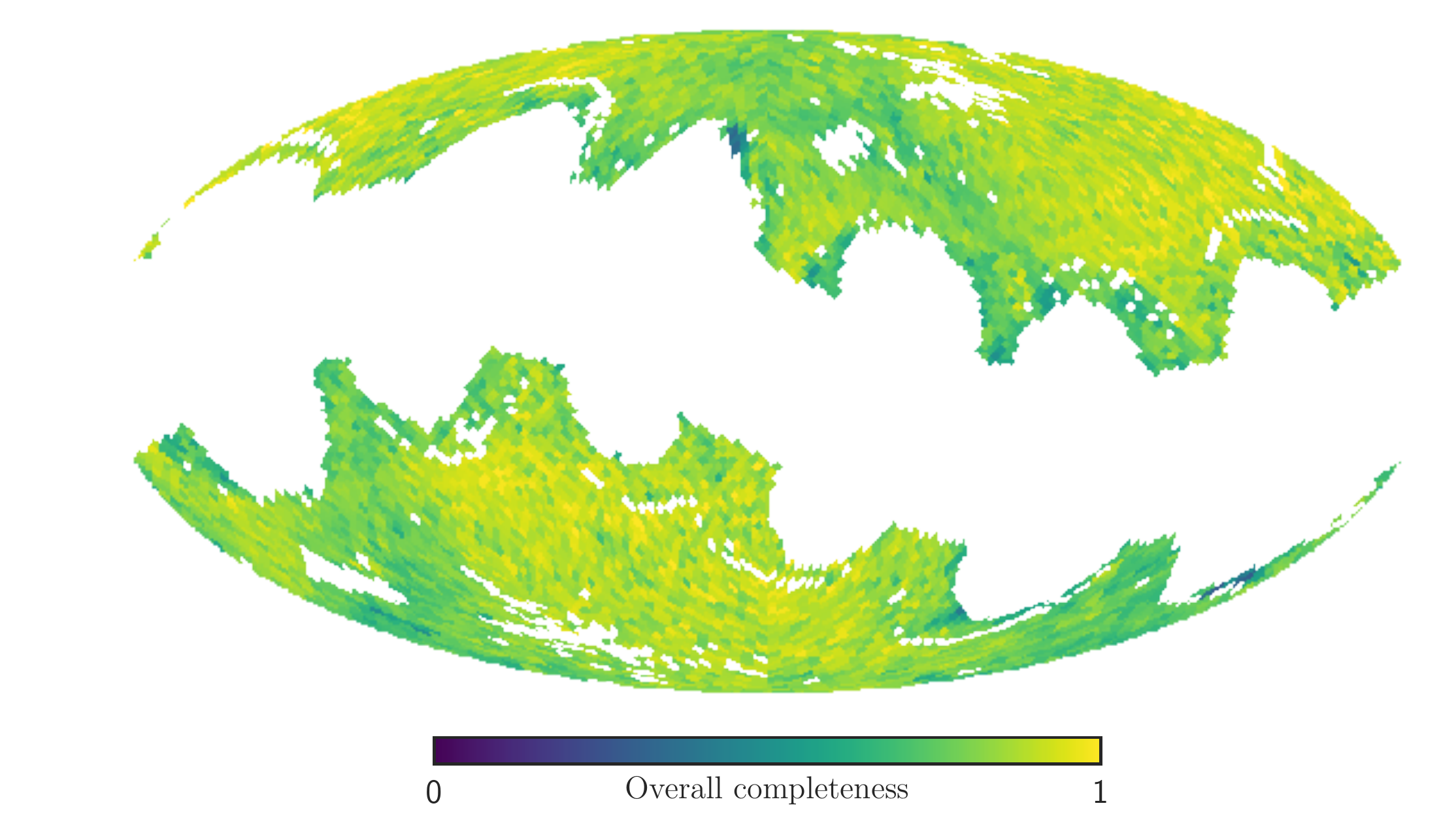}
  \caption{Overall completeness in the $\approx48\,\%$ of the sky that
    satisfies our \tgas\ observational quality cuts. The overall
    completeness in this part of the sky is largely isotropic. The
    Galactic plane has slightly lower overall completeness, because
    its magnitude distribution is skewed toward fainter magnitudes
    where it is affected by the incompleteness of
    \tgas.}\label{fig:overall_completeness}
\end{figure*}

The overall completeness in the `good' part of the sky is displayed in
\figurename~\ref{fig:overall_completeness}. It is clear that the
overall completeness is largely isotropic and does not contain sharp
features. The overall completeness is slightly lower near the Galactic
plane (which runs through much of the light-green parts of this map),
which is because the mean $J$ is somewhat fainter near the plane and
\tgas\ is not complete down to $J=10$ at all colors (see below). The
sky pixels selected by the above cuts are shown as blue dots in
\figurename~\ref{fig:props}. The cut to the `good' part of the sky
selects regions that by and large have the same color and apparent
magnitude distribution as the rest of the sky. The mean parallax
uncertainty in the `good' part of the sky is typically $\sigma_\varpi <
0.45\mas$.

\begingroup \centering
\subsection{Completeness as a function of color and magnitude}
\endgroup

\begin{figure}
\begin{center}
  \includegraphics[width=0.79\textwidth,clip=]{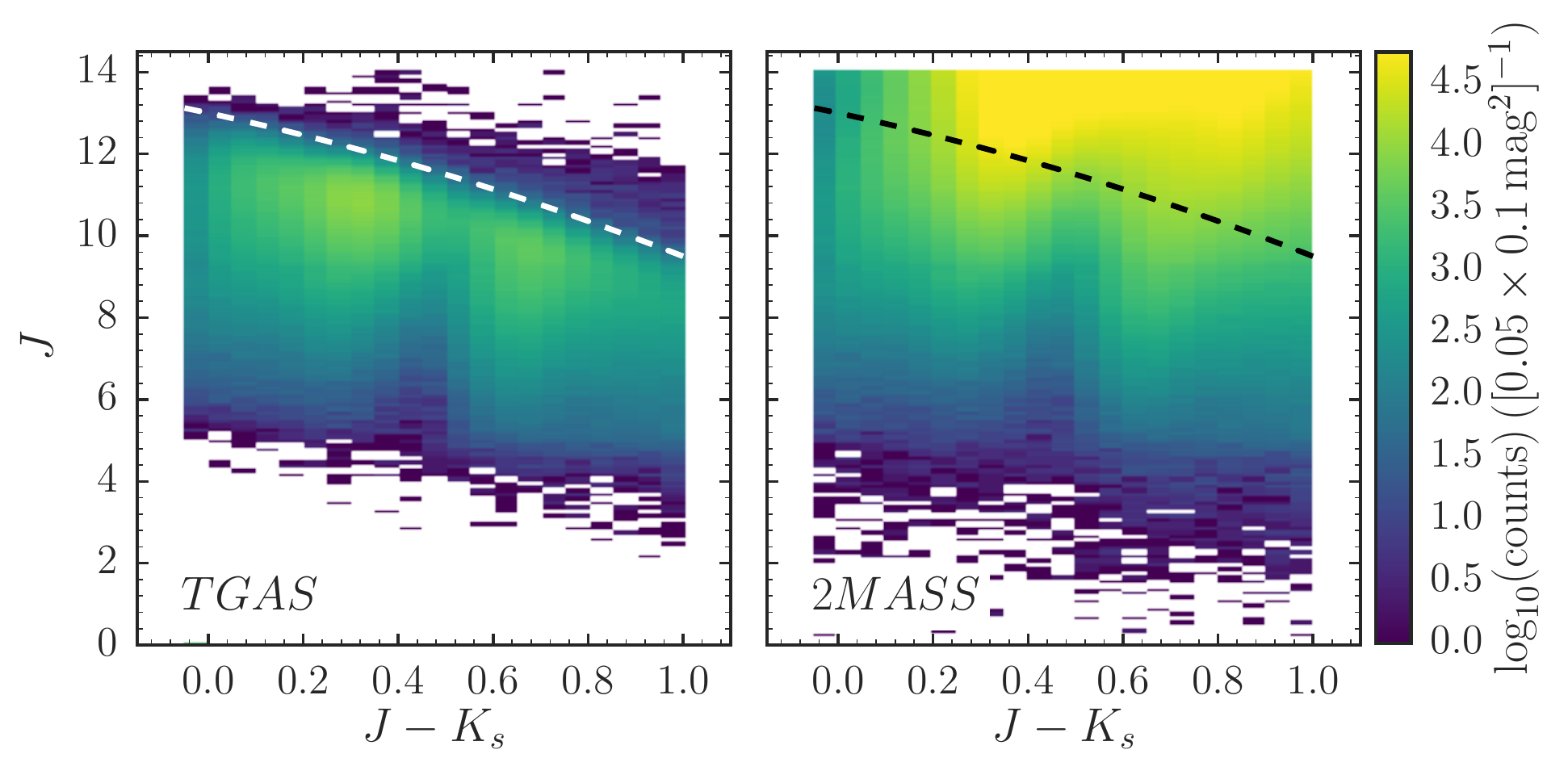}
  \caption{Number counts as a function of $(J,J-K_s)$ for \tgas\ and
    2MASS in the `good' $48\,\%$ of the sky. The dashed line ($J =
    13-(J-K_s)^2-2.5\,(J-K_s)$ indicates the location of a sharp
    drop-off in the \tgas\ number counts that is not present in 2MASS
    and is therefore due to the \tgas\ selection. In $(J,J-K_s)$ the
    \tgas\ selection function depends on
    color.}\label{fig:cmds_2m_tg}
\end{center}
\end{figure}

The overall completeness in the `good' part of the sky selected in the
previous subsection does not appear to have any significant residual
dependence on position on the sky (see
\figurename~\ref{fig:props}). Therefore, we continue under the
assumption that the completeness in the `good' region of the sky does
not depend on sky position. We can then use the large number of stars
in this region of the sky to determine the dependence of the
\tgas\ completeness on color and magnitude in detail.

\figurename~\ref{fig:cmds_2m_tg} displays number counts in \tgas\ and
2MASS in the `good' part of the \tgas\ sky in the range $-0.05 < J-K_s
< 1$ and $J < 14$ as a function of $(J-K_s,J)$. The main features of
the distribution at brighter magnitudes are the same in both catalogs,
but there is both a lack of stars in \tgas\ at the bright and faint
end. At the faint end in particular, there is a steep cut-off in the
\tgas\ number counts that is absent in 2MASS and therefore needs to be
because of the \tgas\ selection. This cut-off has a strong dependence
on color, the approximate shape of which is indicated by the dashed
line. The overall number counts (summed over $J-K_s$ in the $-0.05 <
J-K_s < 1$ range) are shown in \figurename~\ref{fig:magdist_j}. These
number counts demonstrate the same bright- and faint-end cut-offs. The
faint-end cut-off appears broad in this representation because it is
summed over color. It is also clear that the \tgas\ counts nowhere
quite reach the 2MASS counts, they fall short at every magnitude.

\begin{figure}
\begin{minipage}[t]{0.49\textwidth}
  \includegraphics[width=0.99\textwidth,clip=]{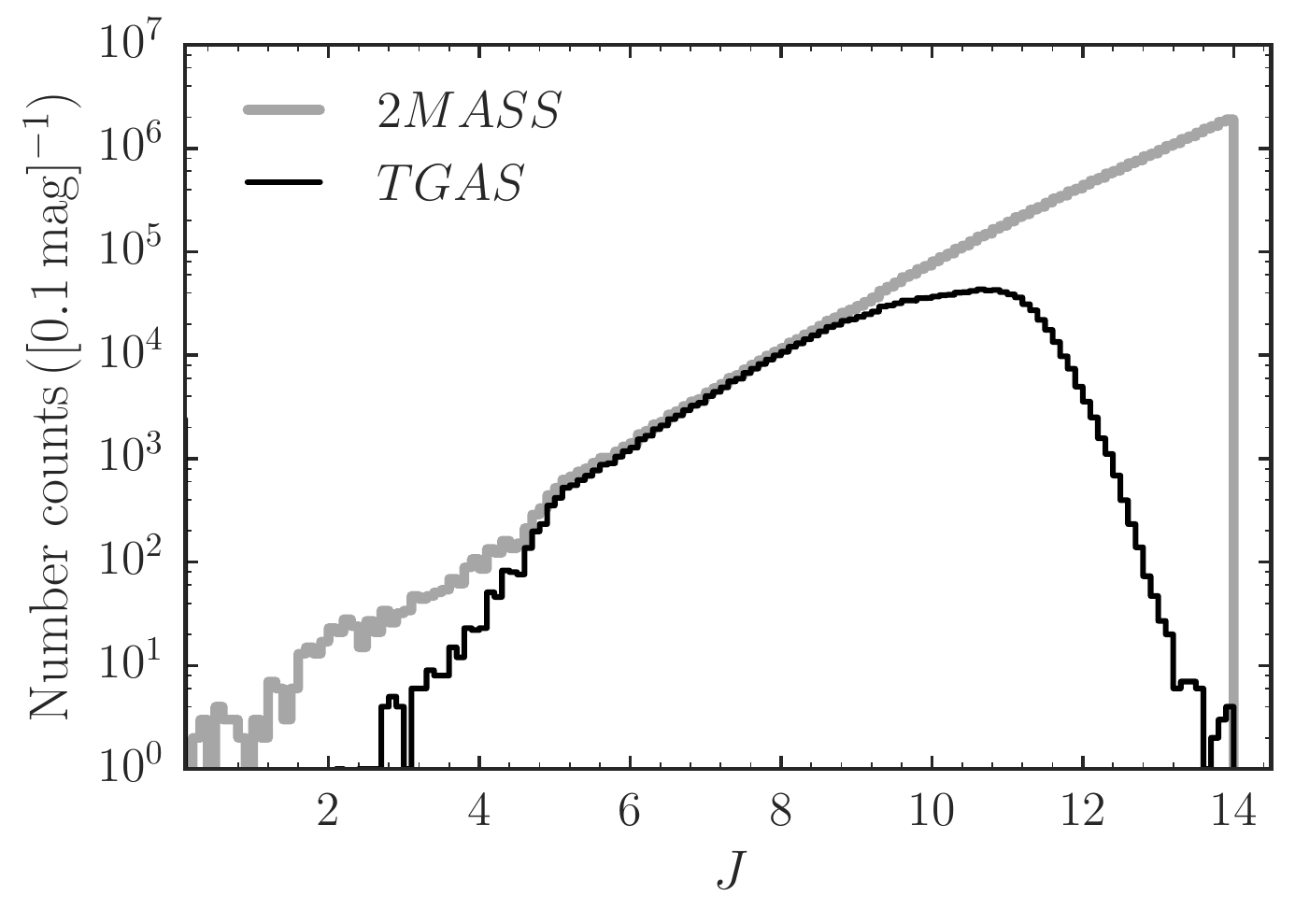}
  \caption{Number counts as a function of $J$ for \tgas\ and 2MASS in
    the `good' $48\,\%$ of the sky. \tgas\ is incomplete at both the
    bright and faint end with broad drop-offs. This broadness is due
    to the color-dependence of the selection function in
    \figurename~\ref{fig:cmds_2m_tg}, which smears out the sharper cut
    in the two-dimensional $(J,J-K_s)$
    plane.}\label{fig:magdist_j}
\end{minipage}\hfill
\addtocounter{figure}{1}
\begin{minipage}[t]{0.49\textwidth}
  \includegraphics[width=0.99\textwidth,clip=]{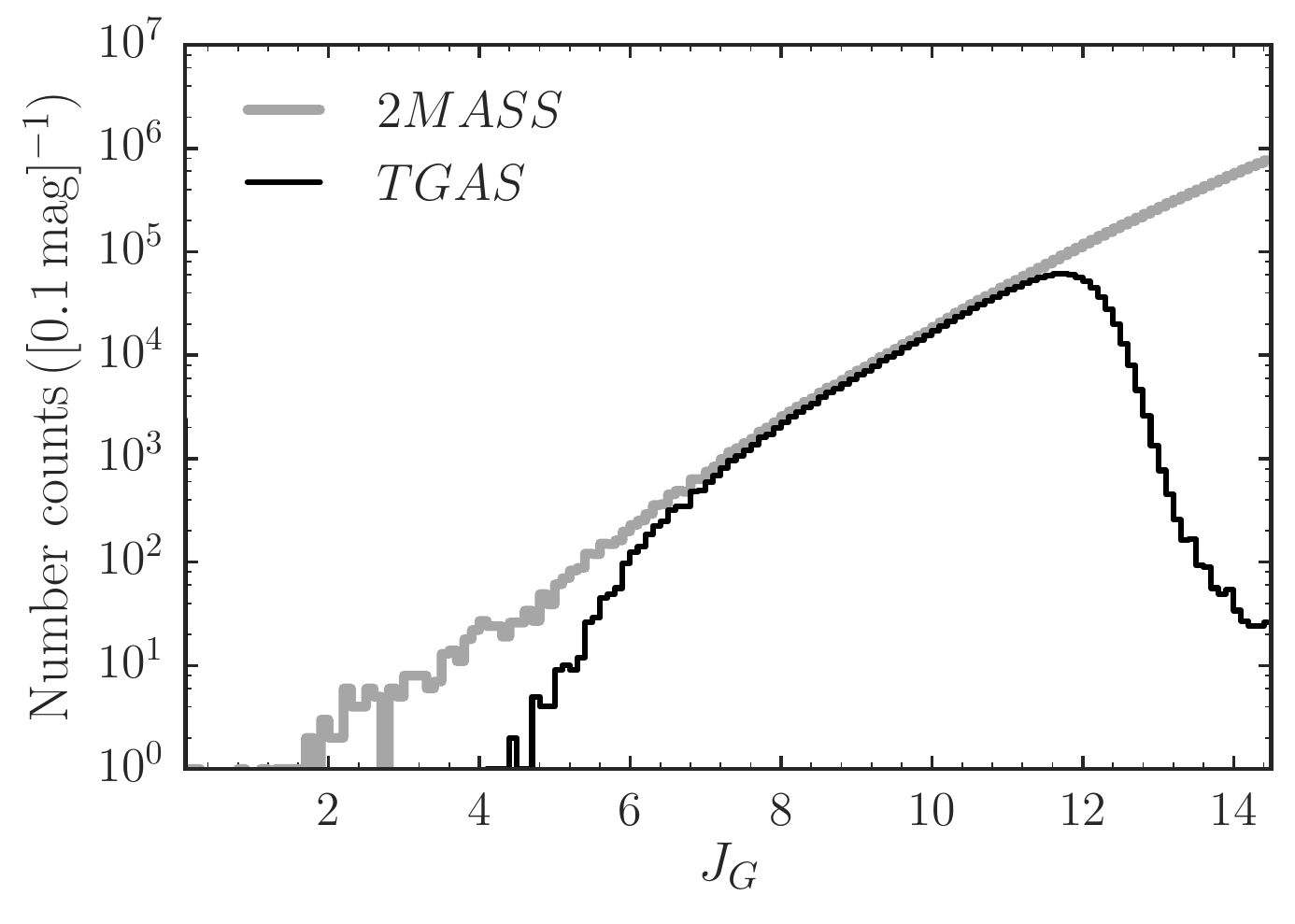}
  \caption{Number counts as a function of $J_G = J + (J-K_s)^2 +
      2.5\,(J-K_s)$ for \tgas\ and 2MASS in the `good' $48\,\%$ of the
      sky. The completeness of \tgas\ is approximately independent of
      color in $J_G$: \tgas\ drops off more sharply than in
      \figurename~\ref{fig:magdist_j}.}\label{fig:magdist_jg}
\end{minipage}
\end{figure}
\addtocounter{figure}{-2}

\begin{figure}
\begin{minipage}[t]{0.49\textwidth}
  \includegraphics[width=0.99\textwidth,clip=]{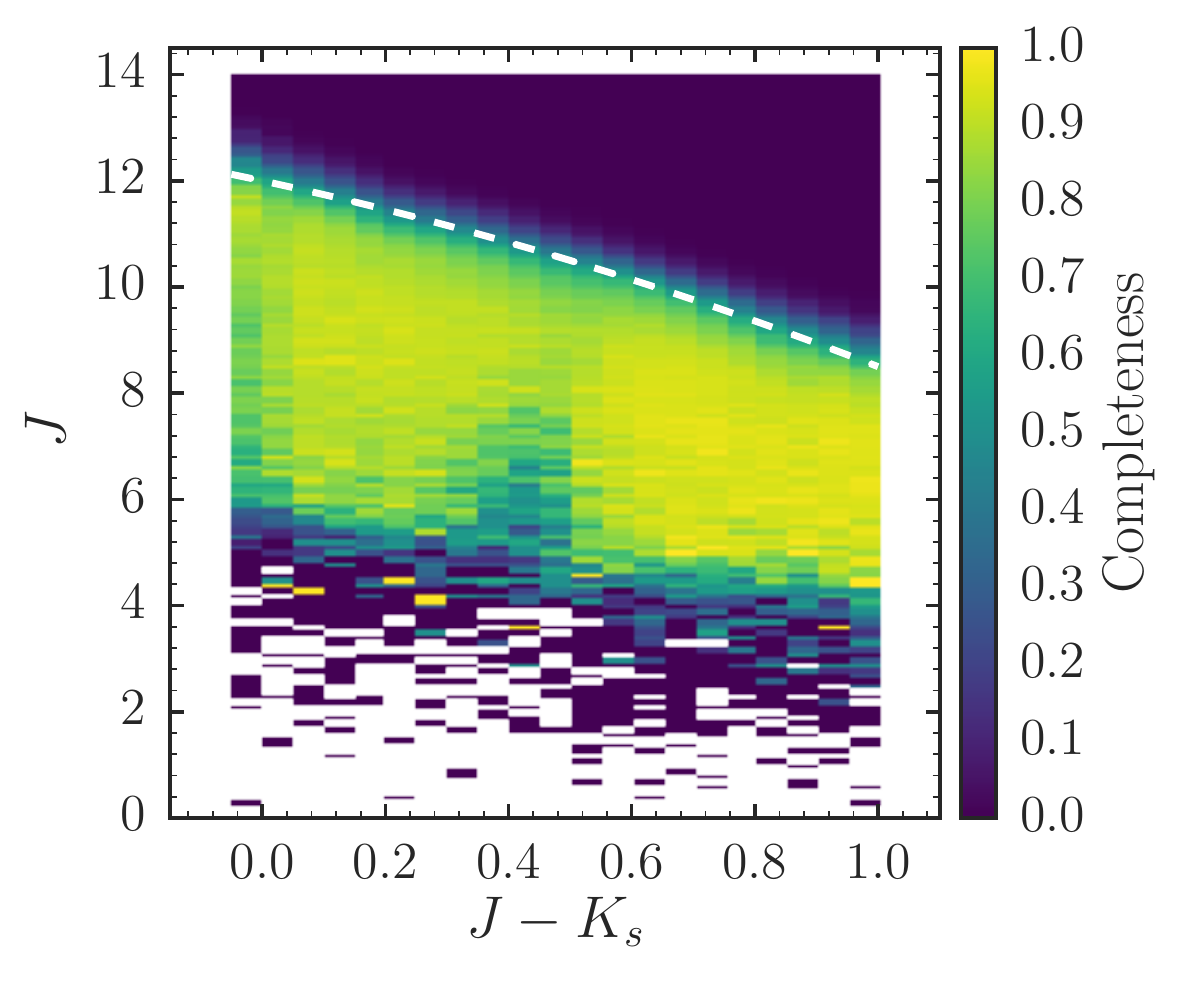}
  \caption{\tgas\ completeness with respect to 2MASS as a function of
    $(J,J-K_s)$. This is the ratio of the left and right panels of
    \figurename~\ref{fig:cmds_2m_tg}. The dashed line is one magnitude
    brighter than in \figurename~\ref{fig:cmds_2m_tg}: $J =
    12-(J-K_s)^2-2.5\,(J-K_s)$. This is the approximate magnitude to
    which \tgas\ is $50\,\%$ complete at the faint
    end.}\label{fig:comp_j_jk}
\end{minipage}
\hfill
\addtocounter{figure}{1}
\begin{minipage}[t]{0.49\textwidth}
  \includegraphics[width=0.99\textwidth,clip=]{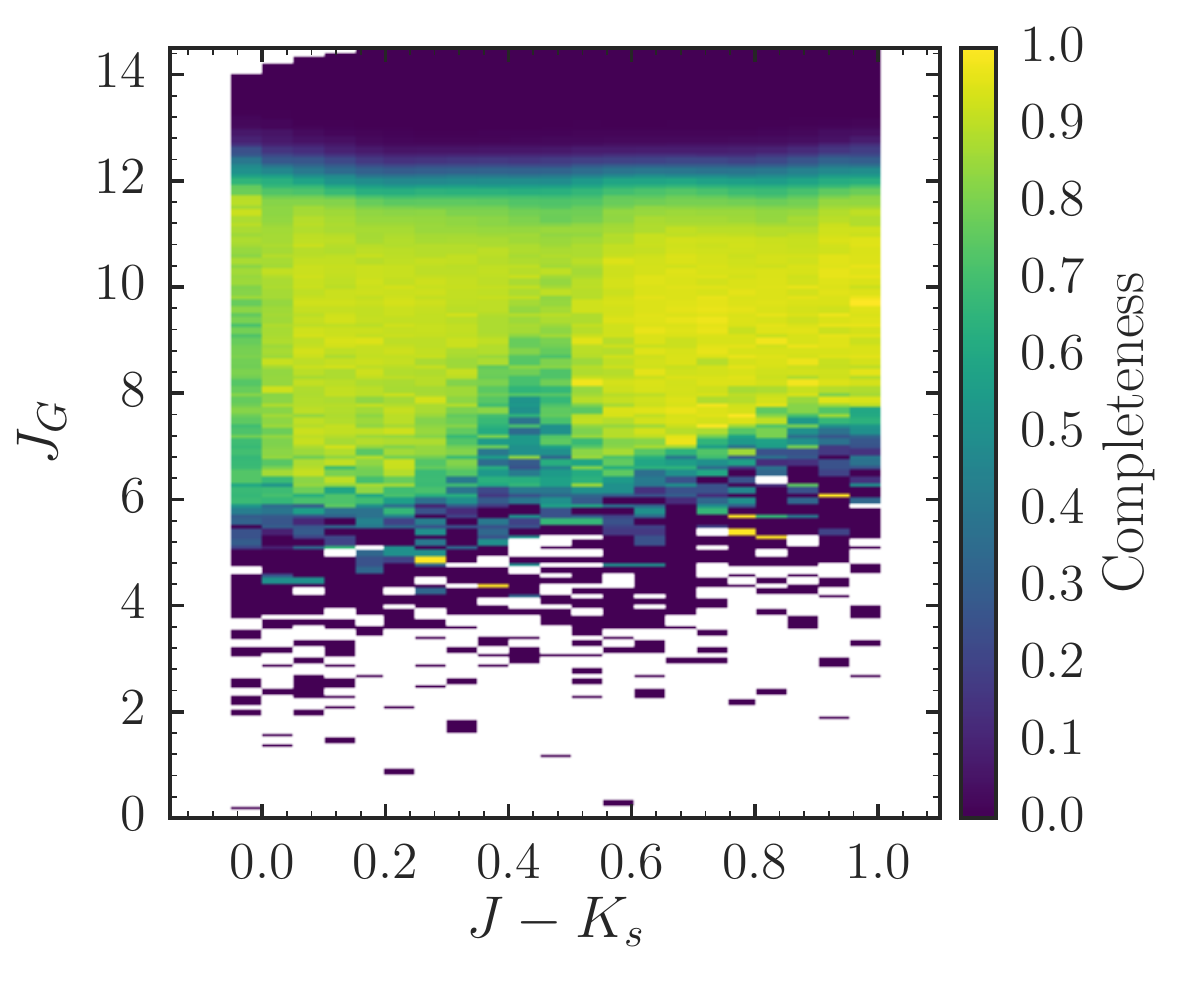}
  \caption{\tgas\ completeness with respect to 2MASS as a function of
    $(J_G,J-K_s)$. The completeness of \tgas\ is approximately
    independent of color in $J_G$. The completeness at redder
    $(J-K_s)$ is slightly higher than at the blue end and we determine
    the selection function in three broad $(J-K_s)$ bins to account
    for this.}\label{fig:comp_jg_jk}
\end{minipage}
\end{figure}

The \tgas\ completeness computed as the number counts in
\tgas\ divided by those in 2MASS are displayed in
\figurename~\ref{fig:comp_j_jk}. This clearly shows the sharp drop in
the completeness at faint magnitudes. To remove the strong color
dependence of the completeness cut-off, we adjust the $J$ magnitude to
a new $J_G = J + (J-K_s)^2 + 2.5\,(J-K_s)$ that runs approximately
parallel to the completeness cut-off. This line for $J_G=12$ is shown
in \figurename~\ref{fig:comp_j_jk}. The $J-K_s$ color dependence is
essentially caused by our use of a near-infrared color and magnitude
for a survey whose completeness is more appropriately a function of an
optical magnitude. The $J \rightarrow J_G$ relation is close to the
relation that maps $J \rightarrow G$ along the stellar locus.

\begin{figure*}
  \includegraphics[width=0.99\textwidth,clip=]{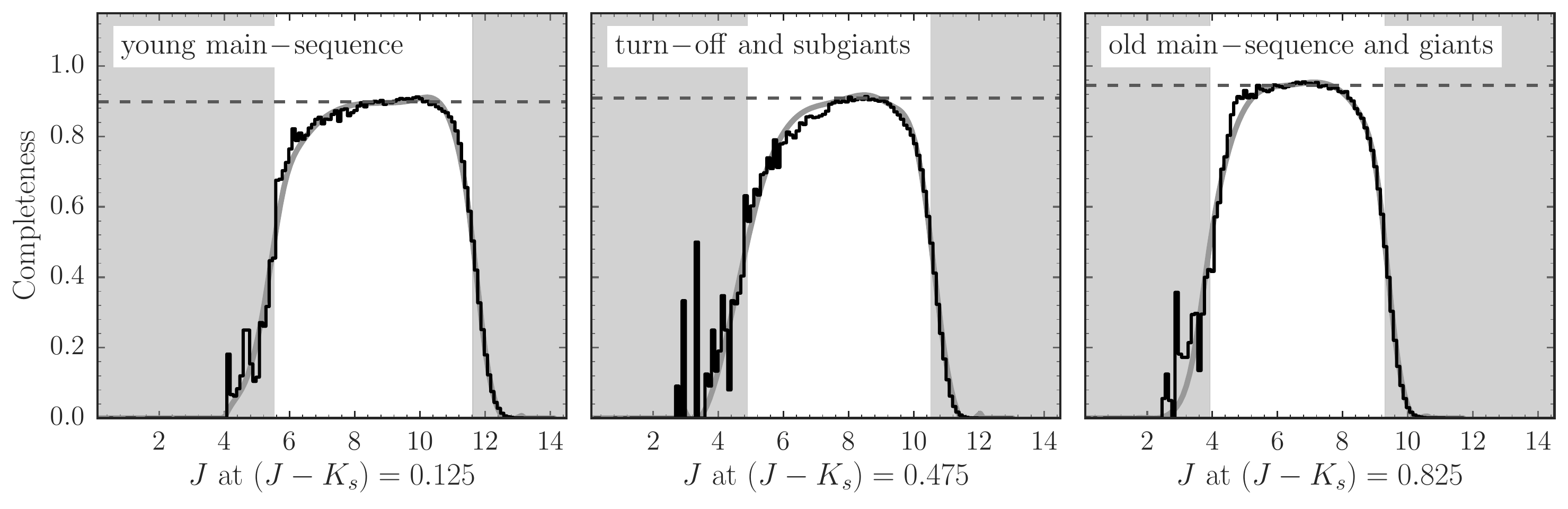}
  \caption{\tgas\ completeness in the three broad $(J-K_s)$ bins in
    which we determine it ($\Delta J-K_s = 0.35$ over $-0.05 < J-K_s <
    1$). Natively, the completeness is a function of $J_G$; we have
    translated this to $J$ for the central color of each bin. The
    dashed line indicates the approximate plateau at intermediate
    magnitudes and the area between the gray bands is where the
    completeness is higher than $50\,\%$. The thick gray curve is a
    smooth interpolation of the histogram that we use as our model for
    the selection function.}\label{fig:comp_3jbins}
\end{figure*}

The number counts in \tgas\ and 2MASS as a function of $J_G$ are shown
in \figurename~\ref{fig:magdist_jg}; the \tgas\ number counts drop
more steeply in $J_G$ than they do in $J$. The completeness (ratio of
\tgas\ to 2MASS number counts) as a function of $(J-K_s,J_G)$ is
displayed in \figurename~\ref{fig:comp_jg_jk}. It is clear that the
definition of $J_G$ has succeeded in removing the color dependence of
the faint-end cut-off. A slight color dependence in both the level of
the completeness around $J_G = 10$ and in the bright cut-off
remains. We ignore the latter, because for star counts there are very
few stars at these bright magnitudes that contribute to the stellar
density. To deal with the former, we divide the color range in three
equal-sized bins with $\Delta J-K_s = 0.35$ in the range $-0.05 <
J-K_s < 1$ and approximate the selection function as being a function
of $J_G$ only. The completeness in these bins as a function of $J_G$
is shown in \figurename~\ref{fig:comp_3jbins} (for easier
interpretability, we have translated $J_G$ to $J$ at the center of
each bin). The gray curve is a smooth spline fit.

The final model for the \tgas\ selection function
$S(J,J-K_s,\ra,\dec)$ is therefore a the function that is (i) zero for
($\ra,\dec$) outside of the `good' part of the sky and (ii) given by
the smooth spline model for the color bin in which $J-K_s$ is located,
evaluated at $J_G$. We do not model stars bluer than $J-K_s = -0.05$
(mainly O and B stars) or redder than $J-K_s = 1$. This model for the
raw \tgas\ selection function is available in the
\texttt{gaia\_tools.select.tgasSelect} class in the
\texttt{gaia\_tools} package available
at\\ \centerline{\url{https://github.com/jobovy/gaia_tools}~.}

\begingroup \centering
\subsection{Comparison with \tyctwo}
\endgroup

\begin{figure*}
  \includegraphics[width=0.79\textwidth,clip=]{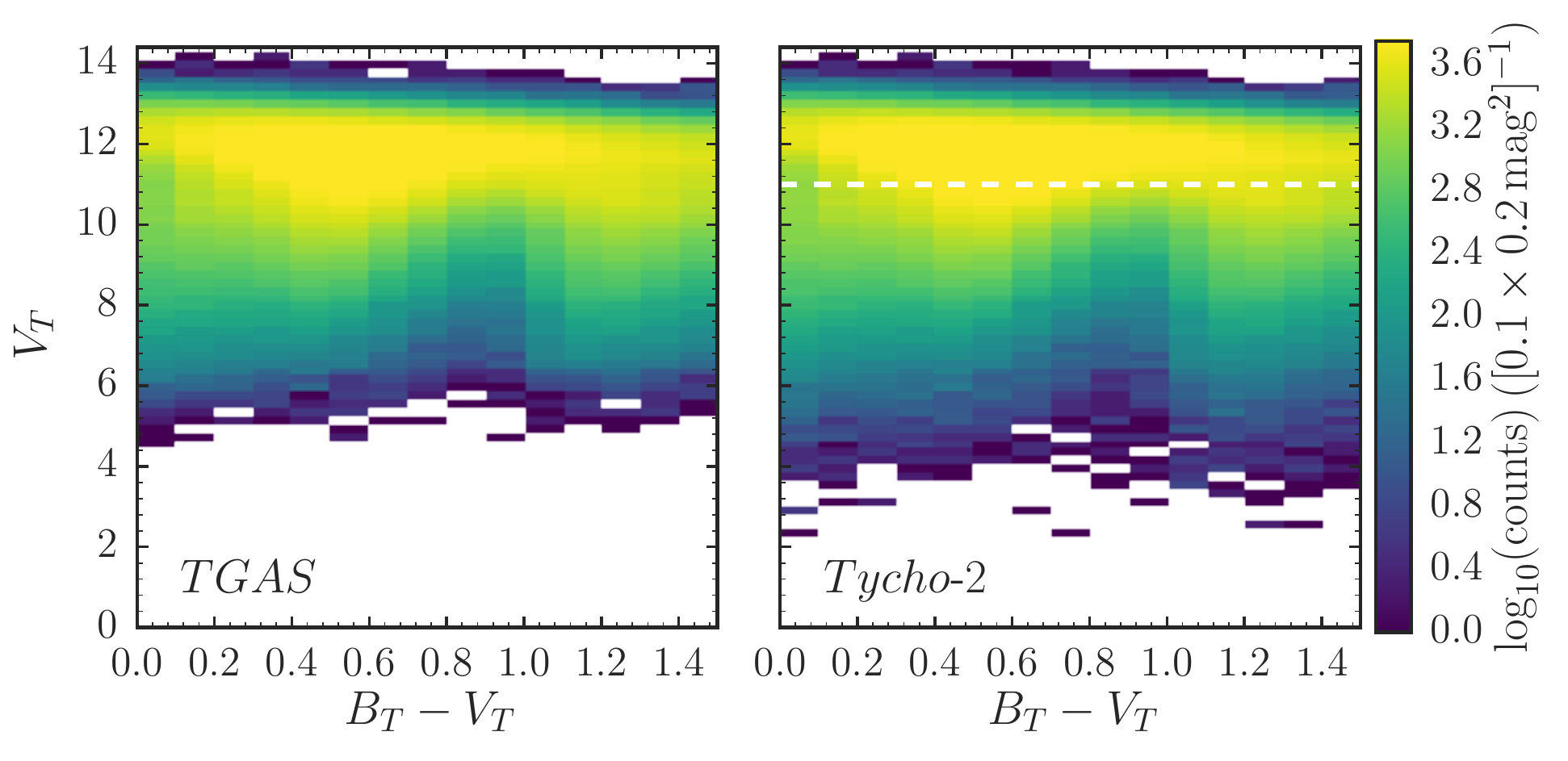}
  \caption{Number counts as a function of $(V_T,B_T-V_T)$ for
    \tgas\ and \tyctwo\ in the `good' $48\,\%$ of the sky. The dashed
    line at $V_T = 11$ indicates the location of the 99\,\%
    completeness limit of \tyctwo. \tgas\ closely traces its parent
    catalog \tyctwo.}\label{fig:cmds_tyc2_tg}
\end{figure*}

\begin{figure*}
  \includegraphics[width=0.79\textwidth,clip=]{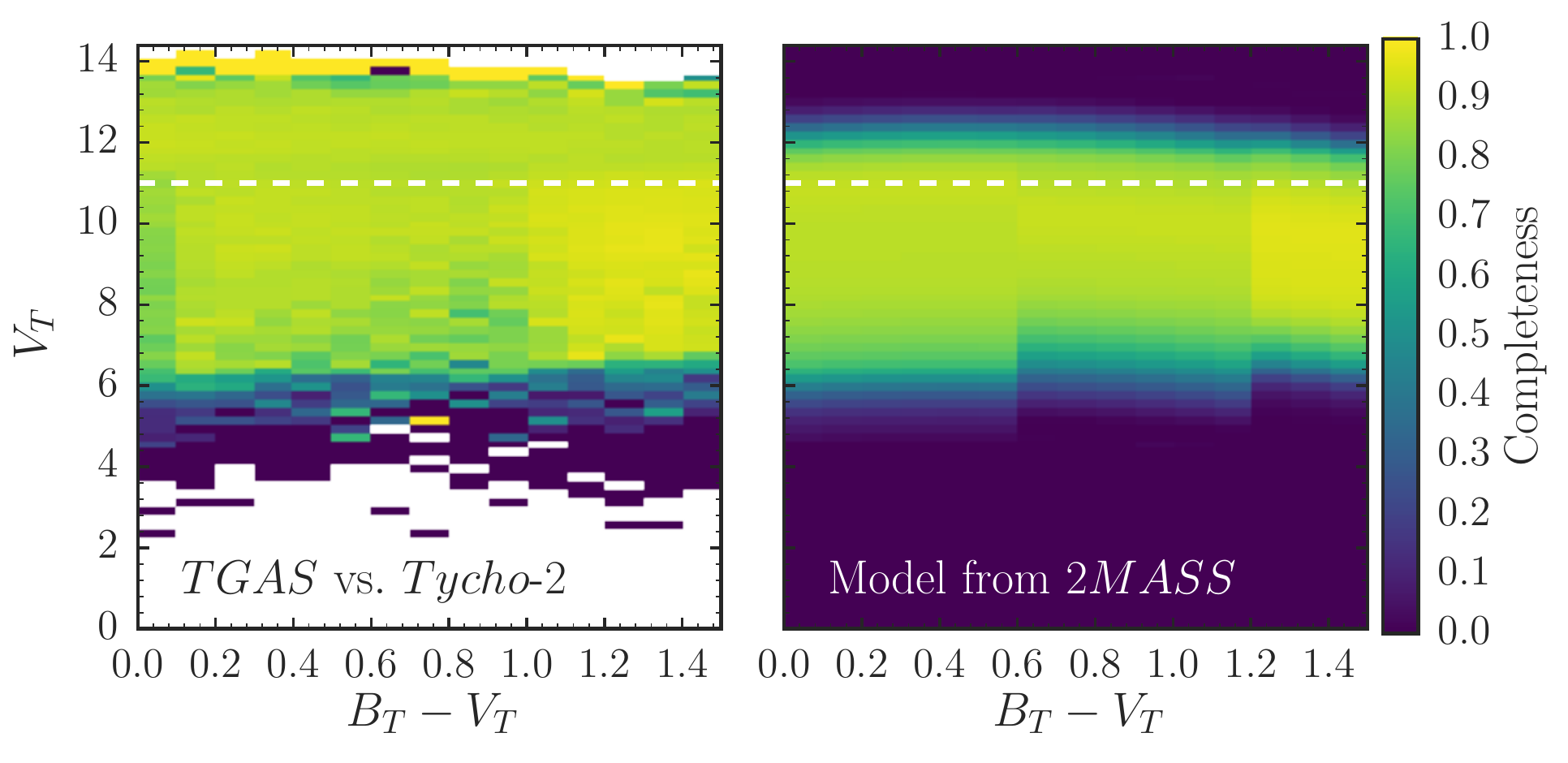}
  \caption{\tgas\ completeness with respect to \tyctwo\ as a function
    of $(V_T,B_T-V_T)$ (left panel). This is the ratio of the left and
    right panels of \figurename~\ref{fig:cmds_tyc2_tg}. The right
    panel represents our model for this completeness derived from (a)
    the \tgas\ selection function derived from comparing \tgas\ to
    2MASS and (b) color-color transformations between the visible and
    near-infrared photometric bands (because we do not include scatter
    in the color transformations, the completeness is constant within
    three color bins as it is in the model). The dashed line in both
    panels is the 99\,\% completeness limit of \tyctwo. Overall, our
    model for the selection function matches the amplitude and the
    lower and upper cut-offs in the
    completeness.}\label{fig:comp_vt_btvt}
\end{figure*}

\begin{figure*}
  \includegraphics[width=0.99\textwidth,clip=]{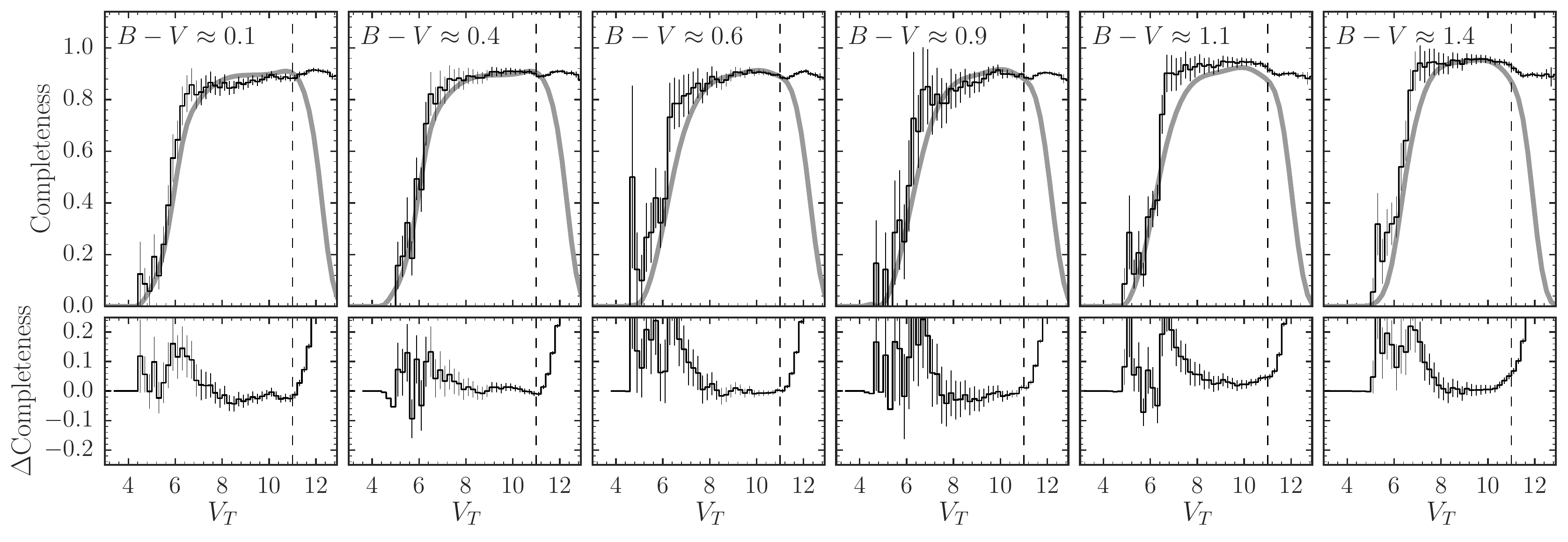}
  \caption{Comparison between the completeness of \tgas\ with respect
    to \tyctwo\ in six $B_T-V_T$ color bins measured directly by
    comparing \tgas\ and \tyctwo\ number counts (histogram with
    uncertainties) and derived from our model for the \tgas\ selection
    function (smooth, gray curve). The top panel directly compares the
    two, the bottom panel displays the residual (direct minus
    model). The dashed line is the 99\,\% completeness limit of
    \tyctwo, beyond which the direct measurement is
    meaningless. Overall the model represents the completeness well,
    with some (expected) deviations at the bright end and around $B-V
    \approx 1.1$.}\label{fig:comp_tyc2}
\end{figure*}

We test our model for the \tgas\ selection function by determining the
completeness of \tgas\ with respect to its parent catalog \tyctwo\ at
apparent magnitudes brighter than the 99\,\% completeness limit of
\tyctwo\ ($V\approx11$). Similar to how we determine the
\tgas\ selection function above by comparing \tgas\ number counts to
number counts in 2MASS, we compute the number counts as a function of
color $B_T-V_T$ and magnitude $V_T$ in the `good' part of the sky
(where our model for the selection function is applicable). These
number counts for \tgas\ and \tyctwo\ are displayed in
\figurename~\ref{fig:cmds_tyc2_tg}. The ratio of these number counts
gives the completeness of \tgas\ with respect to \tyctwo\ and this
ratio is shown in the left panel of
\figurename~\ref{fig:comp_vt_btvt}.

Our model for the \tgas\ selection function is a function of
$(J,J-K_s)$ and to compare it to the \tyctwo\ number counts we need to
translate the model to $(V_T,B_T-V_T)$. We do this using the following
color--color transformations
\begin{align}
  J-K_s & = 0.55\,(B_T-V_T)-0.02\,,\\
  V_T-J & = -0.21\,(B_T-V_T)^2 +1.8\,(B_T-V_T) +0.1\,.
\end{align}
Evaluating our model for the selection function as a function of
$(V_T,B_T-V_T)$ using these relations, we obtain the model displayed
in the right panel of \figurename~\ref{fig:comp_vt_btvt}. Because we
use deterministic color--color relations that do not account for the
scatter in this transformation, our model evaluated as a function of
$B_T-V_T$ is constant in three $B_T - V_T$ ranges (owing to the fact
that this is true for the model as a function of $J-K_s$). Comparing
the measured \tgas/\tyctwo\ completeness to the model in
\figurename~\ref{fig:comp_vt_btvt}, we find that the overall agreement
of our model with the measurement is good. The model captures the
shape and amplitude of the completeness from the bright end to the
faint end and from the blue end to the red end.

A more detailed comparison is shown in
\figurename~\ref{fig:comp_tyc2}. This figure compares the completeness
as a function of $V_T$ of \tgas\ versus \tyctwo\ derived from the
\tgas\ and \tyctwo\ number counts to the model in eight bins in
$B_T-V_T$. The model agrees well with the data in almost all cases,
except (a) at the bright end in almost all color bins and (b) in its
overall amplitude around $B_T-V_T \approx 1.1$. These discrepancies
are expected based on how we determined the selection function
above. To produce a simple model that accurately describes the faint
end of the completeness, we performed a simple transformation
$(J,J-K_s) \rightarrow J_G$ in which the faint end cut-off is
independent of color. However, the completeness at the bright end does
depend on color in $J_G$ and our model does not capture this
dependence. This causes the discrepancy at the bright end of the
completeness in $(V_T,B_T-V_T)$. Clearly, the completeness has a
simpler color dependency in $B_T-V_T$ than in $J-K_s$, so if we had a
complete optical survey we could determine a more accurate selection
function. For the purpose of this paper, a small discrepancy at bright
magnitudes does not influence our results much, because very few stars
have such bright magnitudes.

The discrepancy near $B_T-V_T \approx 1.1$ is caused by our choice of
binning in $J-K_s$. For ease of interfacing with the large 2MASS
database, the $J-K_s$ color bins were chosen to be equal size and our
binning does not perfectly capture the higher completeness at the red
end. This shows up most prominently around $J-K_s \approx 0.6$ or
$B_T-V_T \approx 1.1$, which falls in the middle color bin of the
model, but in reality is closer to the higher completeness of the
reddest color bin of the model. The discrepancy amounts to only a few
percent underestimation of the completeness in a narrow color strip.

\begingroup \centering
\section{Overall completeness of the low-quality portion of the sky}
\endgroup

\begin{figure*}
  \includegraphics[width=0.69\textwidth,clip=]{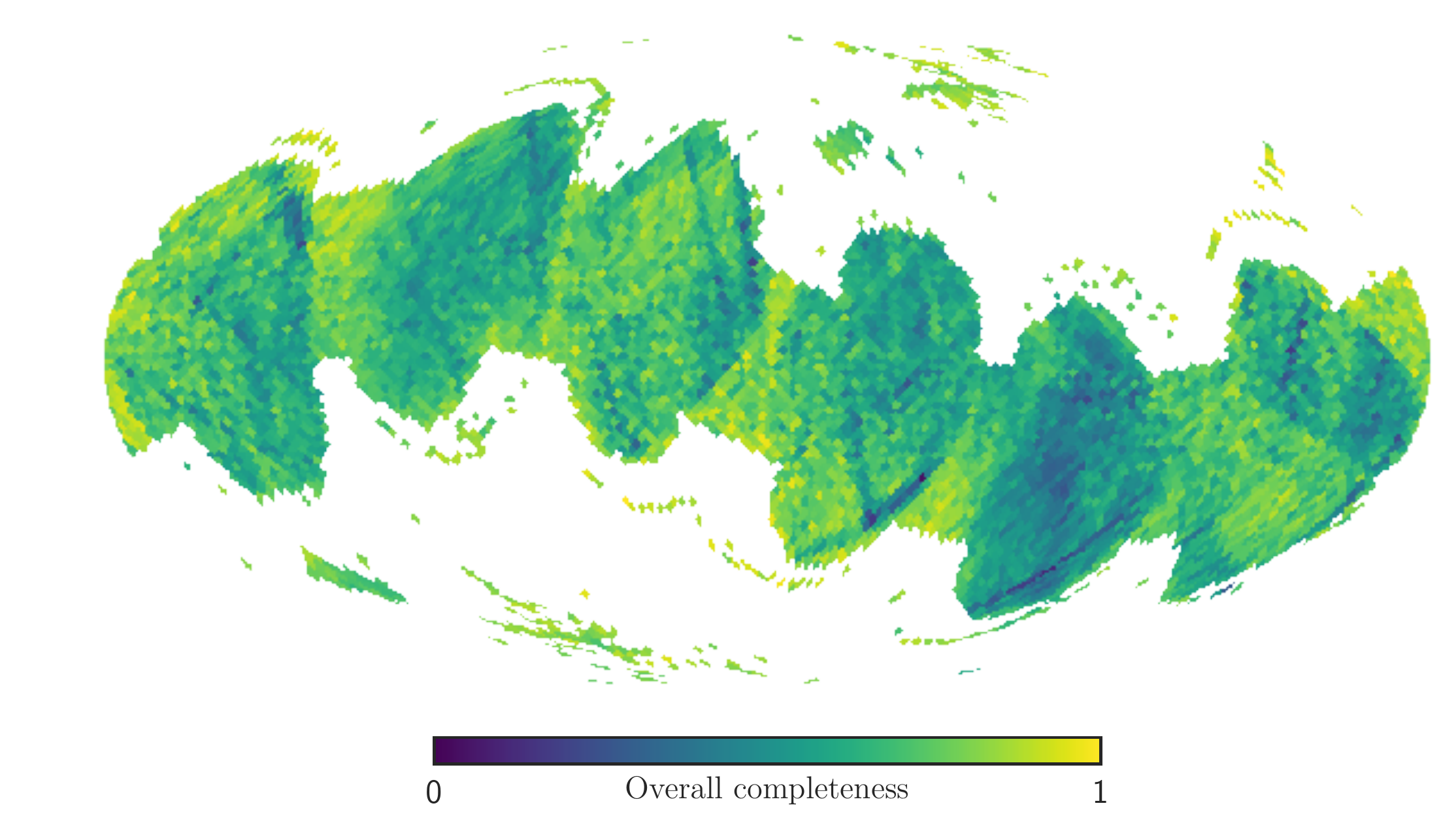}
  \caption{Overall completeness in the $\approx52\,\%$ of the sky that
    fails our \tgas\ observational quality cuts. Compared to
    \figurename~\ref{fig:overall_completeness}, it is clear that the
    completeness in this part of the sky is lower and has more
    artificial structure than in the `good' $48\,\%$ of the
    sky.}\label{fig:overall_completeness_bad}
\end{figure*}

\begin{figure*}
  \includegraphics[width=0.49\textwidth,clip=]{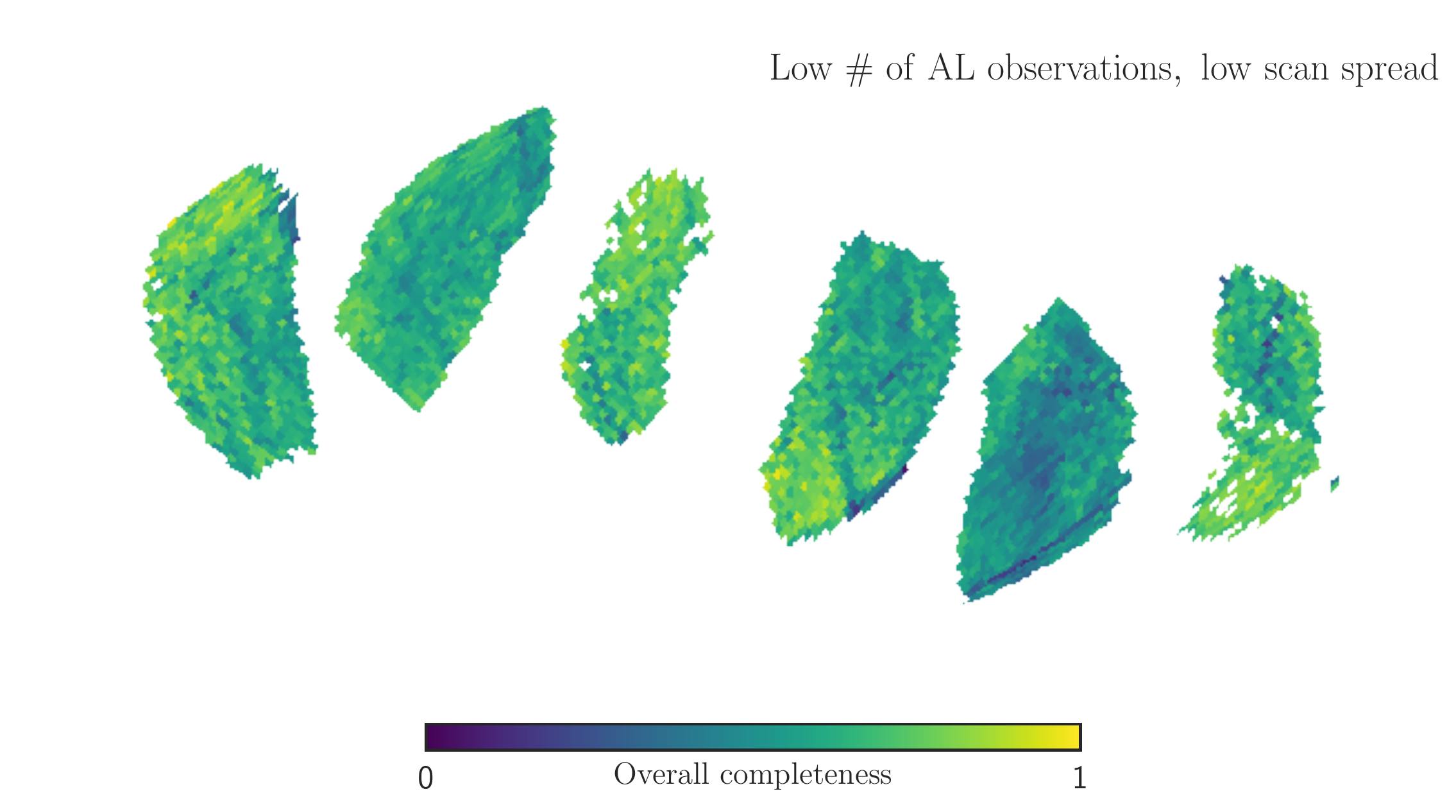}
  \includegraphics[width=0.49\textwidth,clip=]{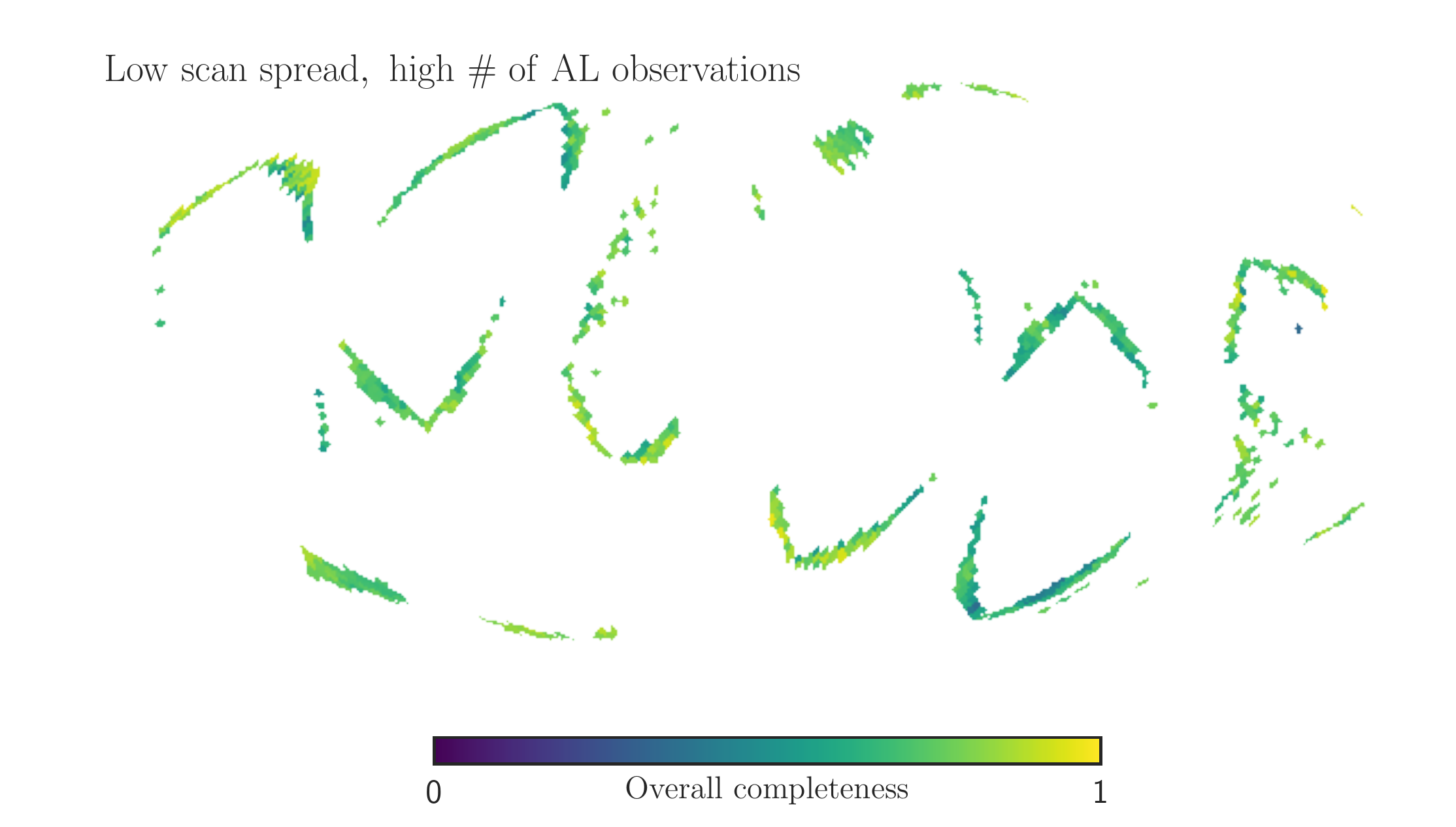}\\
  \includegraphics[width=0.49\textwidth,clip=]{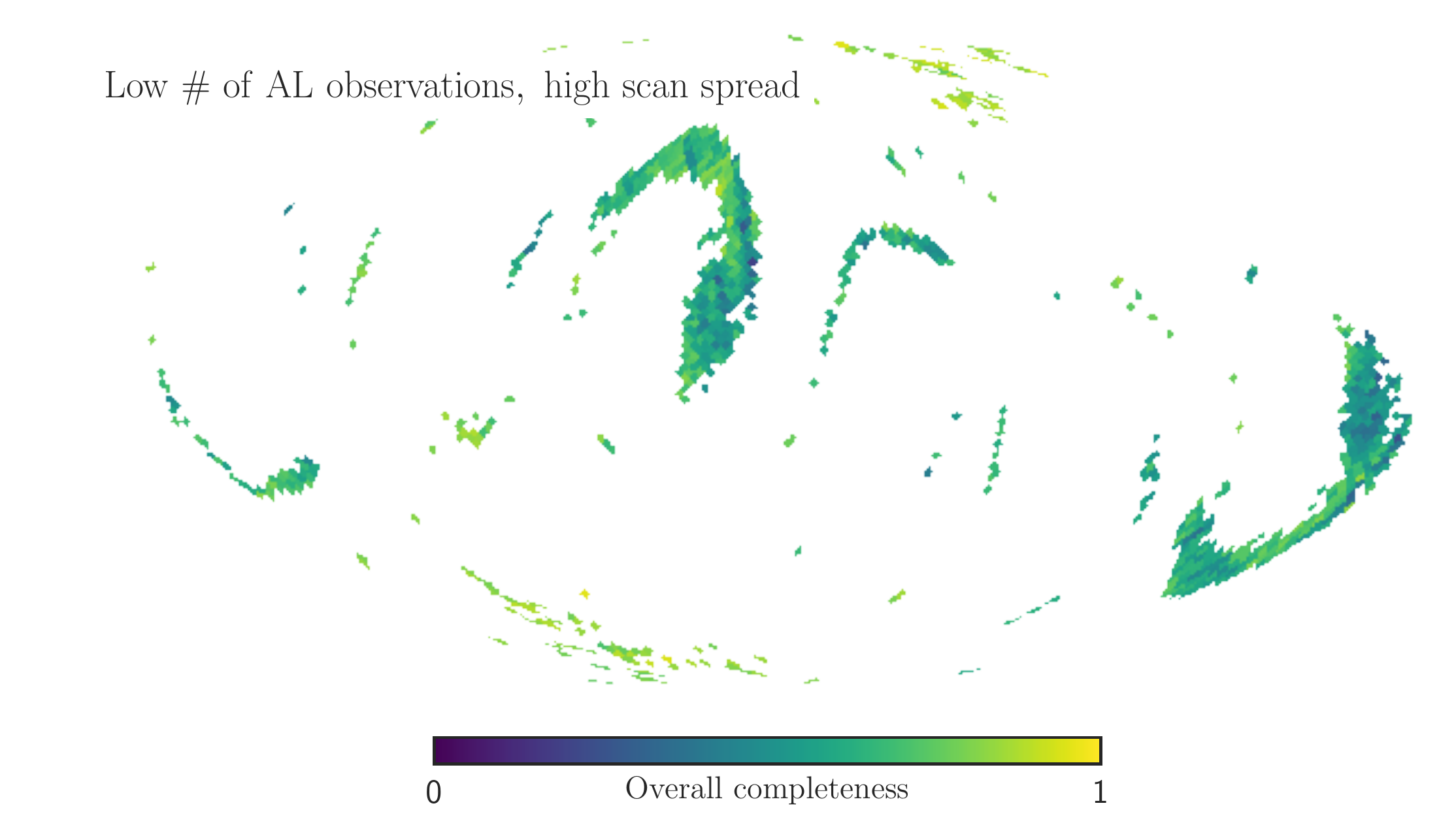}
  \includegraphics[width=0.49\textwidth,clip=]{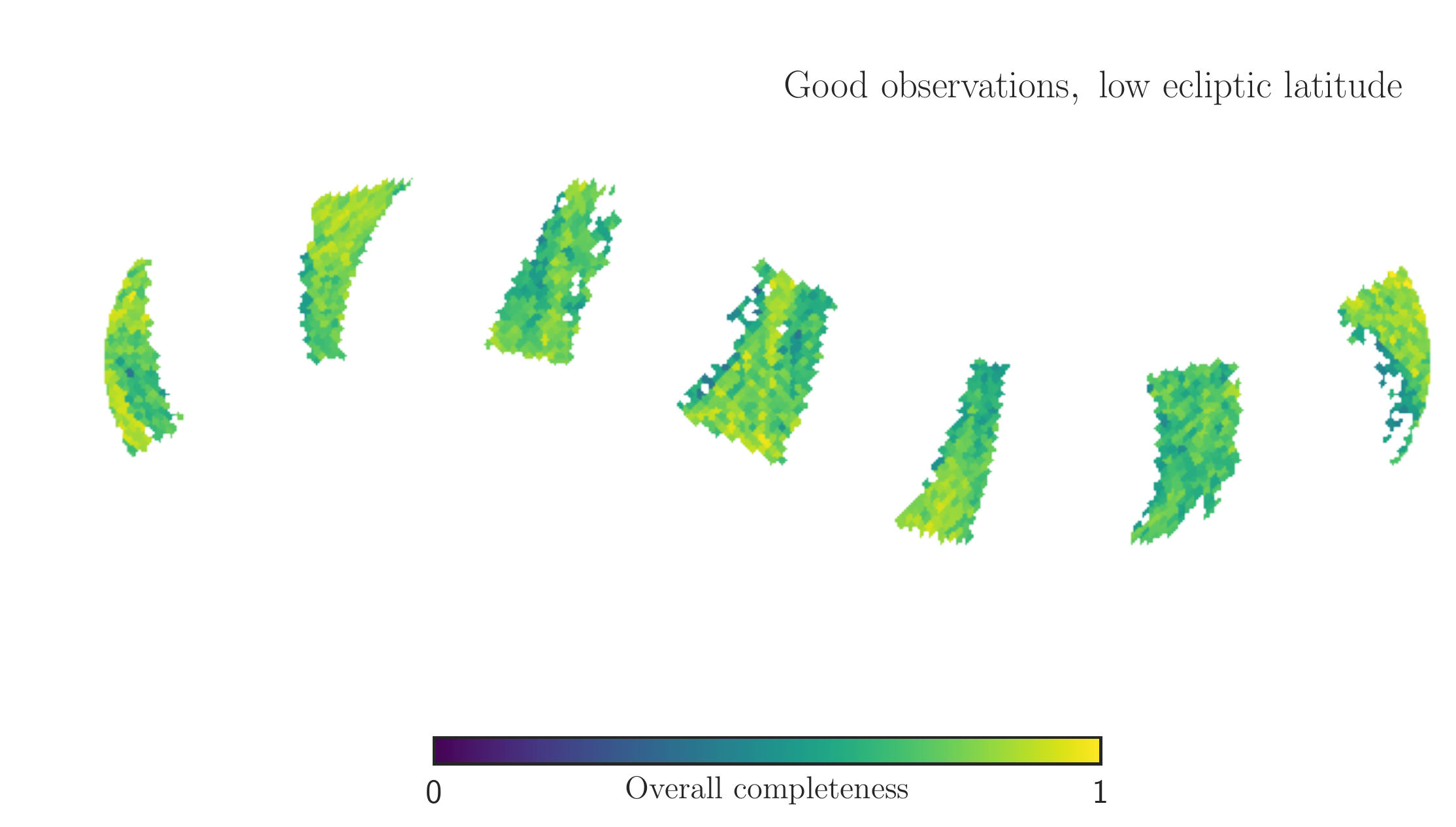}
  \caption{Overall completeness in the $\approx52\,\%$ of the sky that
    fails our \tgas\ observational quality cuts like in
    \figurename~\ref{fig:overall_completeness_bad}, but split into
    four categories of reasons why these regions fail our
    cuts.}\label{fig:overall_completeness_bad_split}
\end{figure*}

We present the overall completeness (see
\sectionname~\ref{sec:overall}) of the badly-observed part of the
\tgas\ sky in
\figurename~\ref{fig:overall_completeness_bad}. Comparing to the
overall completeness of the `good' part of the sky in
\figurename~\ref{fig:overall_completeness}, the completeness in this
badly-observed part is significantly lower and has artifacts due to
the scanning law. The overall completeness of this part of the sky is
also shown in \figurename~\ref{fig:overall_completeness_bad_split}
split into the four main reasons why a part of the sky is not included
in the `good' part.

\end{document}